\documentclass[10pt,conference]{IEEEtran}
\usepackage{nopageno}
\ifCLASSOPTIONcompsoc
  \usepackage[nocompress]{cite}
\else
  \usepackage{cite}
\fi
\ifCLASSINFOpdf
\else
\fi
\usepackage{booktabs}
\usepackage{color, colortbl}
\usepackage{comment}
\usepackage{amssymb}
\usepackage{xcolor}
\usepackage{tikz}
\usepackage{pgfplots}
\pgfplotsset{width=10cm,compat=1.9}
\usetikzlibrary{shapes.geometric, arrows}
\definecolor{mediumGreenAccentColor}{HTML}{A6A68A}
\definecolor{lightGreenAccentColor}{HTML}{B7B7A4}
\definecolor{neutralAccentColor}{HTML}{EDC9B9}

\tikzstyle{methodology} = [rectangle, rounded corners, minimum width=3cm, text width = 3cm, minimum height=1cm,text centered, draw=black, fill=mediumGreenAccentColor!95]

\tikzstyle{methodology2} = [rectangle, rounded corners, minimum width=3cm,  minimum height=.5cm,text centered, draw=black, fill=mediumGreenAccentColor!95]

\tikzstyle{rq} = [rectangle, rounded corners, minimum width=1cm, minimum height=.5cm,text centered, draw=black, fill=neutralAccentColor]

\tikzstyle{arrow} = [thick,->,>=stealth]

\usetikzlibrary{arrows.meta,
                chains,
                positioning}

\usepackage{fontawesome}
\usepackage{xcolor}
\usepackage{lookup}
\usepackage{xspace}
\usepackage{tcolorbox}

\renewenvironment{quote}{%
   \list{}{%
     \leftmargin\parindent%
     \rightmargin0cm
   }
   \item\relax
}
{\endlist}

\newcommand{\myquote}[2]{\begin{quote}{\small\faComment}~\emph{#1} (\lookupGet{#2})\end{quote}}

\newcommand{\myquotex}[2]{\emph{``#1''} (\lookupGet{#2})\xspace}

\definecolor{darkgreen}{rgb}{0.05,0.5,0.05}
\definecolor{darkred}{rgb}{0.5,0.05,0.05}

\newcommand{\tz}[1]{{\color{blue}\bfseries Tom: #1}}

\newcommand{\QteamproductivityWFH}{WFH-Q1\xspace}%
\newcommand{\Qculture}{WFH-Q2\xspace}%
\newcommand{\Qsupport}{WFH-Q3\xspace}%
\newcommand{\Qmilestones}{WFH-Q4\xspace}%

\newcommand{\Qteamproductivity}{Team-Q1\xspace}%

\newcommand{\Qcommunication}{Team-Q2\xspace}%
\newcommand{\Qsocialinteraction}{Team-Q3\xspace}%
\newcommand{\Qsocialconnection}{Team-Q4\xspace}%
\newcommand{\Qchallenges}{Team-Q5\xspace}%

\lookupPut{R_1OHnN0gHtZ5NExB}{S1}
\lookupPut{R_33y9pkd1I7Jza4i}{S2}
\lookupPut{R_2ZCeMk9I3uFJwtl}{S3}
\lookupPut{R_1oifNn9qpoJBduG}{S4}
\lookupPut{R_bOvsqPTQFtsd7qh}{S5}
\lookupPut{R_Abw3T6tSqYDDBYt}{S6}
\lookupPut{R_1gM7t4kxZMlPiO8}{S7}
\lookupPut{R_9uw8tsZnixPY9JT}{S8}
\lookupPut{R_3EWFC0ug16I5u8y}{S9}
\lookupPut{R_3qV9skIgB2q8lRN}{S10}
\lookupPut{R_3F3QRciov7uKVZZ}{S11}
\lookupPut{R_2WHV2QVp8qoSCCO}{S12}
\lookupPut{R_2bP55as0mVTcvHY}{S13}
\lookupPut{R_10uCOOaKJOFiRyU}{S14}
\lookupPut{R_3MEypirrAoPUz2C}{S15}
\lookupPut{R_1Hc6r0Wu5wYIkxf}{S16}
\lookupPut{R_1hZu5xgSHCKcC2D}{S17}
\lookupPut{R_2WHxa8AsCjqTkTp}{S18}
\lookupPut{R_1H8L3MvEUa35bPr}{S19}
\lookupPut{R_120fQj1PLHNLuSi}{S20}
\lookupPut{R_2WA4IrV0sdBFecH}{S21}
\lookupPut{R_3PgErXjGONoBYOV}{S22}
\lookupPut{R_cRTTtw7YFseddJv}{S23}
\lookupPut{R_0Vs4Tl07v9rKlIl}{S24}
\lookupPut{R_w01mZBxw2FEECdj}{S25}
\lookupPut{R_3OcYu1y4jOYSIR0}{S26}
\lookupPut{R_210P7MJiJb3B7QU}{S27}
\lookupPut{R_10PdyQsu1ZTsbI1}{S28}
\lookupPut{R_1jUHpGXGhrvFQCv}{S29}
\lookupPut{R_30dz0EPkl1DsN3q}{S30}
\lookupPut{R_2ONMwM11Hdd8rPf}{S31}
\lookupPut{R_1jfjvXu6t3fTmLz}{S32}
\lookupPut{R_1QiRAywQFUUQRv3}{S33}
\lookupPut{R_3NP3gtwtktGx98e}{S34}
\lookupPut{R_2qsWaeT6tAH2qPl}{S35}
\lookupPut{R_SGBAP3kaNCSyWUF}{S36}
\lookupPut{R_2rG9E0zWpaDlISD}{S37}
\lookupPut{R_211JiNZb2QY6Ssb}{S38}
\lookupPut{R_1kYoekMAZE9ZqH1}{S39}
\lookupPut{R_x49SG8KJszNeldT}{S40}
\lookupPut{R_1LGAmlKWWi2rhva}{S41}
\lookupPut{R_xa4cIZ3rocg6KNb}{S42}
\lookupPut{R_2zRShNegvRe0zFV}{S43}
\lookupPut{R_3itoB14akBzCHAV}{S44}
\lookupPut{R_3m9Lg9kP6oN0U9h}{S45}
\lookupPut{R_tLoll333O2zIuqt}{S46}
\lookupPut{R_2xUWZP08Ns4zZjI}{S47}
\lookupPut{R_1cUDoWs8PgmB9c1}{S48}
\lookupPut{R_2xFx8wcpPiIYjsG}{S49}
\lookupPut{R_5mCsRflDLrlje6J}{S50}
\lookupPut{R_0CH8VqOUhSTaxSV}{S51}
\lookupPut{R_28HfAQB0UsrspzX}{S52}
\lookupPut{R_PTngE1ocYzOvwK5}{S53}
\lookupPut{R_25Za5EnWH9FTpjZ}{S54}
\lookupPut{R_z7GcfYTfwfbNoCl}{S55}
\lookupPut{R_2ebq9T0XLsKGSB8}{S56}
\lookupPut{R_2ZQbO0BoTq0nhgO}{S57}
\lookupPut{R_OxnXH5ii0fdFpGV}{S58}
\lookupPut{R_2t41kH5vdJds7YV}{S59}
\lookupPut{R_vAp7uAxFnKV4Uj7}{S60}
\lookupPut{R_1n7Qe30YPZKJqQQ}{S61}
\lookupPut{R_27OI7AlAD8QsUXD}{S62}
\lookupPut{R_1JRU7OBiy6TSQyv}{S63}
\lookupPut{R_3Pq99eE9ApWwqUK}{S64}
\lookupPut{R_bCMmA2ojq2rJU2d}{S65}
\lookupPut{R_8e8TJvMoRGHpU6R}{S66}
\lookupPut{R_3CVC5kiZg97HiS0}{S67}
\lookupPut{R_1JUZXhGg1q5Qod9}{S68}
\lookupPut{R_qOAKZ3Zv20UZoit}{S69}
\lookupPut{R_1Kmdl1Xefz2oigS}{S70}
\lookupPut{R_Z4aqMIWVMlGv8w9}{S71}
\lookupPut{R_2wvlnpvv4z84JHd}{S72}
\lookupPut{R_1nNJ8mtUfLizyRq}{S73}
\lookupPut{R_3GuJlhq0C8fsNR5}{S74}
\lookupPut{R_2Y4IwZSHtw3AbVX}{S75}
\lookupPut{R_3nNTL6sh8hNMcDF}{S76}
\lookupPut{R_3ht1nexh0HsCD1o}{S77}
\lookupPut{R_3dMV4mZz22FA57w}{S78}
\lookupPut{R_3Enwttj1D3jMKjI}{S79}
\lookupPut{R_2yaBCAMiwX5Hj74}{S80}
\lookupPut{R_1Fxg84HfgSGDzP0}{S81}
\lookupPut{R_0BZDsbKP6S2IRDH}{S82}
\lookupPut{R_3qHUbde584tTzB0}{S83}
\lookupPut{R_3gS2ZvuD47rPOXi}{S84}
\lookupPut{R_2TpA0fz07j2b9Gy}{S85}
\lookupPut{R_21badzCb3cFpURa}{S86}
\lookupPut{R_bp7TYmME2Braf3X}{S87}
\lookupPut{R_3PjijPt09MewuCg}{S88}
\lookupPut{R_20NcPaf0piPSB7u}{S89}
\lookupPut{R_3LgOwjY1vpcpZiG}{S90}
\lookupPut{R_11hgCvrjsowD3JT}{S91}
\lookupPut{R_1LomokUncj84C7a}{S92}
\lookupPut{R_3MbOYLG7w2MSLLu}{S93}
\lookupPut{R_3r0e4W0wpbZakZs}{S94}
\lookupPut{R_bf70cokASuHnpuh}{S95}
\lookupPut{R_2CI0gziiqCEtBJi}{S96}
\lookupPut{R_27eeHa60jqC7yBh}{S97}
\lookupPut{R_2verN7XNFUYy9At}{S98}
\lookupPut{R_1rdGo8Ipaojwbwa}{S99}
\lookupPut{R_uf8u4xLHthxqYj7}{S100}
\lookupPut{R_3Eg6NECvohqb3OB}{S101}
\lookupPut{R_ULRux9ycMrOrD7X}{S102}
\lookupPut{R_10rpUlcMOlnHdSg}{S103}
\lookupPut{R_riqHrREuqhneBuV}{S104}
\lookupPut{R_1i3AGv6S0ma86sy}{S105}
\lookupPut{R_r0Aju2EeLYiCXyp}{S106}
\lookupPut{R_3PnL1Bkr061ckpD}{S107}
\lookupPut{R_1Dr6HH6TAaQt5cL}{S108}
\lookupPut{R_1QnLMrIJagyr6v2}{S109}
\lookupPut{R_zTCKFxGSn0nrM1r}{S110}
\lookupPut{R_302Z4QUvccnt4TQ}{S111}
\lookupPut{R_2UXWGhtgILwzYvw}{S112}
\lookupPut{R_UFHh7WQxXQlfFxT}{S113}
\lookupPut{R_1Kr2o5oswmRFpqd}{S114}
\lookupPut{R_3m8FUl9OApvByJW}{S115}
\lookupPut{R_2B9vtKR67EvP5LJ}{S116}
\lookupPut{R_3HGq3uiNMO6Rbk1}{S117}
\lookupPut{R_2EH8wMmv3XCk9g7}{S118}
\lookupPut{R_3KGAU1nVHFon6Pv}{S119}
\lookupPut{R_1ruCc3xPAND1U3v}{S120}
\lookupPut{R_3Li3FnWcODST7DD}{S121}
\lookupPut{R_1pEjGzgi6coOLNx}{S122}
\lookupPut{R_3GpjnM7MSRrlPZw}{S123}
\lookupPut{R_1qaqrulaeDZhWyY}{S124}
\lookupPut{R_SJAdb9ydetv2Drb}{S125}
\lookupPut{R_3svttzDggdlKDX9}{S126}
\lookupPut{R_2qsD7WtbZs1e8n8}{S127}
\lookupPut{R_RaVEQlp0UoVQzgl}{S128}
\lookupPut{R_3RrOh8fuc3L2hKd}{S129}
\lookupPut{R_1JLyqNuO4pDSCa7}{S130}
\lookupPut{R_4I8g29o1iSJ7P3j}{S131}
\lookupPut{R_21h8H9qw44DPLwN}{S132}
\lookupPut{R_2D665TO44JSp5gG}{S133}
\lookupPut{R_3EnWtTKFwmjVXY6}{S134}
\lookupPut{R_1LTi0wq8Eq3NvLq}{S135}
\lookupPut{R_3nIvXML64ks3Nep}{S136}
\lookupPut{R_1qfgx9Xn6PvJWph}{S137}
\lookupPut{R_3fknZx189IHOYvL}{S138}
\lookupPut{R_2zPbbFL6TXPF4Ba}{S139}
\lookupPut{R_3DhD0dsJfZukYvv}{S140}
\lookupPut{R_pQauv0WKAxHNCbD}{S141}
\lookupPut{R_3Pb2qJBZZk0ZMqd}{S142}
\lookupPut{R_2WGdJ65KIRAz47L}{S143}
\lookupPut{R_A56efmQ9QXIuPrH}{S144}
\lookupPut{R_231RnsXBimmIFnb}{S145}
\lookupPut{R_0jEcBjkri0lZ9bX}{S146}
\lookupPut{R_3EHeggNPuqfzHaZ}{S147}
\lookupPut{R_2WCiAt99bUuvD2V}{S148}
\lookupPut{R_2TFQm5DfDwlyWmV}{S149}
\lookupPut{R_2zdwXnA2QXYXpns}{S150}
\lookupPut{R_2DUFLf1LkF4lQLK}{S151}
\lookupPut{R_9uDJji4IW6qdjxL}{S152}
\lookupPut{R_1pEVRBpgDBTCknm}{S153}
\lookupPut{R_3QWxcgDqzLa1MZE}{S154}
\lookupPut{R_3kuNR7sQrtWCQag}{S155}
\lookupPut{R_29ogNb9dtdafTGg}{S156}
\lookupPut{R_2tpjlRCfBXXK5fr}{S157}
\lookupPut{R_2wmGqF5UYG7tOBV}{S158}
\lookupPut{R_3lLlUoebolhnyzL}{S159}
\lookupPut{R_23fviGKiGTUfOlB}{S160}
\lookupPut{R_DeLyhADY3hzm5dn}{S161}
\lookupPut{R_dbZl7zUsWwxliy5}{S162}
\lookupPut{R_2BhZ9Kj4JryVtCd}{S163}
\lookupPut{R_qQoRYqxuIzdKX6h}{S164}
\lookupPut{R_1nSDB9bZVZl5o1M}{S165}
\lookupPut{R_3GrGwJ7mtQXl87x}{S166}
\lookupPut{R_1GK1Oumc14l3YA2}{S167}
\lookupPut{R_1LCO5uDKRbhHizD}{S168}
\lookupPut{R_2zGbBEbu6wasQqd}{S169}
\lookupPut{R_WcBUA6vjkbFiepb}{S170}
\lookupPut{R_1pSwEx0RH9KSFiW}{S171}
\lookupPut{R_2wauvOi8uIY1vir}{S172}
\lookupPut{R_beZKDu0qxS3bIuB}{S173}
\lookupPut{R_5auHQmw1eL8xrih}{S174}
\lookupPut{R_3fl6a3ympqOG93S}{S175}
\lookupPut{R_11hlLwO6uRs8wyY}{S176}
\lookupPut{R_qKkTI9Lxndt0ak1}{S177}
\lookupPut{R_ctJ29NUtz58zKYF}{S178}
\lookupPut{R_1nOs9PZ1CvRb5ne}{S179}
\lookupPut{R_2SxiJzSnP42aLr6}{S180}
\lookupPut{R_3NDT1jB6A9F93tT}{S181}
\lookupPut{R_2CUJTsuDbTt1W06}{S182}
\lookupPut{R_3RkZJPHKBVPzxti}{S183}
\lookupPut{R_2dviHArK8JN8l31}{S184}
\lookupPut{R_1N3U6dImnhnqO5P}{S185}
\lookupPut{R_3IWVY6XGxaJlSfG}{S186}
\lookupPut{R_3n7MhsW7celc06Q}{S187}
\lookupPut{R_2U9nzTCerJgs9v7}{S188}
\lookupPut{R_1OpElAGVXo9FgIG}{S189}
\lookupPut{R_22WhClZEuUEkBZ9}{S190}
\lookupPut{R_3jflgjdqwKInfus}{S191}
\lookupPut{R_vJr5jg3JzUaFI0F}{S192}
\lookupPut{R_1ojdWtizURNzUSg}{S193}
\lookupPut{R_27eQUtlmA1T0VWV}{S194}
\lookupPut{R_10OA7HVhQnvAGFg}{S195}
\lookupPut{R_1EXTSgorpUjf6uW}{S196}
\lookupPut{R_qyAaaGx4hb0yNIR}{S197}
\lookupPut{R_1j9hejdtmeEppak}{S198}
\lookupPut{R_3FKBu70y9ciVA4Q}{S199}
\lookupPut{R_1r39DdOLDnAXXkm}{S200}
\lookupPut{R_31vDule4VymybEA}{S201}
\lookupPut{R_1hR95FO5WeFFQ2N}{S202}
\lookupPut{R_1feDadfBro102EL}{S203}
\lookupPut{R_2UX7TNySgLhXBUL}{S204}
\lookupPut{R_1LSEH3ZSxl1oMaa}{S205}
\lookupPut{R_2EsYmVTbeYkNCBe}{S206}
\lookupPut{R_1PZ8obwHoCVjLqk}{S207}
\lookupPut{R_338epmQsMaATHC7}{S208}
\lookupPut{R_1JX7H4rVGC53K0V}{S209}
\lookupPut{R_1obePSq0FyIDnBQ}{S210}
\lookupPut{R_cMuIG5C4xFxiW7T}{S211}
\lookupPut{R_6fCbZANc7uNWgkp}{S212}
\lookupPut{R_28IC9Lnsoy51kwM}{S213}
\lookupPut{R_1HpfVzXoEmHF8WF}{S214}
\lookupPut{R_3h6eYx3lnsNTOis}{S215}
\lookupPut{R_2YVYB23FxUByu6O}{S216}
\lookupPut{R_2cedzZ2EPiOUwJX}{S217}
\lookupPut{R_1DSwqAQ6CkrJrHx}{S218}
\lookupPut{R_3Rmd5itds88VzCT}{S219}
\lookupPut{R_3hiHEKSsLUE7qv3}{S220}
\lookupPut{R_3RrR6MCNqHJ066i}{S221}
\lookupPut{R_W8X8l6nwOmeoJk5}{S222}
\lookupPut{R_3KvyhDrbVrdC49q}{S223}
\lookupPut{R_e8nl3VzxYGFxbDb}{S224}
\lookupPut{R_21cyPYwvLlKQOkq}{S225}
\lookupPut{R_zcA5zNJtypUBb4l}{S226}
\lookupPut{R_22Qh12d0p2EqIZh}{S227}
\lookupPut{R_30jEDhfFdgcom52}{S228}
\lookupPut{R_XLDmyxIkkwU8Xu1}{S229}
\lookupPut{R_2EveDQAzgymoeJE}{S230}
\lookupPut{R_T5jS1Ee9dWfqbjb}{S231}
\lookupPut{R_p4tFScS1BtYeYxj}{S232}
\lookupPut{R_3pkZ11pGL510czY}{S233}
\lookupPut{R_2EfP60a1rJu0KzO}{S234}
\lookupPut{R_x9EWmOoSOVhHDjz}{S235}
\lookupPut{R_31pxFUVF4qofpJF}{S236}
\lookupPut{R_24310tiTaPbD2ba}{S237}
\lookupPut{R_1eQ7xjKWsVejcMK}{S238}
\lookupPut{R_1IXH37MOr4mNpEj}{S239}
\lookupPut{R_1fdTlHWhcmOnRww}{S240}
\lookupPut{R_3kEOIxzTybXIQQO}{S241}
\lookupPut{R_ONWL3mLsHp5zuFP}{S242}
\lookupPut{R_3CIWGCsyonXCPTT}{S243}
\lookupPut{R_sBF52s4M9ZqnSW5}{S244}
\lookupPut{R_3EoqhAUqyTecCn6}{S245}
\lookupPut{R_26goonFOu7OY7bd}{S246}
\lookupPut{R_1f9Lsepl4J2XhQH}{S247}
\lookupPut{R_2SueEOXRTV3OPGI}{S248}
\lookupPut{R_2cv0ixyXFVNY7FY}{S249}
\lookupPut{R_vxEY2YQ0lqsyKFr}{S250}
\lookupPut{R_2AZjxz1NOdp4kSR}{S251}
\lookupPut{R_29cX3PQtKA6Mssj}{S252}
\lookupPut{R_2wc5aEHA1aeo79U}{S253}
\lookupPut{R_3CPNnbi4SLuwdKS}{S254}
\lookupPut{R_2B5nrPwDJCMPqya}{S255}
\lookupPut{R_2wuTnPSSwMjIIXa}{S256}
\lookupPut{R_1kSmCfQ5LnznQi1}{S257}
\lookupPut{R_2BsuAlq95JEGWOt}{S258}
\lookupPut{R_1Ej4G0TrBFgddz7}{S259}
\lookupPut{R_3saKrnXYggeasN6}{S260}
\lookupPut{R_2xCQtJWLDUekAfO}{S261}
\lookupPut{R_XY3EdrfMKxyrazD}{S262}
\lookupPut{R_1IXCFy45YOmx2ZQ}{S263}
\lookupPut{R_ABZeMOsUeUcPsOZ}{S264}
\lookupPut{R_cvYgflYjkL7JYZj}{S265}
\lookupPut{R_1FKdNvf1TtRRX2A}{S266}
\lookupPut{R_2QujNx6qs0EJ04e}{S267}
\lookupPut{R_3Hn16rxfcLIx5Vu}{S268}
\lookupPut{R_1RM7MLpFWM4T5MR}{S269}
\lookupPut{R_SOfASkf9Na3gJzj}{S270}
\lookupPut{R_22yHzX5ATtrAEY4}{S271}
\lookupPut{R_12JeC1IoCcwYLxa}{S272}
\lookupPut{R_248XjqYZfSG5sp3}{S273}
\lookupPut{R_1eUx1yUqdDdhBXk}{S274}
\lookupPut{R_20TFRuG8dcPHeG1}{S275}
\lookupPut{R_tQis0ZVuGQauYZb}{S276}
\lookupPut{R_3HYdxC1m1T86ked}{S277}
\lookupPut{R_3dX5EIlD79rHAtg}{S278}
\lookupPut{R_2bP3vrez2bwUwjA}{S279}
\lookupPut{R_25TlUzkZLx2JQ6z}{S280}
\lookupPut{R_1FFyFoRbInpUtEn}{S281}
\lookupPut{R_12bCckB2BrRXapt}{S282}
\lookupPut{R_vJQvwUbJbByRyjT}{S283}
\lookupPut{R_295NM9aVWbi5cze}{S284}
\lookupPut{R_25vCpD1WcAjBADF}{S285}
\lookupPut{R_xhcG8auRDHlKAcV}{S286}
\lookupPut{R_3CH3CjJ4w8fVGKh}{S287}
\lookupPut{R_1PXIxgC6of7XP3P}{S288}
\lookupPut{R_2S6upGEVoekDGLH}{S289}
\lookupPut{R_2QZ2iGaHKH7loL5}{S290}
\lookupPut{R_1hXwoqgfZeoOix1}{S291}
\lookupPut{R_3s1faWofXS6k0EC}{S292}
\lookupPut{R_28Y6CabgFz6cz64}{S293}
\lookupPut{R_2bIvBGWrJuNTWlz}{S294}
\lookupPut{R_XnCJVDYt4gDtDQR}{S295}
\lookupPut{R_10I7Z8FvmRSzf1d}{S296}
\lookupPut{R_29hx0SiRg8rL9sM}{S297}
\lookupPut{R_2E06kb0uaH4Rc7F}{S298}
\lookupPut{R_2YJLnnIbKWknXvK}{S299}
\lookupPut{R_2SDefwqoHkymRhh}{S300}
\lookupPut{R_RxI8nIXGtyhfKh3}{S301}
\lookupPut{R_3QDwhhNVKpcb1ir}{S302}
\lookupPut{R_3n3l5E8Pt4kxfnq}{S303}
\lookupPut{R_1igSy5ONT1vpjg5}{S304}
\lookupPut{R_W2vwQP6oo67092x}{S305}
\lookupPut{R_2uyw8q3mwZ3GYtq}{S306}
\lookupPut{R_20Zsy4kzhkS3gib}{S307}
\lookupPut{R_3sv1TpiBjJGXq6R}{S308}
\lookupPut{R_3CUCNFnApYr9dAz}{S309}
\lookupPut{R_e4jZm1PIxulkb0l}{S310}
\lookupPut{R_3JE20ZZpLFGQmWT}{S311}
\lookupPut{R_WqBIvU0vmTJuH3b}{S312}
\lookupPut{R_3MfKgf11sNsURNJ}{S313}
\lookupPut{R_3elE4Y3nmJFGmhW}{S314}
\lookupPut{R_3EXamuwdHoa6z2Z}{S315}
\lookupPut{R_2axXpay0ZzUc7Xr}{S316}
\lookupPut{R_zUKOyhPQnNZa9Wx}{S317}
\lookupPut{R_2D63IoQooRMyvum}{S318}
\lookupPut{R_3ISMmKxaY4oFs8M}{S319}
\lookupPut{R_2TXy7WlMsotIYnj}{S320}
\lookupPut{R_1E4lPJAu0bt1XRC}{S321}
\lookupPut{R_1ezMGNXmCVgiIHI}{S322}
\lookupPut{R_1eRz8YtCe0Dmjst}{S323}
\lookupPut{R_oZBlfXwxmWlGIcp}{S324}
\lookupPut{R_28TGF7KNPzKeNK7}{S325}
\lookupPut{R_3FKiCJEcU3eihii}{S326}
\lookupPut{R_0rFZsZYKl232gcF}{S327}
\lookupPut{R_1GN6xGneXr0cV9t}{S328}
\lookupPut{R_1rJ3zxj9dnhf8x4}{S329}
\lookupPut{R_3inc8OgdUO08fwX}{S330}
\lookupPut{R_NWm6tOtkbD0PZC1}{S331}
\lookupPut{R_2fCnimXT87UbjAg}{S332}
\lookupPut{R_2w62agTiAiuw4tm}{S333}
\lookupPut{R_2ATFCNWWMiqSb3l}{S334}
\lookupPut{R_2zBeRyDhEv52idY}{S335}
\lookupPut{R_Or2KFbWMnMS6dCp}{S336}
\lookupPut{R_2cCyMykqjXdUQaM}{S337}
\lookupPut{R_2sYED5Mq9BxcDPi}{S338}
\lookupPut{R_239Q23zT2y6ANhb}{S339}
\lookupPut{R_3Do6SvomWSU5lQa}{S340}
\lookupPut{R_2rpU8TXjeMNe3QZ}{S341}
\lookupPut{R_3CJW1om6qL8PSX1}{S342}
\lookupPut{R_2Qm7NHFYNs2xJCc}{S343}
\lookupPut{R_1mlXQSPKWQubc2b}{S344}
\lookupPut{R_3j8SOLAjutFWYuJ}{S345}
\lookupPut{R_SPMV0kS5hPEXOeZ}{S346}
\lookupPut{R_vJ3LEd1q7nS4mJj}{S347}
\lookupPut{R_tLEhD7Gy213LOLL}{S348}
\lookupPut{R_0DMQbd8vzY7lXB7}{S349}
\lookupPut{R_3gT162w9yS1m6ne}{S350}
\lookupPut{R_WjkwjskkwhUFbtD}{S351}
\lookupPut{R_XUnqaVZEJmSJmYp}{S352}
\lookupPut{R_3irudQWdtE0DqIz}{S353}
\lookupPut{R_3G3XRso28hpXYa9}{S354}
\lookupPut{R_1hQBr6T5vnF1o5J}{S355}
\lookupPut{R_T1wzdpqwThvCyiJ}{S356}
\lookupPut{R_10oMkU0eEhIc24o}{S357}
\lookupPut{R_qV0ljCQsUjmRwQ1}{S358}
\lookupPut{R_3OdBH85UCIo6SmL}{S359}
\lookupPut{R_sp9a9FqXEw4LdVD}{S360}
\lookupPut{R_PYDixW5jn0mB185}{S361}
\lookupPut{R_29s19BKCLUxbXwQ}{S362}
\lookupPut{R_2S7hV8Ar5ShA54P}{S363}
\lookupPut{R_8HaXIHisZc6wA7f}{S364}
\lookupPut{R_1i58SR7NWSgS47e}{S365}
\lookupPut{R_3iPhXeWLZpzHHS7}{S366}
\lookupPut{R_uk14Sa7L9zs28hz}{S367}
\lookupPut{R_XChQgbVcdUGs6yt}{S368}
\lookupPut{R_3dQcoX0i3uOSZrU}{S369}
\lookupPut{R_2rCIiWPpsMTo7eI}{S370}
\lookupPut{R_3iXxydCh3xU7OfH}{S371}
\lookupPut{R_D8AednmjoSIHkB3}{S372}
\lookupPut{R_28ZMmEagctxTkfO}{S373}
\lookupPut{R_cPcVcrqrIoKPtSx}{S374}
\lookupPut{R_3PEMKHKGLtkbR5B}{S375}
\lookupPut{R_QhbHVdELy9H0BSV}{S376}
\lookupPut{R_1F9cGz4Ew96ZSSk}{S377}
\lookupPut{R_BCZtqrk84QvkjGF}{S378}
\lookupPut{R_1jiyts0M7YU8JSg}{S379}
\lookupPut{R_1Clv6YfDqOb54yL}{S380}
\lookupPut{R_10u6MzdvcEU7yjA}{S381}
\lookupPut{R_2vlfrOi3TyXDQbs}{S382}
\lookupPut{R_rkGXZRSdpbWd4Ah}{S383}
\lookupPut{R_88hJS2d0kILpxhD}{S384}
\lookupPut{R_DjJd2NOuACDhDUt}{S385}
\lookupPut{R_2zDguqX3mjeE3Sp}{S386}
\lookupPut{R_SDx7R6VBeOIWcaB}{S387}
\lookupPut{R_3MSizZGUTET8IgH}{S388}
\lookupPut{R_2xDDjQWVMmVgKUv}{S389}
\lookupPut{R_2SGZiJCFZwFq2dS}{S390}
\lookupPut{R_1pzIqwHXkRG79zn}{S391}
\lookupPut{R_SZaywyvizQ4WSop}{S392}
\lookupPut{R_3EantaU99KZehaL}{S393}
\lookupPut{R_1eql0sb4GmFjrgH}{S394}
\lookupPut{R_2PpBxkhNS7wUKvF}{S395}
\lookupPut{R_PUQJqlOOCHYofp7}{S396}
\lookupPut{R_1mDHqlhVOHEYxgf}{S397}
\lookupPut{R_30f1o42VBU72o8q}{S398}
\lookupPut{R_voCstpLP1jitTZT}{S399}
\lookupPut{R_3fE5buZtKzfzsjw}{S400}
\lookupPut{R_pggE7vKO5ilv08F}{S401}
\lookupPut{R_3lYATwcrSFzXiIE}{S402}
\lookupPut{R_1IZfsImA9frASz6}{S403}
\lookupPut{R_3emYpct7xfR5e3R}{S404}
\lookupPut{R_2ZEoXLavAmeSpqj}{S405}
\lookupPut{R_1JWvvwtCHkhNcqs}{S406}
\lookupPut{R_1oz0ORGFMoy2Ifb}{S407}
\lookupPut{R_3OjJ71Gxw8mwaYP}{S408}
\lookupPut{R_3MrqvR7DBP9u6OR}{S409}
\lookupPut{R_XBYxxrWg7j0OHgl}{S410}
\lookupPut{R_siXIs3KoEFyKdYB}{S411}
\lookupPut{R_1k1sTcwTool0Eh4}{S412}
\lookupPut{R_1LCO4rK8MFAr9zE}{S413}
\lookupPut{R_5vgBRUN1z8sdfeV}{S414}
\lookupPut{R_20OkYzNEfWjhdpw}{S415}
\lookupPut{R_3nJbJ2DtBRRCKbG}{S416}
\lookupPut{R_2rTv0S0l8TczeW6}{S417}
\lookupPut{R_1OZcAs3mAX9sy3W}{S418}
\lookupPut{R_1FafGLmPrb6hPZT}{S419}
\lookupPut{R_2VwQCgiDG4z3dZx}{S420}
\lookupPut{R_2wH0eYn94HDCh8D}{S421}
\lookupPut{R_1OW8KvHIhkmIP4u}{S422}
\lookupPut{R_3eLC780CvQ9S7lv}{S423}
\lookupPut{R_2t9ZKcoOl87rL2D}{S424}
\lookupPut{R_3GvTqepKtJmRM1a}{S425}
\lookupPut{R_29giT3b2X0ZFS3H}{S426}
\lookupPut{R_2V3GnVuwuJwbAUQ}{S427}
\lookupPut{R_2BfjKm32MDoSfz5}{S428}
\lookupPut{R_RDdudnxbXemd0ch}{S429}
\lookupPut{R_3CH3glvVOqUzf0l}{S430}
\lookupPut{R_31aNpGdGIK8WtPm}{S431}
\lookupPut{R_3rOFSkST3S0GmBJ}{S432}
\lookupPut{R_UstN3AmSXMhXkSl}{S433}
\lookupPut{R_OOl0W3K1Bjn1v4R}{S434}
\lookupPut{R_3NTZmmHyqvfIUcK}{S435}
\lookupPut{R_2X5Pdss7KcUtkPa}{S436}
\lookupPut{R_b7ORPwd9nBUs1G1}{S437}
\lookupPut{R_3EB7FF1nJXrVKfG}{S438}
\lookupPut{R_cwsL3oZOznKC7It}{S439}
\lookupPut{R_2ZQ75VEBCXYAaWm}{S440}
\lookupPut{R_25Nlv9Z3ePZUtBl}{S441}
\lookupPut{R_2DUYgWfh5kLaho9}{S442}
\lookupPut{R_2wum3x9E2I65U26}{S443}
\lookupPut{R_2S6rusx0XXRgWX7}{S444}
\lookupPut{R_2zp7CYGxZrfZXb7}{S445}
\lookupPut{R_110ZQ00ZcBBP1bX}{S446}
\lookupPut{R_1oirfRMila6h9kg}{S447}
\lookupPut{R_1N3sfoxwgsskPcc}{S448}
\lookupPut{R_10vhxBkC2mRJPgc}{S449}
\lookupPut{R_1msbb25NLzyH5oe}{S450}
\lookupPut{R_3p2LxV5bfLo5ekk}{S451}
\lookupPut{R_4SinwfYl10xBavv}{S452}
\lookupPut{R_pLV2r2toLZ1yksV}{S453}
\lookupPut{R_OkVvfr3mp8g354Z}{S454}
\lookupPut{R_24iWydeMXEBYohd}{S455}
\lookupPut{R_3EolbtJZLnQBbP7}{S456}
\lookupPut{R_UfoXz8HvwnfEEz7}{S457}
\lookupPut{R_24I9OiQ4PhUExzv}{S458}
\lookupPut{R_2zZx8ioliOto2BT}{S459}
\lookupPut{R_1EZNfHnDMGb4aad}{S460}
\lookupPut{R_1OIhlzeMjTwtjnU}{S461}
\lookupPut{R_2doshYoTmMouGfj}{S462}
\lookupPut{R_1LTZPRM0wUNU6Bz}{S463}
\lookupPut{R_2tzXQUVNEVIbzS9}{S464}
\lookupPut{R_32K1gZlVQJTbTvH}{S465}
\lookupPut{R_2X1E29BrNeuMDOv}{S466}
\lookupPut{R_1cTIzfogKbSUKXX}{S467}
\lookupPut{R_21AHksV3F19tygb}{S468}
\lookupPut{R_31vp4EtqmHBp3XE}{S469}
\lookupPut{R_9Wv5EAqhM1DK58d}{S470}
\lookupPut{R_2B4h1mrGQo8ns6M}{S471}
\lookupPut{R_3pcoepCsTtaKB2O}{S472}
\lookupPut{R_1pohJEvNGGFC9t6}{S473}
\lookupPut{R_1g5g1QRITSLOSan}{S474}
\lookupPut{R_vZY36aTuLeMLUgF}{S475}
\lookupPut{R_2DTzfUxstsK40SD}{S476}
\lookupPut{R_Ug54UdUVQe5zEfT}{S477}
\lookupPut{R_zVxkemCQ81KlJYt}{S478}
\lookupPut{R_1N96p4Lu0vqNK5W}{S479}
\lookupPut{R_bfJjkvM1XV1zfJT}{S480}
\lookupPut{R_3MrDKbiIWsjetTB}{S481}
\lookupPut{R_324LZSMKmlOXso1}{S482}
\lookupPut{R_2VswmHPSu4oCieP}{S483}
\lookupPut{R_2wRHulKda52ODAk}{S484}
\lookupPut{R_2wtzk61CfsahDUD}{S485}
\lookupPut{R_PHizQlN9U6vyheF}{S486}
\lookupPut{R_u7YjqkAcYfb6TVD}{S487}
\lookupPut{R_5iEwXAhXt59vLB7}{S488}
\lookupPut{R_3CJwZUY1Ki8y0GK}{S489}
\lookupPut{R_1Qhd8nptj4h4a1g}{S490}
\lookupPut{R_3lEr3d44WhthkyZ}{S491}
\lookupPut{R_1dL8SAZV5kFePci}{S492}
\lookupPut{R_31MWksgbIcxtoAN}{S493}
\lookupPut{R_3MMBCWnidhPMT8f}{S494}
\lookupPut{R_2XjU3VxxlZ8briL}{S495}
\lookupPut{R_6nZygJPUGwFMzuh}{S496}
\lookupPut{R_3Eu6Cp87vNIBGX9}{S497}
\lookupPut{R_3kAhZwgU6laoIAw}{S498}
\lookupPut{R_2YA2VpyVDkk2Qiz}{S499}
\lookupPut{R_2wtaAtGQXi705eG}{S500}
\lookupPut{R_Qix8iYO3NCBhu93}{S501}
\lookupPut{R_1QMbfPJ593h7qJJ}{S502}
\lookupPut{R_1JFPfrYm5G02M6Y}{S503}
\lookupPut{R_3rUitWl3lQgRr7d}{S504}
\lookupPut{R_1FE9G37qbA01nrh}{S505}
\lookupPut{R_5uIudLcLzmuZRpn}{S506}
\lookupPut{R_2V2vqd5FRE9TkQy}{S507}
\lookupPut{R_1BRiUz5ZPELSj5k}{S508}
\lookupPut{R_3h0BAiNBtZAYW4W}{S509}
\lookupPut{R_2bN7jLPZugwSlYZ}{S510}
\lookupPut{R_3Poe8oKLigmKTwv}{S511}
\lookupPut{R_2QSd0kFyA12RUhH}{S512}
\lookupPut{R_2B8m1I0MtWxMZWH}{S513}
\lookupPut{R_3s6YLLKrlYkfGmP}{S514}
\lookupPut{R_3rSsfKSYveJw6al}{S515}
\lookupPut{R_vliafKLNQYfi5Nv}{S516}
\lookupPut{R_2Wur8hw1cYYc8hd}{S517}
\lookupPut{R_UKONZQW65OP6c9j}{S518}
\lookupPut{R_2fEfaVLYJMq34Ai}{S519}
\lookupPut{R_wNX2OKYsJ4LKHex}{S520}
\lookupPut{R_3fI8np42kf4DkGn}{S521}
\lookupPut{R_20MO825nNu5q01C}{S522}
\lookupPut{R_QfwTOa6Esc5gmMF}{S523}
\lookupPut{R_1jBrtRIlPUWKX4l}{S524}
\lookupPut{R_1Fna59mqGoYYy7O}{S525}
\lookupPut{R_1eCZUkmJYe639nu}{S526}
\lookupPut{R_rlJ3IIKJVZQODVT}{S527}
\lookupPut{R_1hKz5LY15PpIXq4}{S528}
\lookupPut{R_25zQvFSfy8bnBSj}{S529}
\lookupPut{R_1QfulLIvfrXFqFd}{S530}
\lookupPut{R_22Rp79JApM8tR5I}{S531}
\lookupPut{R_27jtUdvka1MBuz5}{S532}
\lookupPut{R_2aaIa5oiO646hEw}{S533}
\lookupPut{R_2eOtHD99RNN2b6N}{S534}
\lookupPut{R_1cUOqa27SrcuQHG}{S535}
\lookupPut{R_32V18D8NtoJFybE}{S536}
\lookupPut{R_5tqqzpPHhLTvsf7}{S537}
\lookupPut{R_1DSa5pfPxwz2KQM}{S538}
\lookupPut{R_2uIihWGnTxkX0vs}{S539}
\lookupPut{R_BWF6SuRLmKWrIvT}{S540}
\lookupPut{R_12rIq3LSClmBQZS}{S541}
\lookupPut{R_7TXxadS0t2a4afD}{S542}
\lookupPut{R_1IusNXzYh5HupC5}{S543}
\lookupPut{R_10xf0ORnFOyDO5r}{S544}
\lookupPut{R_3j0QxyP0hUwzbob}{S545}
\lookupPut{R_WiKER04AR4imEeZ}{S546}
\lookupPut{R_DoWeANhoUQfXqyR}{S547}
\lookupPut{R_2fqwqx1wWzpg6i7}{S548}
\lookupPut{R_1KwosxB2o2iTPuo}{S549}
\lookupPut{R_1ifizdyzMgZy1bf}{S550}
\lookupPut{R_25uooa6Dy1gQ7fw}{S551}
\lookupPut{R_3oLL25zC2xKiY8z}{S552}
\lookupPut{R_3HHsmPYfNXzlMeq}{S553}
\lookupPut{R_3ltAR7LPW36kIJo}{S554}
\lookupPut{R_9BLPyk8t49M3ljX}{S555}
\lookupPut{R_1gi46BEt0ARoZFM}{S556}
\lookupPut{R_3rNX0AumURuWDyN}{S557}
\lookupPut{R_RmLb3c9EAnsGSkN}{S558}
\lookupPut{R_3qEbevrhXSamSkn}{S559}
\lookupPut{R_PFosBIWpxUnTPOh}{S560}
\lookupPut{R_2CN8UAQD5tBcEoX}{S561}
\lookupPut{R_1mQTkgm3yeAquWI}{S562}
\lookupPut{R_1fkBIiyWkGPyBeJ}{S563}
\lookupPut{R_6WPHuVnkYKDkz7z}{S564}
\lookupPut{R_dmqQqqqW50lIArD}{S565}
\lookupPut{R_wNsdpGWBYDbUMO5}{S566}
\lookupPut{R_3GBBITWUeqgFpHD}{S567}
\lookupPut{R_3nIP3O6nMw0VTdi}{S568}
\lookupPut{R_DGLNtnf0BBHvuaR}{S569}
\lookupPut{R_11ZoZi0f2liQ2HD}{S570}
\lookupPut{R_wO77pEjZdmSuC5P}{S571}
\lookupPut{R_1rHWQTfspk92SxO}{S572}
\lookupPut{R_1MS9IX9MOSCY8GT}{S573}
\lookupPut{R_1jUkiDYUAI1FfPf}{S574}
\lookupPut{R_3mdTzKARAzifZMB}{S575}
\lookupPut{R_6yy1GJLTNEm2PLP}{S576}
\lookupPut{R_bvzddb7l9RFyaDT}{S577}
\lookupPut{R_3rJwEFyx55anlik}{S578}
\lookupPut{R_1C2yFN6LpwHEiAM}{S579}
\lookupPut{R_3nw8VhKQOS87V2G}{S580}
\lookupPut{R_2SiEplq7NwDlUzY}{S581}
\lookupPut{R_1PTbfdOCiGgbKf6}{S582}
\lookupPut{R_2Yy9rU1NScLmveQ}{S583}
\lookupPut{R_2ql7WIta0FCRbBP}{S584}
\lookupPut{R_3IRV2ab5m8unJYK}{S585}
\lookupPut{R_3KvDUMUAH66ROZq}{S586}
\lookupPut{R_qFlCIppiKKp4Bjj}{S587}
\lookupPut{R_3lQB53b4Sj3AEO9}{S588}
\lookupPut{R_2TXHtd5gQsKtTnE}{S589}
\lookupPut{R_2WtZuvc3wvYcqGn}{S590}
\lookupPut{R_3KC6qiaVujTGGJr}{S591}
\lookupPut{R_1jSYEtjODsUGsKP}{S592}
\lookupPut{R_PS4J8NI0jvkV1Bv}{S593}
\lookupPut{R_SK9nhAkX9vngYYF}{S594}
\lookupPut{R_yElwoD6cWynGEXD}{S595}
\lookupPut{R_42QcbxNUNHSxcoV}{S596}
\lookupPut{R_tLssygbMcM3WkA9}{S597}
\lookupPut{R_1Nn7zBWgC41mdCk}{S598}
\lookupPut{R_2QWkJ2UpVlSs6U8}{S599}
\lookupPut{R_1KmdII9iS80cgRX}{S600}
\lookupPut{R_2QETbsfJ8RvNLnB}{S601}
\lookupPut{R_ZxHMlQ62IrKZkzf}{S602}
\lookupPut{R_9zzbO3pSdp7FNe1}{S603}
\lookupPut{R_1kY6o77Ji4AXxWX}{S604}
\lookupPut{R_2b3FtoK82VSyeL6}{S605}
\lookupPut{R_3PjK5w02f0gVzKI}{S606}
\lookupPut{R_1F3wgkHx9lPH7Uv}{S607}
\lookupPut{R_4Gh8BpFi3x8ZvvX}{S608}
\lookupPut{R_3PRzDk1YFj1F2Vc}{S609}
\lookupPut{R_1100alPBNH3orVs}{S610}
\lookupPut{R_1MVZ4Wh0LWwBCMP}{S611}
\lookupPut{R_31XjgTdIzjmQhd8}{S612}
\lookupPut{R_116edAzCZiRzs6Y}{S613}
\lookupPut{R_3L0j6mvCtBIi3w6}{S614}
\lookupPut{R_UinFHSt53iE4Hi9}{S615}
\lookupPut{R_tDPwqk0DqxugeJj}{S616}
\lookupPut{R_3jfHhLosLHAR7f5}{S617}
\lookupPut{R_32VKEaNr2wFTI1T}{S618}
\lookupPut{R_1PcTUmS7Oz34w0f}{S619}
\lookupPut{R_2xQOHr7HhceA7kO}{S620}
\lookupPut{R_3CTs5w7muKjiu2c}{S621}
\lookupPut{R_2Btjn03BVaJhavV}{S622}
\lookupPut{R_2QVWEdVwi7fCFIM}{S623}
\lookupPut{R_2TAXktDoIVVIOwr}{S624}
\lookupPut{R_10xsN8k3gd5NIAm}{S625}
\lookupPut{R_1CwhXblaAA3NP80}{S626}
\lookupPut{R_2BbULGQF9Brakdf}{S627}
\lookupPut{R_1hQngCtw8XwGy6J}{S628}
\lookupPut{R_1Kkg2JFiNGEXdwU}{S629}
\lookupPut{R_qVeEWz9ayybAGkx}{S630}
\lookupPut{R_BDhuQe7pXnv1Q5z}{S631}
\lookupPut{R_1CINDyhpxy5f8gW}{S632}
\lookupPut{R_2BqHfdWw0aaJA1L}{S633}
\lookupPut{R_1LAReuE8eT0zqPm}{S634}
\lookupPut{R_qC4C3Fmd6KLERrz}{S635}
\lookupPut{R_3ffLfRA2hz4liKG}{S636}
\lookupPut{R_3HAGk2ygFNzE9ZR}{S637}
\lookupPut{R_1gtrGKjjfj8zhRv}{S638}
\lookupPut{R_3HoAtN9sQxLhIvE}{S639}
\lookupPut{R_2UhpG49ZtQf6qHH}{S640}
\lookupPut{R_2w4TTi1n2k7ub0I}{S641}
\lookupPut{R_2fvYHQBmCrQ8eCL}{S642}
\lookupPut{R_3Eg2nCAM3k9vRJx}{S643}
\lookupPut{R_prCDm5TdTH5yiJz}{S644}
\lookupPut{R_3aS2lqscrpmCWIh}{S645}
\lookupPut{R_29mghK9wX3qGcjw}{S646}
\lookupPut{R_Ab5OH0V3VNiHibn}{S647}
\lookupPut{R_PGuXyi9ctWi4MQp}{S648}
\lookupPut{R_25XdKRGgQEndC0Y}{S649}
\lookupPut{R_33s8kjbDeRvt4jc}{S650}
\lookupPut{R_2EgIHzXpYH7dkdK}{S651}
\lookupPut{R_RqQxvAkCHsX1225}{S652}
\lookupPut{R_ZJhVxFFLps5ILqV}{S653}
\lookupPut{R_2yfIYDimaFhSAdI}{S654}
\lookupPut{R_3M4NfE0McCiNi4e}{S655}
\lookupPut{R_1eKVrTIICE8lpGC}{S656}
\lookupPut{R_27QGZY2dI8oKRmR}{S657}
\lookupPut{R_esNaBhkvmfsepzz}{S658}
\lookupPut{R_1Fxm3gsflxyJR2I}{S659}
\lookupPut{R_XM1asjIPrTWnZSx}{S660}
\lookupPut{R_1dsrIIZ6DfKXaLw}{S661}
\lookupPut{R_3lxdaFlrK0bqlkP}{S662}
\lookupPut{R_tMyJx6yO8ls5ULT}{S663}
\lookupPut{R_3xtCDRKNvtLP7qh}{S664}
\lookupPut{R_ZJqOOdTmLNiXk53}{S665}
\lookupPut{R_2ya74bTUwRzV9Af}{S666}
\lookupPut{R_vTz2wGeectOWodX}{S667}
\lookupPut{R_2ZOBAtQhuVuNDcV}{S668}
\lookupPut{R_9NZAkDuUsWcGNz3}{S669}
\lookupPut{R_2Y8QqAMLZm3fPII}{S670}
\lookupPut{R_2y71n3dBsu5iAOI}{S671}
\lookupPut{R_06YzURhg9Gx0fFT}{S672}
\lookupPut{R_2z6byCA4IW5tWod}{S673}
\lookupPut{R_2P7sWkLTGjaNs9Q}{S674}
\lookupPut{R_3M5O1tZSEEtpie7}{S675}
\lookupPut{R_Opbf6gn4XfXZ9M5}{S676}
\lookupPut{R_30tX2b7hzejVFTv}{S677}
\lookupPut{R_33EgSa8ncOLz1CW}{S678}
\lookupPut{R_3qm1vFfZRpoR1P6}{S679}
\lookupPut{R_1jDJNzWwPYWCBVc}{S680}
\lookupPut{R_reaQqEMFU2nkCAh}{S681}
\lookupPut{R_1r6zG8QBCWyOoU8}{S682}
\lookupPut{R_2zd74m5pi44xP3i}{S683}
\lookupPut{R_32QpFMkOZvpGmIQ}{S684}
\lookupPut{R_11Y0H5KxcWnMu9L}{S685}
\lookupPut{R_1ghJY384PmZG9yS}{S686}
\lookupPut{R_2BrkqCF7MoqSaaH}{S687}
\lookupPut{R_6P76RxCt81CaN9f}{S688}
\lookupPut{R_1mtvfHrP5WYjigR}{S689}
\lookupPut{R_s81641pTC5LA2lz}{S690}
\lookupPut{R_2zu6oCpazHI4KGW}{S691}
\lookupPut{R_3iqyCRg9a1hAkQR}{S692}
\lookupPut{R_11XYjgvEDIa5eVu}{S693}
\lookupPut{R_x3p7YGO9N8odS7f}{S694}
\lookupPut{R_1mVt2iRd9NkNdTn}{S695}
\lookupPut{R_bHD9coR7SToHhkd}{S696}
\lookupPut{R_3PdlP7QwzZhmpyM}{S697}
\lookupPut{R_4MUEBOUPyyoGgy5}{S698}
\lookupPut{R_2zcnAEk4bCcREaI}{S699}
\lookupPut{R_54nE3n5rzYU7oFr}{S700}
\lookupPut{R_2DNRKsrJ3GWXteQ}{S701}
\lookupPut{R_e4nEoB7yYJPp2j7}{S702}
\lookupPut{R_2AEfADqKFkPze5q}{S703}
\lookupPut{R_1HcznJ357x2l4GW}{S704}
\lookupPut{R_2SddxxylexVbQ8J}{S705}
\lookupPut{R_10Pl2OofNmrG9el}{S706}
\lookupPut{R_BW094QRf9AGdWp3}{S707}
\lookupPut{R_3NInDhi3ThkOczN}{S708}
\lookupPut{R_2wyr2M2S44MHc4N}{S709}
\lookupPut{R_2Ejl5TJfjJk8beo}{S710}
\lookupPut{R_2QZ0cSgRSJAJsMF}{S711}
\lookupPut{R_1hYW14NGyGb18md}{S712}
\lookupPut{R_1GVpiUni7AHmLUL}{S713}
\lookupPut{R_11cHWxGVnOs7Hzl}{S714}
\lookupPut{R_z1YYJ4zSch5AoRH}{S715}
\lookupPut{R_2ab2pGyzPqWI2Vg}{S716}
\lookupPut{R_2UgEh4w6wwrURtX}{S717}
\lookupPut{R_2f3eh1WiEvf7o73}{S718}
\lookupPut{R_RnwZSq7PV0G0cbT}{S719}
\lookupPut{R_2xFY3caD9hCgUC6}{S720}
\lookupPut{R_2sZmCqDw6WKFOJs}{S721}
\lookupPut{R_1gAfAdYmeLnPqo3}{S722}
\lookupPut{R_1jCYlYZHzHMpa1Z}{S723}
\lookupPut{R_3QnCqpRSBxX1xFT}{S724}
\lookupPut{R_3CINKvlr5S385IE}{S725}
\lookupPut{R_2whpzcsbXzv5wPV}{S726}
\lookupPut{R_1C7rRaUJ9BG9Jpu}{S727}
\lookupPut{R_3QQXQnlyNv1TxZA}{S728}
\lookupPut{R_33x43XrXhuQTD4j}{S729}
\lookupPut{R_1ocXVaatDVa6R70}{S730}
\lookupPut{R_2SAhbxU8vpJjsdA}{S731}
\lookupPut{R_3Rf8AWQV9vBIv4J}{S732}
\lookupPut{R_2pLYgWkoXtvu86f}{S733}
\lookupPut{R_bjX4yOjCUUw6u0F}{S734}
\lookupPut{R_Q6p311eGwmYoSKR}{S735}
\lookupPut{R_1d79CWi4Hh9ZSfH}{S736}
\lookupPut{R_3L0WWuzmTsQACcS}{S737}
\lookupPut{R_3nPKFxwDEfWpLhD}{S738}
\lookupPut{R_3ikwBn7NQSMzrkp}{S739}
\lookupPut{R_1CgrOtELkcLgYFi}{S740}
\lookupPut{R_2t8Rn9Gs5TPrVrh}{S741}
\lookupPut{R_Ck9gNe2Iaf2Qgp3}{S742}
\lookupPut{R_3k6w1Kb1eLKbrUU}{S743}
\lookupPut{R_3JaxsKW0KeGrJUo}{S744}
\lookupPut{R_3sAZd4ZWa7MnCYe}{S745}
\lookupPut{R_C49j4jpTqYChawN}{S746}
\lookupPut{R_3jYBLVeBLx0zM7E}{S747}
\lookupPut{R_2WHg2IxpkEXkni8}{S748}
\lookupPut{R_1LjJRw7mbojz50W}{S749}
\lookupPut{R_3mf4uGgDThfPMkD}{S750}
\lookupPut{R_25WRrjFFAKgGKqT}{S751}
\lookupPut{R_8cSBPKRJclLmMDv}{S752}
\lookupPut{R_2P5mtqCV1g0EKMl}{S753}
\lookupPut{R_10Dc3xGWmtR3hhN}{S754}
\lookupPut{R_2a8iRq8DeBHlwYh}{S755}
\lookupPut{R_8oJLDZj0rsG80s9}{S756}
\lookupPut{R_2v8hRayTPFCWjSe}{S757}
\lookupPut{R_W8Vj7CsJBtpav6x}{S758}
\lookupPut{R_1hRpExaPKvRZsVD}{S759}
\lookupPut{R_AhrlBARGgYFrgoF}{S760}
\lookupPut{R_1FhWAP0h5ZR7KCY}{S761}
\lookupPut{R_XGm2RvclfceTbXj}{S762}
\lookupPut{R_3hgV0gZNYa4EMmJ}{S763}
\lookupPut{R_1JUZKqyHtjnLqVc}{S764}
\lookupPut{R_2AMXDqCpXZgdAPd}{S765}
\lookupPut{R_2CNi8JDqV7g8qSn}{S766}
\lookupPut{R_24bHc3CS70uxcMO}{S767}
\lookupPut{R_1hYKjvytP1Pel82}{S768}
\lookupPut{R_12y6D2IxV6FAOom}{S769}
\lookupPut{R_1K7OORVl7HP4Uek}{S770}
\lookupPut{R_1qaJEI4rlu9GCqj}{S771}
\lookupPut{R_1CJBLoDDYMm30CG}{S772}
\lookupPut{R_AvamgQRhDgm3PS9}{S773}
\lookupPut{R_ve76C4hfBIrPsYx}{S774}
\lookupPut{R_5sUlcPqpMmD2moF}{S775}
\lookupPut{R_1CpwX2xKU5uP5aM}{S776}
\lookupPut{R_25F2fMZQBRnhR8R}{S777}
\lookupPut{R_1mye1CSrPTq7GEM}{S778}
\lookupPut{R_3CQoBesOmsPycLX}{S779}
\lookupPut{R_RrFBbv8pRcRVbHj}{S780}
\lookupPut{R_ahjTiNZczmxTrMd}{S781}
\lookupPut{R_2QRSS9ZvW02jquk}{S782}
\lookupPut{R_2dXOMlxHlEkCFal}{S783}
\lookupPut{R_2EhO7r24mIAixFt}{S784}
\lookupPut{R_ZI5bdiwk8bqT1bX}{S785}
\lookupPut{R_sC45MyuoFRyPgFr}{S786}
\lookupPut{R_2ysicsP0X2E2Pnz}{S787}
\lookupPut{R_2foE22pN6WCashz}{S788}
\lookupPut{R_6D5Tk8VE1A4ggzT}{S789}
\lookupPut{R_2rvWdZEuN876cz5}{S790}
\lookupPut{R_Rbng2uxtKSVKTxT}{S791}
\lookupPut{R_0lwXQXm8zlDzeYV}{S792}
\lookupPut{R_bgBFAaOkc7WM0lr}{S793}
\lookupPut{R_3D7VuOBfqVCniQh}{S794}
\lookupPut{R_1H2B1nvNjqADnlw}{S795}
\lookupPut{R_24e1mZh9VqE2B7G}{S796}
\lookupPut{R_UDQUTFwE3mubYVH}{S797}
\lookupPut{R_3oNn099ygCTYLCR}{S798}
\lookupPut{R_31tGtShpjHTNqLF}{S799}
\lookupPut{R_2Tv3kweSiWq3GPH}{S800}
\lookupPut{R_a3PFNlQqPDUmHFD}{S801}
\lookupPut{R_2wM6xegOcTVz9I5}{S802}
\lookupPut{R_3lzYiaWU6K5Il5T}{S803}
\lookupPut{R_12heUjOpXyJdSNW}{S804}
\lookupPut{R_2xWx73kjA55gOpc}{S805}
\lookupPut{R_2VkeYaZsTjtNvZb}{S806}
\lookupPut{R_3JqFmUVsAn6T30x}{S807}
\lookupPut{R_12mZCgcg2qCbE19}{S808}
\lookupPut{R_3r0xngn6AAlwWdv}{S809}
\lookupPut{R_0VWEqrxZkkp3F1n}{S810}
\lookupPut{R_27r0WraG1vVVTnc}{S811}
\lookupPut{R_1LukXVbp4UzgNpk}{S812}
\lookupPut{R_OjpvkMtqMbRgKlP}{S813}
\lookupPut{R_12kUWGzXtlcWnKi}{S814}
\lookupPut{R_1EchdTaqPY7sVcv}{S815}
\lookupPut{R_1jdisZ6fUkOcyaf}{S816}
\lookupPut{R_3dLhSd6Z6h1d8Ns}{S817}
\lookupPut{R_w5zavRxfdig5eIV}{S818}
\lookupPut{R_3KpFjsIIy0YZ0YS}{S819}
\lookupPut{R_3qJ6tdtQEf7kTfI}{S820}
\lookupPut{R_0dmuZdaxDn7RwjL}{S821}
\lookupPut{R_29izQNG7DTDO4pb}{S822}
\lookupPut{R_T8fsACG5hqNPYmR}{S823}
\lookupPut{R_3m1xFdLsfj34uSG}{S824}
\lookupPut{R_2zvCAwYvdRICokd}{S825}
\lookupPut{R_1riEPBd7Myqu3Hb}{S826}
\lookupPut{R_vAyBRPpoAfTuoXD}{S827}
\lookupPut{R_xnJKoqf7s9bfPRT}{S828}
\lookupPut{R_2B39zXglou8BOeR}{S829}
\lookupPut{R_1TSRFl0verDAKIx}{S830}
\lookupPut{R_RIkmqHzgKedylDH}{S831}
\lookupPut{R_1hRo7TLnFrXdfOC}{S832}
\lookupPut{R_3hbSrrpCvH1snfc}{S833}
\lookupPut{R_sXaTNZfQPXFKnPX}{S834}
\lookupPut{R_1rojnV1ntpeqYnm}{S835}
\lookupPut{R_2uDttDPrt8IMsCE}{S836}
\lookupPut{R_1QLdwCX4jivSXBG}{S837}
\lookupPut{R_3iFpzOrZmGQmrqD}{S838}
\lookupPut{R_z7M3xJDcC6Yam7T}{S839}
\lookupPut{R_21ib1Uo9Lj1oLBr}{S840}
\lookupPut{R_2TN74nZSqqRiiOS}{S841}
\lookupPut{R_2yl9RW0DiDUP1FI}{S842}
\lookupPut{R_1rCh7Kkegfx15F5}{S843}
\lookupPut{R_3g5k1rw8hSKRYXI}{S844}
\lookupPut{R_1LGOOzxGb0wCTnw}{S845}
\lookupPut{R_2znHP47JmQ5hOul}{S846}
\lookupPut{R_3fUBkENRQhSUU98}{S847}
\lookupPut{R_3sbpNNBefy8NiYN}{S848}
\lookupPut{R_2tlN4plUzawnutP}{S849}
\lookupPut{R_3Km1Cg3Qgi7IqSB}{S850}
\lookupPut{R_1gnYzzv7pES5N3Z}{S851}
\lookupPut{R_yxo0i7spjO5N0vD}{S852}
\lookupPut{R_tGIfcVei48G1nuV}{S853}
\lookupPut{R_0eLeRHpBTo8SdWx}{S854}
\lookupPut{R_3nor6WKildMzmyr}{S855}
\lookupPut{R_2VkI3bh3coRa3r0}{S856}
\lookupPut{R_2fIBTe5CBV9wdMT}{S857}
\lookupPut{R_3FWCLZQ7e2QzXPO}{S858}
\lookupPut{R_2wn9DyjSW0d2btw}{S859}
\lookupPut{R_1eqxfOakhsqROCf}{S860}
\lookupPut{R_OIh1lg11sqhyJuV}{S861}
\lookupPut{R_2WTjZ6NcSpIekCI}{S862}
\lookupPut{R_Qol7Pu9VvD7DVF7}{S863}
\lookupPut{R_2wMrmYFvneOb68x}{S864}
\lookupPut{R_2VQQGsQiUiG23v6}{S865}
\lookupPut{R_3M9WjVPDGqscBRN}{S866}
\lookupPut{R_265sMr3oHUNiP6H}{S867}
\lookupPut{R_3HHR9BcbcnHyCwz}{S868}
\lookupPut{R_238In0o6NlpmSS5}{S869}
\lookupPut{R_2qldZMaGUDkWKXp}{S870}
\lookupPut{R_vObIYOKFqcFTYhb}{S871}
\lookupPut{R_WjwYhrfUQukOF7H}{S872}
\lookupPut{R_3p5NLpDvKNmlcOW}{S873}
\lookupPut{R_2R2MjirqhVCeD5j}{S874}
\lookupPut{R_2wHm65STTqJPX8r}{S875}
\lookupPut{R_YRENZRCaHxFhY7D}{S876}
\lookupPut{R_xydnaQTj5qwMLfj}{S877}
\lookupPut{R_3rMRoD7DYqcmdsB}{S878}
\lookupPut{R_3ir6toYmkeuj68H}{S879}
\lookupPut{R_ezX3ScsQjqKtPtD}{S880}
\lookupPut{R_bHMFRExIjWrAmoV}{S881}
\lookupPut{R_ysfq9hMICnMt5Yt}{S882}
\lookupPut{R_W1k4WFKe1zYdkop}{S883}
\lookupPut{R_27wfGh5N1oOlFwM}{S884}
\lookupPut{R_1qewVSb8GOcInHQ}{S885}
\lookupPut{R_1mvXjRtgqIUKl3G}{S886}
\lookupPut{R_2dWpecDRqxBOmm4}{S887}
\lookupPut{R_eExmkgLsaXEwp1L}{S888}
\lookupPut{R_1hMQTm86S5Ih4J8}{S889}
\lookupPut{R_2bHkMPkzwYdUewL}{S890}
\lookupPut{R_3hiiXoUhmFEEBIB}{S891}
\lookupPut{R_RE5w8b4ry8Y8A5H}{S892}
\lookupPut{R_2f34j26XEHFXuse}{S893}
\lookupPut{R_3Rr9s1dDRBGaaN3}{S894}
\lookupPut{R_2xWyWMOdPb4qXUo}{S895}
\lookupPut{R_3QQyYw5sHsf2PIv}{S896}
\lookupPut{R_3JFbJmj3JBrGxEM}{S897}
\lookupPut{R_3RacNZG9Ict0IsN}{S898}
\lookupPut{R_2dvc1n5jDanEYRM}{S899}
\lookupPut{R_urVTi0mdEfWgbmN}{S900}
\lookupPut{R_UQlIB1JziuZMg2B}{S901}
\lookupPut{R_3npWNj1BAXcbkn7}{S902}
\lookupPut{R_2YEv64xIiRo53t6}{S903}
\lookupPut{R_1g25WkeM1nMGkdR}{S904}
\lookupPut{R_3CBmmR9q6PmnO2H}{S905}
\lookupPut{R_1C22nh6esyyJ0T2}{S906}
\lookupPut{R_27NBxM1pgNlXon2}{S907}
\lookupPut{R_5BhWK9q2bGC3Qk1}{S908}
\lookupPut{R_XwjjaUyOIH2sf9D}{S909}
\lookupPut{R_3oSYcABwiXpxzWN}{S910}
\lookupPut{R_eghuQ0SXxLyOAuJ}{S911}
\lookupPut{R_3CT4bcW0zC15yTH}{S912}
\lookupPut{R_1giO6aH3qvzLH5i}{S913}
\lookupPut{R_3m1weI2KlQfKHW5}{S914}
\lookupPut{R_3L22qDSWIK42E3X}{S915}
\lookupPut{R_3OitLOdRzRsAOlO}{S916}
\lookupPut{R_3emnyh4LmEdg95W}{S917}
\lookupPut{R_ZeOUMTQQ1eN2LOp}{S918}
\lookupPut{R_1f3LiEhVHSfH0Hw}{S919}
\lookupPut{R_2w0M8iKRR0dFasN}{S920}
\lookupPut{R_1da97bz4O0BK8NY}{S921}
\lookupPut{R_sAss5nnAiw95Wvv}{S922}
\lookupPut{R_zSiqYF5rdScaPKh}{S923}
\lookupPut{R_VWRIo9dFdl8oj5f}{S924}
\lookupPut{R_3j6Q7AkPvm8KT0d}{S925}
\lookupPut{R_2Ec7yS1ll6jqJ6i}{S926}
\lookupPut{R_2eUzmCbO7etlrHV}{S927}
\lookupPut{R_247BUft6w6YE7zH}{S928}
\lookupPut{R_3hsyFzKJv1tLwuZ}{S929}
\lookupPut{R_2VmZHMXxApguQSl}{S930}
\lookupPut{R_2Qfeznh8Yp0CRDo}{S931}
\lookupPut{R_2SelEWkLpS6O67O}{S932}
\lookupPut{R_33kIQuKaZYAZazy}{S933}
\lookupPut{R_PSZiOHHifXZPWJb}{S934}
\lookupPut{R_tS5vXNczajzJAPv}{S935}
\lookupPut{R_1rrkbHsNDF5sOaV}{S936}
\lookupPut{R_3kug4669n4y2y0W}{S937}
\lookupPut{R_2qEpRxT64E7i8Ak}{S938}
\lookupPut{R_41qWfFL4zsTj7y1}{S939}
\lookupPut{R_1hyt5zKWnc17qgW}{S940}
\lookupPut{R_2rDvHXjs2WE1X7Y}{S941}
\lookupPut{R_D5JXMcxtSYw1ymt}{S942}
\lookupPut{R_24utuUQOSBv4iro}{S943}
\lookupPut{R_3EMPG2qrv0IdDQV}{S944}
\lookupPut{R_2aFxZxyy6OSndLJ}{S945}
\lookupPut{R_O6FRRCxbMUZcGIN}{S946}
\lookupPut{R_3JxskK1n41krfaO}{S947}
\lookupPut{R_1DNJFTcvujiVZcp}{S948}
\lookupPut{R_3ktybskWKbnJIc3}{S949}
\lookupPut{R_807ooVuBNcEEaUV}{S950}
\lookupPut{R_3lVUjYZY61YmD0P}{S951}
\lookupPut{R_28Y9YNk7Z98bulr}{S952}
\lookupPut{R_De29SeBOspzKDSx}{S953}
\lookupPut{R_9RmOGjYFUQWWCoV}{S954}
\lookupPut{R_1fcQ2BGrq2rb7Ug}{S955}
\lookupPut{R_1mgXDx4S0isA5t7}{S956}
\lookupPut{R_2DN7iQBHEkbROis}{S957}
\lookupPut{R_2EufRcO7SCgW5Mr}{S958}
\lookupPut{R_OHaqqmYkonEm169}{S959}
\lookupPut{R_1lhgGcxf4NBWHRM}{S960}
\lookupPut{R_sckY02EyKlMTeBr}{S961}
\lookupPut{R_5j6r6MZLJwrYgzD}{S962}
\lookupPut{R_1DogVVOAe1xyvgV}{S963}
\lookupPut{R_2uW20nBM50J1X2X}{S964}
\lookupPut{R_VPeJwwRRrWjGcTf}{S965}
\lookupPut{R_1eF39hOxiVZ8Pmi}{S966}
\lookupPut{R_3P69sOPq3K8FAej}{S967}
\lookupPut{R_2UbVDZNEE8KTo41}{S968}
\lookupPut{R_300OcxR2byRaEFo}{S969}
\lookupPut{R_xbgtCLdM4H96BR7}{S970}
\lookupPut{R_3DvyjTAt3UmODDp}{S971}
\lookupPut{R_3CCxnucLmDEAtvs}{S972}
\lookupPut{R_2PBt6u7jBGkZZXx}{S973}
\lookupPut{R_XAkUmcTONe6yEcp}{S974}
\lookupPut{R_2azTI1AfMq6Zyh2}{S975}
\lookupPut{R_3EaOWSavJp4FCRw}{S976}
\lookupPut{R_2aqVQihxm0nMKeY}{S977}
\lookupPut{R_sU7yhsQZDlHkUoh}{S978}
\lookupPut{R_27JZL0vIekZK4EE}{S979}
\lookupPut{R_3L8aljSPovAkTiP}{S980}
\lookupPut{R_0e387YIKAhbYcY9}{S981}
\lookupPut{R_UYhRdWn4GkKC4Qp}{S982}
\lookupPut{R_Xn2Nwnd1mDykRkB}{S983}
\lookupPut{R_1MLgtMcobsS9nsm}{S984}
\lookupPut{R_XLEEG1MpNR1hEK5}{S985}
\lookupPut{R_3hDtH9n5sIOBeHp}{S986}
\lookupPut{R_3szF13aNCPMkDG4}{S987}
\lookupPut{R_3l3mBFuDaFXerAJ}{S988}
\lookupPut{R_2QFIHKykFXs2z3n}{S989}
\lookupPut{R_2VJdBIZhwUvsQLX}{S990}
\lookupPut{R_6PAig6DCtn0bRPX}{S991}
\lookupPut{R_1OB601i8PzNqOzz}{S992}
\lookupPut{R_2qqIA7s4TJqlwB4}{S993}
\lookupPut{R_3kfnjThCs3QE3fz}{S994}
\lookupPut{R_2aWdngdnea0VvBU}{S995}
\lookupPut{R_2QugmYj7PHnPjIU}{S996}
\lookupPut{R_1qgKPtA67lJIBX3}{S997}
\lookupPut{R_2xKRYsJAFx7V5Ba}{S998}
\lookupPut{R_Xqfe9ULv0OLa1iN}{S999}
\lookupPut{R_2TTACcsbp9OcmQx}{S1000}
\lookupPut{R_125Zm6kIqPcCBhv}{S1001}
\lookupPut{R_3FR9786jI2bjjgL}{S1002}
\lookupPut{R_3hEEORceQXqFWoZ}{S1003}
\lookupPut{R_214bfVe4w9YZE8x}{S1004}
\lookupPut{R_3nDwJr23PwkizV6}{S1005}
\lookupPut{R_3nIRmYN2rdfcDIy}{S1006}
\lookupPut{R_1rpEsoqPYU0yiPS}{S1007}
\lookupPut{R_3iwj03qlCsmg1xv}{S1008}
\lookupPut{R_CkSxbUiZCikxmSd}{S1009}
\lookupPut{R_3PpgmJPW3b9NTzC}{S1010}
\lookupPut{R_3kMUvMPTad4wACd}{S1011}
\lookupPut{R_2X6Y2yeDRbJAwkG}{S1012}
\lookupPut{R_2yg6zHQCFuRLDDr}{S1013}
\lookupPut{R_1qV4B5dvjZYqNpv}{S1014}
\lookupPut{R_2e8LX4lZqM3Uhiv}{S1015}
\lookupPut{R_2U3jnTkEq4PFh3E}{S1016}
\lookupPut{R_1DTJkjMTupIORPX}{S1017}
\lookupPut{R_3F59a0rhhkxEaF9}{S1018}
\lookupPut{R_2OQ1C2NC8aLrLOQ}{S1019}
\lookupPut{R_1OI84UkgSxv3BVp}{S1020}
\lookupPut{R_24pXLbd7WeXePQ0}{S1021}
\lookupPut{R_XgJwE33uSGMFgpH}{S1022}
\lookupPut{R_12MwXA5B7DWxkBq}{S1023}
\lookupPut{R_1ilH5TqiybgGbfK}{S1024}
\lookupPut{R_1lspSSpPJQF5LSi}{S1025}
\lookupPut{R_1QKpzSSdsQLxkSJ}{S1026}
\lookupPut{R_1yQSy0VLMRimEet}{S1027}
\lookupPut{R_1Q3Gv2Qys3FMudr}{S1028}
\lookupPut{R_2TKMDKgnkNFCXAK}{S1029}
\lookupPut{R_10BDf7rjwczq1g4}{S1030}
\lookupPut{R_2t9xEUSB8YSA5AY}{S1031}
\lookupPut{R_1InjzqqYKgqkaCW}{S1032}
\lookupPut{R_9vHLEwnImfeS921}{S1033}
\lookupPut{R_3R7AfvGy4tbhfIl}{S1034}
\lookupPut{R_3nVQuQz9z2YzFOh}{S1035}
\lookupPut{R_1FP7962ZJsBVMvR}{S1036}
\lookupPut{R_2cABa3g9SGR5el9}{S1037}
\lookupPut{R_3QQoKXQnpIFKJy8}{S1038}
\lookupPut{R_20MdpZapfFqL7My}{S1039}
\lookupPut{R_1DomTcO7tsDsqnX}{S1040}
\lookupPut{R_8oDTSmskZFRXM8F}{S1041}
\lookupPut{R_smOIu8paNf9ecA9}{S1042}
\lookupPut{R_2V9Kxda0P3ukvi9}{S1043}
\lookupPut{R_1jOoXPdMG1ojqYg}{S1044}
\lookupPut{R_21cFYGhv7B0SqVW}{S1045}
\lookupPut{R_2zO9gkEo0FPpGmB}{S1046}
\lookupPut{R_1EZDXIkc5ijRC6e}{S1047}
\lookupPut{R_2WvbIKolROmDV2L}{S1048}
\lookupPut{R_1lsLKpcQTGfeKnm}{S1049}
\lookupPut{R_1gbgvnmUuo6u5VH}{S1050}
\lookupPut{R_1BOTt9mzpR5wNH4}{S1051}
\lookupPut{R_1IQS8Izk3zRmhn6}{S1052}
\lookupPut{R_2eQYoRRjHxYTTgG}{S1053}
\lookupPut{R_Q5MBqbUHCqjwDkd}{S1054}
\lookupPut{R_1f2ENpcKvRDlAfl}{S1055}
\lookupPut{R_2Y4xULUbaoMCu1k}{S1056}
\lookupPut{R_1Qsd8SQOIl9lyRd}{S1057}
\lookupPut{R_2q47Z4UNPDXdWvS}{S1058}
\lookupPut{R_1CmuqA2bmdVDzkr}{S1059}
\lookupPut{R_Qna8CysjQhsd0fn}{S1060}
\lookupPut{R_2giVGL7qi7v5rvr}{S1061}
\lookupPut{R_24pmU4Bh15HGVDU}{S1062}
\lookupPut{R_3EZlmkIhP9c5p8q}{S1063}
\lookupPut{R_3FKEf2YtZQnCHSw}{S1064}
\lookupPut{R_yyiZEKJmNNB3vj3}{S1065}
\lookupPut{R_2V3YsPkVfR6L5uK}{S1066}
\lookupPut{R_1rJvP8vbHareGet}{S1067}
\lookupPut{R_C8lXwe4dyetCJIR}{S1068}
\lookupPut{R_2c7T05RvcUcbbTY}{S1069}
\lookupPut{R_1CJzKibREl31EdL}{S1070}
\lookupPut{R_1jJ6V0VTgm8otUu}{S1071}
\lookupPut{R_zd4LRlJc0MoNEbL}{S1072}
\lookupPut{R_1mQ2gSURlcoTrTD}{S1073}
\lookupPut{R_1IQnFgbXi5aRCcl}{S1074}
\lookupPut{R_2wsjEyCPqSYk2v3}{S1075}
\lookupPut{R_RyOPPSmGzoi4J8d}{S1076}
\lookupPut{R_1K0vq3ySu1bu5jD}{S1077}
\lookupPut{R_3MFHX7qhVihgRPH}{S1078}
\lookupPut{R_SI4zgVp8XjcIvC1}{S1079}
\lookupPut{R_29mk3lchXGt6w8v}{S1080}
\lookupPut{R_111itVgJ03TrDua}{S1081}
\lookupPut{R_Y9xX2lYNXAIMOUp}{S1082}
\lookupPut{R_2xVmwhXWj6FDnBy}{S1083}
\lookupPut{R_2Ebo0YZzsk0XuHs}{S1084}
\lookupPut{R_2zAMUaHsdVuf8QF}{S1085}
\lookupPut{R_2SJEffUsK5LuIR3}{S1086}
\lookupPut{R_27rpxxD6jkwcEcG}{S1087}
\lookupPut{R_tSVVArwPGypQVZn}{S1088}
\lookupPut{R_33p2S1xxmaHclyb}{S1089}
\lookupPut{R_28Gv2Rmo1ZM5niV}{S1090}
\lookupPut{R_1P0r99WnjiJ4HAQ}{S1091}
\lookupPut{R_2RPOxCOTElpfncJ}{S1092}
\lookupPut{R_25A6Jgm8NyQqs6l}{S1093}
\lookupPut{R_2duLEQ5dhRfyXmg}{S1094}
\lookupPut{R_2fKdAWUzCXS1KOF}{S1095}
\lookupPut{R_PIorMCQBbvbJWg1}{S1096}
\lookupPut{R_211k1CCHouVWrdc}{S1097}
\lookupPut{R_2qeok8Ya9Grvn1e}{S1098}
\lookupPut{R_2VKEwP6Gv2K4vPF}{S1099}
\lookupPut{R_3jSWAMzT2RBCFHc}{S1100}
\lookupPut{R_1QixrDlbsjJ7Ebc}{S1101}
\lookupPut{R_1pX6PjKDvoMhLVm}{S1102}
\lookupPut{R_3qvw6sdzLyVf4gm}{S1103}
\lookupPut{R_1lzwcgDXkz867tW}{S1104}
\lookupPut{R_1C9eZh360PDg4UM}{S1105}
\lookupPut{R_3QGnzTfMQY2rFiR}{S1106}
\lookupPut{R_3FWCf7nhUSXHHjQ}{S1107}
\lookupPut{R_31KrPD3DVWH8xic}{S1108}
\lookupPut{R_ABgNZyTOIOJv3IB}{S1109}
\lookupPut{R_1DvbNyywRn4XZtq}{S1110}
\lookupPut{R_3EPVTsNqw0GlV2p}{S1111}
\lookupPut{R_2y9I5Lx5oBlD5nB}{S1112}
\lookupPut{R_3lzUZT94S1KZ6nR}{S1113}
\lookupPut{R_1JQQ8iiZF25Y7RQ}{S1114}
\lookupPut{R_1l6mwwF0GTop841}{S1115}
\lookupPut{R_2aqVe0zkK6Rvrid}{S1116}
\lookupPut{R_1DtZUjap8RYa3nM}{S1117}
\lookupPut{R_Z923ebJNJbUss8h}{S1118}
\lookupPut{R_1Ljgvi5RAqG0KMV}{S1119}
\lookupPut{R_5buk9vYL3Xlepu9}{S1120}
\lookupPut{R_1NzxxYhy0qA24ct}{S1121}
\lookupPut{R_1jpm8lGlUVV5I6y}{S1122}
\lookupPut{R_2aVQMqAILcmesh5}{S1123}
\lookupPut{R_1qaCv6GfVtmd7re}{S1124}
\lookupPut{R_2YPMII2TXuHyqBc}{S1125}
\lookupPut{R_abH0H5f6JyDdHUJ}{S1126}
\lookupPut{R_2QM2DNHEszIRtR3}{S1127}
\lookupPut{R_bCxwKbRbLLlM0pP}{S1128}
\lookupPut{R_1K8cwhm6UEO99aG}{S1129}
\lookupPut{R_SJLzezkpOkknAfT}{S1130}
\lookupPut{R_AjTCq7QwkCZa7zr}{S1131}
\lookupPut{R_1ND3Fk3jfC14IoK}{S1132}
\lookupPut{R_12rvhCTKvcaffha}{S1133}
\lookupPut{R_eR8kG2BROto6ZZ7}{S1134}
\lookupPut{R_C46h5RNwecEAvU5}{S1135}
\lookupPut{R_1ePPNYFoXjinuin}{S1136}
\lookupPut{R_2OPXhAhrAHekwBn}{S1137}
\lookupPut{R_3NCAzWe8vUh8D5n}{S1138}
\lookupPut{R_3NxL595UM5U1A1d}{S1139}
\lookupPut{R_u51q1DIkd8n4tXz}{S1140}
\lookupPut{R_21ERBHwHldrgWXV}{S1141}
\lookupPut{R_3R9m7yGhcoyyaGG}{S1142}
\lookupPut{R_2SH7Qj32NFVcmPJ}{S1143}
\lookupPut{R_eY8U5H9PRMvSGMF}{S1144}
\lookupPut{R_ZkrX08j0YP1juZr}{S1145}
\lookupPut{R_1cSS2WGSsJorHbY}{S1146}
\lookupPut{R_3HSOqZjH4PAWSrA}{S1147}
\lookupPut{R_2e2mCxDx535UZ29}{S1148}
\lookupPut{R_2cA7E4AmFbHFCaR}{S1149}
\lookupPut{R_1jjnU10bhGqlGxp}{S1150}
\lookupPut{R_1lmbz2ZULbfX7ol}{S1151}
\lookupPut{R_2OINkwWM2RGCPPZ}{S1152}
\lookupPut{R_26l2QK2CcSj1PG0}{S1153}
\lookupPut{R_3M9ENoKsV3v9s19}{S1154}
\lookupPut{R_2CKacmzjUorirBR}{S1155}
\lookupPut{R_3D0cTma1lqJFuI7}{S1156}
\lookupPut{R_2fwjnzyY1U7ZhYD}{S1157}
\lookupPut{R_2EgEeYGLonebDRW}{S1158}
\lookupPut{R_r2NrDXWmeOmE4Jr}{S1159}
\lookupPut{R_UgbhTI9TepjP6AF}{S1160}
\lookupPut{R_3KwXJCYLqD4SYEx}{S1161}
\lookupPut{R_UQrtyPlJFLbNYD7}{S1162}
\lookupPut{R_8c8o6yxxkDldCq5}{S1163}
\lookupPut{R_1EZ0kZbaCszAs1C}{S1164}
\lookupPut{R_2SHOOR2UM32n7dv}{S1165}
\lookupPut{R_1P6GxYXd9F2Vvpo}{S1166}
\lookupPut{R_1Q4qaXJtiGD3yWn}{S1167}
\lookupPut{R_QbqFpnKax1LJsCB}{S1168}
\lookupPut{R_ddp4FJ5QA2Ymwpj}{S1169}
\lookupPut{R_2pKskvI6sth22Ja}{S1170}
\lookupPut{R_21oFIH4Cea7k0Vv}{S1171}
\lookupPut{R_3qfBiH54XFMFOAB}{S1172}
\lookupPut{R_qFtoQXOA9wASTQJ}{S1173}
\lookupPut{R_3MbsVN4QTCKbf9Q}{S1174}
\lookupPut{R_3lGEmCq2VNcfZ10}{S1175}
\lookupPut{R_2z8f8t8nF9kL0i8}{S1176}
\lookupPut{R_1kTfa3rSS2inx6j}{S1177}
\lookupPut{R_2Wx6D81FqMOB6TM}{S1178}
\lookupPut{R_3G2kD2ySY5WLa6A}{S1179}
\lookupPut{R_2bOCxC4AJiPHkzW}{S1180}
\lookupPut{R_1LLpAL7ZwnfGCYW}{S1181}
\lookupPut{R_24MWY4S4AbMgPH2}{S1182}
\lookupPut{R_3HtRID85yJakDkW}{S1183}
\lookupPut{R_3iyWYEUgnXtXGmJ}{S1184}
\lookupPut{R_1oulkl2iF0CcYuK}{S1185}
\lookupPut{R_1PXGyjN5B36OkfS}{S1186}
\lookupPut{R_3DkFi1JXsGHXRtQ}{S1187}
\lookupPut{R_30eITD2spw4yVmj}{S1188}
\lookupPut{R_vrEOWXuITOMCrYJ}{S1189}
\lookupPut{R_2v1XFPQaIXJCDgV}{S1190}
\lookupPut{R_OBggnX4P1PKbjGh}{S1191}
\lookupPut{R_RQyg0OZvsMSlmIV}{S1192}
\lookupPut{R_3hhY2VeMMyF2GHv}{S1193}
\lookupPut{R_1E6zC1zIzD8Q0ne}{S1194}
\lookupPut{R_3DoJi43DtUfbmLJ}{S1195}
\lookupPut{R_sBzaOSqEJWI4eaJ}{S1196}
\lookupPut{R_2RP0ODIX5ayqAf3}{S1197}
\lookupPut{R_1rcqXpEpfzIkiLy}{S1198}
\lookupPut{R_3iCiNxAuYRdZSDx}{S1199}
\lookupPut{R_sbdnr6yMEVQUAal}{S1200}
\lookupPut{R_3CZZxErxs1lDz1D}{S1201}
\lookupPut{R_3DhccRQrKk6UpLn}{S1202}
\lookupPut{R_D8zQRXkuzgAuGnD}{S1203}
\lookupPut{R_2XpagI1ayL4LmIW}{S1204}
\lookupPut{R_2ZOlLGHRG1hdlKL}{S1205}
\lookupPut{R_1gNoUZLXgjBDSQQ}{S1206}
\lookupPut{R_24Bmb0oNv1Kc8tD}{S1207}
\lookupPut{R_1ILOJyUxvbToyYQ}{S1208}
\lookupPut{R_305X4By8PsdjY6h}{S1209}
\lookupPut{R_2DLytPJp8jbkMRu}{S1210}
\lookupPut{R_3EtIW58hZFNdT0r}{S1211}
\lookupPut{R_20VFgAs3W8WUNty}{S1212}
\lookupPut{R_3gN5x8KbgJLk5h7}{S1213}
\lookupPut{R_10UBb9aIRqdEolF}{S1214}
\lookupPut{R_ZJBaIatjyWEuIyR}{S1215}
\lookupPut{R_a5aWIAsGINnGRvr}{S1216}
\lookupPut{R_2waZVz8akR5YLB8}{S1217}
\lookupPut{R_3GBWphNpMdTzbAr}{S1218}
\lookupPut{R_3KGz9OpBKkMMa3z}{S1219}
\lookupPut{R_1CfRgog2OAh29XQ}{S1220}
\lookupPut{R_Uom0fD9qzPx4f7P}{S1221}
\lookupPut{R_3fZWBx41ZQNtVZB}{S1222}
\lookupPut{R_2TGcQ3VsoYcOvdv}{S1223}
\lookupPut{R_3JzqxIp7XREGGso}{S1224}
\lookupPut{R_qVgY9VOCtG8y5ot}{S1225}
\lookupPut{R_cAqJo5Qj7axA1Pz}{S1226}
\lookupPut{R_2bH8I8DLWRI8CLa}{S1227}
\lookupPut{R_scDrUdv9puCNfCF}{S1228}
\lookupPut{R_sGOi5NXCUrQgGuB}{S1229}
\lookupPut{R_2TMCNDwLus5hoAp}{S1230}
\lookupPut{R_b2USa2uzCgWiIMh}{S1231}
\lookupPut{R_116lAh2yNwFqXns}{S1232}
\lookupPut{R_2R86mgHuo0YwIft}{S1233}
\lookupPut{R_2X68HvMxoPPNTG4}{S1234}
\lookupPut{R_3qlqmiuStkVYRGh}{S1235}
\lookupPut{R_RhpHiTZQJFtvLIl}{S1236}
\lookupPut{R_1qUpuIXCXjb0DjX}{S1237}
\lookupPut{R_1IRydWGaFQJbC0J}{S1238}
\lookupPut{R_zUoGIpsZXBn11Pb}{S1239}
\lookupPut{R_1ON9ZuOAT5UY3iw}{S1240}
\lookupPut{R_1hQq6tlV9pFFUgQ}{S1241}
\lookupPut{R_2XpnF0bSUqycR9o}{S1242}
\lookupPut{R_1C1IoV90pA94ejA}{S1243}
\lookupPut{R_1LFElX4cWWc1MWN}{S1244}
\lookupPut{R_1MRSkhs0POiimE6}{S1245}
\lookupPut{R_1ImwXOizbF9bebe}{S1246}
\lookupPut{R_26ngz45epB15SR1}{S1247}
\lookupPut{R_268OysytE2SXp9U}{S1248}
\lookupPut{R_1ruFWKbsP00hKqk}{S1249}
\lookupPut{R_2VNHfifHPaW9Zoj}{S1250}
\lookupPut{R_237bZ83eLxO3xZ3}{S1251}
\lookupPut{R_Yc5nERts633xVlL}{S1252}
\lookupPut{R_2zklVWfIhkmDLbr}{S1253}
\lookupPut{R_2WSvAV1Ae8DRULo}{S1254}
\lookupPut{R_3kcbftkg8qS6mYc}{S1255}
\lookupPut{R_1M5GLXFRhfOslsA}{S1256}
\lookupPut{R_DudPRkWrv5zUBEt}{S1257}
\lookupPut{R_DuHQUvm6pm704Pn}{S1258}
\lookupPut{R_1eUGaQi14N5E8Wn}{S1259}
\lookupPut{R_XpHujgD7xbrP33P}{S1260}
\lookupPut{R_6Lvphov0erHREDT}{S1261}
\lookupPut{R_1nSAQc4XIXMRaAp}{S1262}
\lookupPut{R_2AHIr9lYI4BGEUs}{S1263}
\lookupPut{R_1LhxJ0VwvuYACF0}{S1264}
\lookupPut{R_zfjUS0Wanm4fB4t}{S1265}
\lookupPut{R_3Dj5uCs1NYSHSl9}{S1266}
\lookupPut{R_VQDbTQZZNEGCdBn}{S1267}
\lookupPut{R_1H11MNgfFIAA1EY}{S1268}
\lookupPut{R_1M0baEf8QPeNsYK}{S1269}
\lookupPut{R_1Q4K53pErM23POb}{S1270}
\lookupPut{R_31QrodGHe4RfMxx}{S1271}
\lookupPut{R_3RedqoeAwTqaxWT}{S1272}
\lookupPut{R_2Tq6GGA7m0Qu4YC}{S1273}
\lookupPut{R_26gvWMScDPlUwGU}{S1274}
\lookupPut{R_3qe76venpsAOwPi}{S1275}
\lookupPut{R_2EypeHZDv6DZ3E8}{S1276}
\lookupPut{R_3KO6nuUKPAcRlXo}{S1277}
\lookupPut{R_urxOpTOmwN7WZoJ}{S1278}
\lookupPut{R_1GTiksoMeBgoklB}{S1279}
\lookupPut{R_r7QWDc95UDMHlfP}{S1280}
\lookupPut{R_77ptNkg4t5OlhgR}{S1281}
\lookupPut{R_3gZq2yXQ0Hx02Xk}{S1282}
\lookupPut{R_3qQbCWo3h3h1zzZ}{S1283}
\lookupPut{R_1puLmiU7VwZ1mqI}{S1284}
\lookupPut{R_2ylrJJ9mxGTgxmk}{S1285}
\lookupPut{R_2YrXKeCn8W2X146}{S1286}
\lookupPut{R_2q32J96w35uHoPU}{S1287}
\lookupPut{R_3j7D0c2tPcziAzU}{S1288}
\lookupPut{R_1JRccjsY1jARntY}{S1289}
\lookupPut{R_1nOUpVyd7Hkh0Dk}{S1290}
\lookupPut{R_3rS7WYb9lt08Ipt}{S1291}
\lookupPut{R_9ssrf3OiqPvFVbr}{S1292}
\lookupPut{R_w05dAjaKAbrjKgx}{S1293}
\lookupPut{R_29oCoSVadjYbLmO}{S1294}
\lookupPut{R_2Qss9QBDgnbtPTB}{S1295}
\lookupPut{R_1g5yN2qkZ5Wvv7Z}{S1296}
\lookupPut{R_2rwPS8T2H8aKAYu}{S1297}
\lookupPut{R_2CEfmAMsXAU9Z5x}{S1298}
\lookupPut{R_AHUjN3e0Q4gJfBD}{S1299}
\lookupPut{R_zUaegz44d7aajLj}{S1300}
\lookupPut{R_31b0o0YkfyVXxkU}{S1301}
\lookupPut{R_2tmLwLaFmGxJUO5}{S1302}
\lookupPut{R_1jO47FAngz1693D}{S1303}
\lookupPut{R_2B8q8YfoImKo4PH}{S1304}
\lookupPut{R_2Y5PjkSIkAKcjwY}{S1305}
\lookupPut{R_d5yczpU4vSrqSrL}{S1306}
\lookupPut{R_2ePYZXVf4srvZ0f}{S1307}
\lookupPut{R_Or1oMmJPbVlK5Tb}{S1308}
\lookupPut{R_2CKARJcFhsbn1Sg}{S1309}
\lookupPut{R_1diXoAqnecpzEsO}{S1310}
\lookupPut{R_3HzdhunHdNZTysT}{S1311}
\lookupPut{R_1EfZxiN8z2EZt26}{S1312}
\lookupPut{R_rkK22ltN6bMOPU5}{S1313}
\lookupPut{R_AAqbZs9FDfX28KJ}{S1314}
\lookupPut{R_2WOUmgSUIaX3kRk}{S1315}
\lookupPut{R_24COiWqfnv4rfNQ}{S1316}
\lookupPut{R_WwfVLEskWPPusYp}{S1317}
\lookupPut{R_21ms8OJ1rEJ3rJL}{S1318}
\lookupPut{R_21h0ovkd1nQXe41}{S1319}
\lookupPut{R_1eQxOU8bScdPlUD}{S1320}
\lookupPut{R_3Mu7rjnKk8sPejs}{S1321}
\lookupPut{R_3QFzuRo3xxUKD8J}{S1322}
\lookupPut{R_zTMSOsbG6Qsf1cZ}{S1323}
\lookupPut{R_3Gwi2NPnqUoQNRV}{S1324}
\lookupPut{R_1MXtXkYj0B9YqTv}{S1325}
\lookupPut{R_3j2ID9Z0RHZJShy}{S1326}
\lookupPut{R_XGtfuANVDZcromR}{S1327}
\lookupPut{R_25RcTVvRFxRFl1m}{S1328}
\lookupPut{R_ZHVk4oq1fy2gHK1}{S1329}
\lookupPut{R_3PpjITcolp4QfBZ}{S1330}
\lookupPut{R_2Sdy9uFKDF7EEVa}{S1331}
\lookupPut{R_tM4aL4iScwBznMZ}{S1332}
\lookupPut{R_2D2n85egfH0dPoU}{S1333}
\lookupPut{R_0VNVF4meLdBL4Xv}{S1334}
\lookupPut{R_3EZ72D6ZFMxO36s}{S1335}
\lookupPut{R_2X1ICkxi5SFjaZQ}{S1336}
\lookupPut{R_1E7q73ZNkdOZ0x0}{S1337}
\lookupPut{R_3oMkK1vNIDeKm2W}{S1338}
\lookupPut{R_3qqByerSFrL21KK}{S1339}
\lookupPut{R_T5uhGXCdvvelxCx}{S1340}
\lookupPut{R_r0Iw36U0LgUFJYZ}{S1341}
\lookupPut{R_1mRu0BboDETOlYC}{S1342}
\lookupPut{R_3dRVJJUbvbeshu9}{S1343}
\lookupPut{R_VIHh63OmT5dXoZP}{S1344}
\lookupPut{R_12mOaXu5rcne0as}{S1345}
\lookupPut{R_3PFXh7V7p7ClMIe}{S1346}
\lookupPut{R_1pFp1MC8TvcsdzQ}{S1347}
\lookupPut{R_306jmkRcpFE0v3U}{S1348}
\lookupPut{R_3PmBBSvAnRHOiBZ}{S1349}
\lookupPut{R_sdS9tur2My5Bycp}{S1350}
\lookupPut{R_3gYJq4JQfogxPE3}{S1351}
\lookupPut{R_2dsZwUDZBytuIYa}{S1352}
\lookupPut{R_2WT2f5cJmC67Paj}{S1353}
\lookupPut{R_3PKLTyKcOrsV58j}{S1354}
\lookupPut{R_6J4EBdvEaGqfXcR}{S1355}
\lookupPut{R_3OrjjoKpEOYuSck}{S1356}
\lookupPut{R_3KqrIB3cyZgK8y8}{S1357}
\lookupPut{R_2SfzpHmIsYg0r21}{S1358}
\lookupPut{R_2cc1D2lNvpeLmjg}{S1359}
\lookupPut{R_1pDwtRuXChbLaM4}{S1360}
\lookupPut{R_3nGbWz8zYa9LT43}{S1361}
\lookupPut{R_3NKu75eEbL263qM}{S1362}
\lookupPut{R_2DOcT8z6nKyOlt3}{S1363}
\lookupPut{R_2VF8570cPUij5MX}{S1364}
\lookupPut{R_vZz5K8TjzZXcO2Z}{S1365}
\lookupPut{R_2PBci06Vm3gDtcN}{S1366}
\lookupPut{R_1GNVzDnJX0dCTpP}{S1367}
\lookupPut{R_3pfmTTK4iS1t8bq}{S1368}
\lookupPut{R_11dBt7cPtxukXlz}{S1369}
\lookupPut{R_1FPPq8Wg2saiLOS}{S1370}
\lookupPut{R_1Kr3QB7GNxgjubO}{S1371}
\lookupPut{R_2tnC1woCzxPtD1H}{S1372}
\lookupPut{R_2zGcfhqTJGbWSP8}{S1373}
\lookupPut{R_1ePJfwQO6P3EA8i}{S1374}
\lookupPut{R_20ToiupSLgUByRk}{S1375}
\lookupPut{R_2t3TDnWj2DGa4zA}{S1376}
\lookupPut{R_RIdLefgJb3ePfW1}{S1377}
\lookupPut{R_1PaUjVbQ1oGSKTC}{S1378}
\lookupPut{R_6icUEbnZmWZvuXn}{S1379}
\lookupPut{R_116oFiypfh3B8f3}{S1380}
\lookupPut{R_1lhopOxKw4dpxH3}{S1381}
\lookupPut{R_3MzdrDfWpa2wrv0}{S1382}
\lookupPut{R_1IQKqwHfZ62zSmA}{S1383}
\lookupPut{R_UEoCJWvIHNWojvz}{S1384}
\lookupPut{R_1K0H2mkbKwFJCE9}{S1385}
\lookupPut{R_3hh9iuihPEGw2x7}{S1386}
\lookupPut{R_3w4EcDOnpHeIwxj}{S1387}
\lookupPut{R_2rrntiQKLMXIZBZ}{S1388}
\lookupPut{R_2bi0p9g953E9ASR}{S1389}
\lookupPut{R_2VNK6yLNhRxcfZ2}{S1390}
\lookupPut{R_UnFx78pNRlVCxa1}{S1391}
\lookupPut{R_1lxEvkF2VHJmX2W}{S1392}
\lookupPut{R_O2vwkqlPLYr6lZn}{S1393}
\lookupPut{R_0fEQSTBRLVOELzX}{S1394}
\lookupPut{R_3efu3YIZfR89rUW}{S1395}
\lookupPut{R_2RaWMlrWmeYgKRc}{S1396}
\lookupPut{R_2TnQ0t8eFStKQ82}{S1397}
\lookupPut{R_bKt41gHKAqjQsff}{S1398}
\lookupPut{R_2gxMGUDpIKgfynf}{S1399}
\lookupPut{R_RWDJlv8BuAssZ8J}{S1400}
\lookupPut{R_2dGePOzugv4D0Us}{S1401}
\lookupPut{R_3j9EF8LjrlrXEtW}{S1402}
\lookupPut{R_2R500FJeUWYrGfs}{S1403}
\lookupPut{R_8faYLW5EKVnMX97}{S1404}
\lookupPut{R_si1dNsuLhefPLGN}{S1405}
\lookupPut{R_2e5nIlXU4fVqtOq}{S1406}
\lookupPut{R_a3iCR321W5MtAGZ}{S1407}
\lookupPut{R_1Ldmn5n23jy8dVL}{S1408}
\lookupPut{R_BtPnKVQC2URrngZ}{S1409}
\lookupPut{R_1poIPsyNkb6siCI}{S1410}
\lookupPut{R_1N3KuFmaXul3iMn}{S1411}
\lookupPut{R_1LNmO8pTI75T1cI}{S1412}
\lookupPut{R_2Skt0J9RQX3rvHY}{S1413}
\lookupPut{R_28MzYYB0mUD3FMY}{S1414}
\lookupPut{R_3GiG6JMtcoBEDNJ}{S1415}
\lookupPut{R_2tA5fIVNICb3AkR}{S1416}
\lookupPut{R_2SuWsYvLuNyKCUb}{S1417}
\lookupPut{R_1nOVLasPUuP9hKv}{S1418}
\lookupPut{R_e5SDcBtwvcdJIS5}{S1419}
\lookupPut{R_2upSb5PWRBJ6QQQ}{S1420}
\lookupPut{R_XC8VX44vyWj24PD}{S1421}
\lookupPut{R_215pkJBwvnVqiA2}{S1422}
\lookupPut{R_RqsoRgMbKcuHTTX}{S1423}
\lookupPut{R_2dEhCKYpIXLUBJi}{S1424}
\lookupPut{R_2U4Jyvh1V2fJ3lC}{S1425}
\lookupPut{R_baoHjxreiQ9czfj}{S1426}
\lookupPut{R_SVspwC7Vzt2v7PP}{S1427}
\lookupPut{R_3fleMBQQo46z4lf}{S1428}
\lookupPut{R_eXxACuWhbkJ2DCx}{S1429}
\lookupPut{R_31AjpK6YhZa8SL5}{S1430}
\lookupPut{R_2WVcQeGtPIrxXrv}{S1431}
\lookupPut{R_3lK5MgPNj9JLvCU}{S1432}
\lookupPut{R_2TyOTrFpQnKfAXY}{S1433}
\lookupPut{R_27esq13IIKFN34c}{S1434}
\lookupPut{R_38jk4IYWBR23jTH}{S1435}
\lookupPut{R_1Cm3Xm7YHEbUrUw}{S1436}
\lookupPut{R_3KUfpmEzZy9LkKz}{S1437}
\lookupPut{R_9st9Cfn4oORBZn3}{S1438}
\lookupPut{R_1mWOPt1bJULe6qn}{S1439}
\lookupPut{R_1FtFCprvyru1tS5}{S1440}
\lookupPut{R_1PUGUwRZIvezh5Y}{S1441}
\lookupPut{R_1LBGiQagT6PIF45}{S1442}
\lookupPut{R_2PzAtLooQIv4r8f}{S1443}
\lookupPut{R_2EF6ni1mpQbCuJd}{S1444}
\lookupPut{R_3svsEqmbYLyW2no}{S1445}
\lookupPut{R_3NDN9x23Ul2V1QF}{S1446}
\lookupPut{R_1hGywgGzEAzW78l}{S1447}
\lookupPut{R_2rw2JEqBaHXvSEK}{S1448}
\lookupPut{R_1TfpXZXckOuUSA1}{S1449}
\lookupPut{R_4IVZgmu25UwijD3}{S1450}
\lookupPut{R_2UgcP5Q8NVFsgJM}{S1451}
\lookupPut{R_D11ybebMAbfC4kF}{S1452}
\lookupPut{R_1QrvK229LF9l0SI}{S1453}
\lookupPut{R_swlR6zV9iB82vtv}{S1454}
\lookupPut{R_1GP6wxJa7BQ8KQy}{S1455}
\lookupPut{R_Oy4weJHI2AK8ceZ}{S1456}
\lookupPut{R_T7rWwtykhMi1h7z}{S1457}
\lookupPut{R_1IXluXGijbUJbo5}{S1458}
\lookupPut{R_sdlaWnvjDnKtxrb}{S1459}
\lookupPut{R_cCQw9JZulObFJOV}{S1460}
\lookupPut{R_3Ra2lgK75gsbhPu}{S1461}
\lookupPut{R_1mgYLs5BwpA2set}{S1462}
\lookupPut{R_1gLQ8vF5UK8fYyG}{S1463}
\lookupPut{R_ZgT9sEzXK7VDMSB}{S1464}
\lookupPut{R_302ZlQkfZn1Kuro}{S1465}
\lookupPut{R_1C28cLf9szRAMCs}{S1466}
\lookupPut{R_301rLpIvsdSCavq}{S1467}
\lookupPut{R_sdNKTbs89yhPTcR}{S1468}
\lookupPut{R_1JDzPa534INjq2S}{S1469}
\lookupPut{R_AnhisexeHW1rh1T}{S1470}
\lookupPut{R_2ckDlSf20Rk08U8}{S1471}
\lookupPut{R_vcpjsDW2b0u9LTb}{S1472}
\lookupPut{R_3gUFUTkwinUSBfl}{S1473}
\lookupPut{R_31Ru5IXAsQtwVM0}{S1474}
\lookupPut{R_12QLH9t2G1qGXS3}{S1475}
\lookupPut{R_cIK9S4K5dyxXDgZ}{S1476}
\lookupPut{R_3JmlkgD5Knfygnr}{S1477}
\lookupPut{R_3Rf3trS8oNhtxwc}{S1478}
\lookupPut{R_2D1WyPEILTkc6u2}{S1479}
\lookupPut{R_1mw12sJjww404I1}{S1480}
\lookupPut{R_30u6YZTnwdZwVsl}{S1481}
\lookupPut{R_2uhQxNmuBXQs4xz}{S1482}
\lookupPut{R_xaY8P38DgsorBC1}{S1483}
\lookupPut{R_247VxjE14LWgKOD}{S1484}
\lookupPut{R_2vddXaExxATdEmA}{S1485}
\lookupPut{R_12hG4sBNHrvUTYm}{S1486}
\lookupPut{R_2uPwi8t96THJ2zb}{S1487}
\lookupPut{R_1IKoHiOAFOIbiN4}{S1488}
\lookupPut{R_3DceVFH7OPnwsQn}{S1489}
\lookupPut{R_1LjkFSrr4UMcBDW}{S1490}
\lookupPut{R_BRHB3cDZiSaYULL}{S1491}
\lookupPut{R_BS3GZ1MI2jdVGaR}{S1492}
\lookupPut{R_1LGmZtjJwTKgfw1}{S1493}
\lookupPut{R_3sdHoQekcjOurOx}{S1494}
\lookupPut{R_2c5fxAS4gSarZdd}{S1495}
\lookupPut{R_2DTRU1Z8goZewWW}{S1496}
\lookupPut{R_10I9SmdQlGyaC88}{S1497}
\lookupPut{R_rjyTxRbWzPocj2F}{S1498}
\lookupPut{R_3Ec02xrHZxbVYH2}{S1499}
\lookupPut{R_1KMW2OPUokm1WvL}{S1500}
\lookupPut{R_3hmrbJouTTsOiAe}{S1501}
\lookupPut{R_2PyR8K9EEffMuMQ}{S1502}
\lookupPut{R_2YlCn6sqOMSML7e}{S1503}
\lookupPut{R_1k1s3KhARehHlql}{S1504}
\lookupPut{R_z7jRpDFa3E1oe8V}{S1505}
\lookupPut{R_1Ne6KUpL13pZjDm}{S1506}
\lookupPut{R_2X1iR4KJQbweDfU}{S1507}
\lookupPut{R_3jU0bQTqIcnYM5F}{S1508}
\lookupPut{R_1jNeahexwLuUf1C}{S1509}
\lookupPut{R_3HSUzNYGxFml2xn}{S1510}
\lookupPut{R_1Qobc8z5RXfs01d}{S1511}
\lookupPut{R_BLnuuhFvCISYftL}{S1512}
\lookupPut{R_3ewu90wWq8InZ7y}{S1513}
\lookupPut{R_12RTlSQS9XKtota}{S1514}
\lookupPut{R_302Hdtixsh9bLak}{S1515}
\lookupPut{R_3C6v3sb3AyViBaN}{S1516}
\lookupPut{R_3qfYBf469P2unsk}{S1517}
\lookupPut{R_2uKBVfKcXsNsQsm}{S1518}
\lookupPut{R_2rC4yZ0gAmT0Kqo}{S1519}
\lookupPut{R_u395W2nvTXwiHXH}{S1520}
\lookupPut{R_3EsB81iJ7dXLOBl}{S1521}
\lookupPut{R_1KveREGEiNjYKxQ}{S1522}
\lookupPut{R_2E06kwMiyfsXVuA}{S1523}
\lookupPut{R_3M0x75VElhw6fbv}{S1524}
\lookupPut{R_2QyB3uSFyGnen42}{S1525}
\lookupPut{R_1eJmDt44M0H4peW}{S1526}
\lookupPut{R_2YLEYDWYUOKkVx2}{S1527}
\lookupPut{R_DDk253yi4vLaxBT}{S1528}
\lookupPut{R_3WxDH538kA2cXWV}{S1529}
\lookupPut{R_3qEvxUJ1Fk0QBAh}{S1530}
\lookupPut{R_cMUNL2lCYynHFNn}{S1531}
\lookupPut{R_28YzJ6BdJkOXEan}{S1532}
\lookupPut{R_3oUhkBlsp82DdQ1}{S1533}
\lookupPut{R_238fXmI1Gh4V6PX}{S1534}
\lookupPut{R_2v5GVVlnifqojor}{S1535}
\lookupPut{R_28FE46r6TUEOgcl}{S1536}
\lookupPut{R_3iP7aNdgY9nZZ84}{S1537}
\lookupPut{R_6nAdZOIRXpTRaCJ}{S1538}
\lookupPut{R_Q0KGgvRjDVzzC2B}{S1539}
\lookupPut{R_eEFoJ5WvSj5NgMF}{S1540}
\lookupPut{R_1QseRVVhKHCrICz}{S1541}
\lookupPut{R_1mUANPadSgSczlp}{S1542}
\lookupPut{R_2doLjOhCX2doFHt}{S1543}
\lookupPut{R_3GvNikgfh12wRf9}{S1544}
\lookupPut{R_3RmabvesfTw28e4}{S1545}
\lookupPut{R_30rhaQGiAOD4PT1}{S1546}
\lookupPut{R_3jT2j9vrH1zcs8F}{S1547}
\lookupPut{R_1eXeFG9ncYDdJNW}{S1548}
\lookupPut{R_3G338mDuY7iPoxV}{S1549}
\lookupPut{R_ZypOsmJJ7mE19dL}{S1550}
\lookupPut{R_1NfbPLxwFrWl8uA}{S1551}
\lookupPut{R_1jvTwMYSunLcgUT}{S1552}
\lookupPut{R_1odmZFrZiDzKQAO}{S1553}
\lookupPut{R_3kvle4EoWAlj7gV}{S1554}
\lookupPut{R_PBUIvAZLPkD4Mrn}{S1555}
\lookupPut{R_3HImnIpT696LTQc}{S1556}
\lookupPut{R_bCtmfNydXSpkBLb}{S1557}
\lookupPut{R_3MzhLjRcsLQdp8B}{S1558}
\lookupPut{R_2dF3uL7cOiZ5N5I}{S1559}
\lookupPut{R_3gXzQDdPmmJ3Bmd}{S1560}
\lookupPut{R_2QJVrw1SnUJo5mE}{S1561}
\lookupPut{R_25RUhPXZbCQzRQi}{S1562}
\lookupPut{R_323m3mPnCaElT99}{S1563}
\lookupPut{R_pguL5Be114SHMXv}{S1564}
\lookupPut{R_4UsHMpoWoGZyCOZ}{S1565}
\lookupPut{R_30xgSkOVGnfD8P6}{S1566}
\lookupPut{R_10ra0JMqE35KrbJ}{S1567}
\lookupPut{R_2eatIUOphZaFm1g}{S1568}
\lookupPut{R_30i5qdZ97U2hrkr}{S1569}
\lookupPut{R_2Tw9ud9QOStmFYq}{S1570}
\lookupPut{R_3CPfdNyJRt2LSUf}{S1571}
\lookupPut{R_cZPkJUCAZGprhXH}{S1572}
\lookupPut{R_AAnld6p73HAvEUp}{S1573}
\lookupPut{R_2CHTATzIdrQwyrC}{S1574}
\lookupPut{R_w7aM2eqztXc7amd}{S1575}
\lookupPut{R_2Ubedb24Qjt6e0y}{S1576}
\lookupPut{R_RrVUi5JVIzxC8SZ}{S1577}
\lookupPut{R_3Kx7MPgI9ztsZIC}{S1578}
\lookupPut{R_294ivVoQ5dZ9FTx}{S1579}
\lookupPut{R_3EnAkfmywF6K410}{S1580}
\lookupPut{R_1KrMSR51MafWfx8}{S1581}
\lookupPut{R_1MYCcJfqw0DmnGV}{S1582}
\lookupPut{R_2VEWcInk97uCCzC}{S1583}
\lookupPut{R_3lyOf2f73PqyUrH}{S1584}
\lookupPut{R_AdLyT0mO9sgHqaR}{S1585}
\lookupPut{R_2pYnZv9kYzdE6Jq}{S1586}
\lookupPut{R_2wgeyHJJNN0ZcMX}{S1587}
\lookupPut{R_2EslYPeCgXdgDUj}{S1588}
\lookupPut{R_eDHtqJgnCRXO9Al}{S1589}
\lookupPut{R_3O8uIBP7om4JWKE}{S1590}
\lookupPut{R_2S78Wma22HLBVJh}{S1591}
\lookupPut{R_55qe6mO0xVOoWRP}{S1592}
\lookupPut{R_2sT0HLRSqLAAouK}{S1593}
\lookupPut{R_STp3CUmjuPtftVT}{S1594}
\lookupPut{R_PFcLF0WmgutY6xH}{S1595}
\lookupPut{R_XIj55a4JMQcopt7}{S1596}
\lookupPut{R_1gRD65IKTZ4NM76}{S1597}
\lookupPut{R_3KD4NjjkhRHsHJZ}{S1598}
\lookupPut{R_VXCu1fqJsFTLHmp}{S1599}
\lookupPut{R_3gNLjrrsD1ur51Z}{S1600}
\lookupPut{R_2uHTjwSVfdvRFaT}{S1601}
\lookupPut{R_Y6zMlEx82cYsLAt}{S1602}
\lookupPut{R_2BbqvoXMVpttNlY}{S1603}
\lookupPut{R_C9PkioDGKLMaFTH}{S1604}
\lookupPut{R_WrjyyeK16cGX0pr}{S1605}
\lookupPut{R_2rJGRs3dp0QDMLI}{S1606}
\lookupPut{R_AsqcVgvjUh2bRLj}{S1607}
\lookupPut{R_6kZym2UhxFdSSbL}{S1608}
\lookupPut{R_yRdGDKrs7cmuF8J}{S1609}
\lookupPut{R_29sP21vBNoqKG2r}{S1610}
\lookupPut{R_1hFOkbiN6tPEVv5}{S1611}
\lookupPut{R_10SMcDcGw0nCGHn}{S1612}
\lookupPut{R_3nk1GWoeBRMq8Ej}{S1613}
\lookupPut{R_3fcO6GG6M0crK9v}{S1614}
\lookupPut{R_ezeFz0MhiVzvLQ5}{S1615}
\lookupPut{R_2YKCtzFfV2lVCZI}{S1616}
\lookupPut{R_1M0Qs718r8Fp3Hg}{S1617}
\lookupPut{R_25Za4xCN8cNdJyY}{S1618}
\lookupPut{R_1K3EX452Skfa6vC}{S1619}
\lookupPut{R_2BmuwPQOjBV7IMc}{S1620}
\lookupPut{R_3ND7bnGVtSDCHns}{S1621}
\lookupPut{R_3L5CEPm3zPpRwQK}{S1622}
\lookupPut{R_tMpc4iZjQq5PEwp}{S1623}
\lookupPut{R_3flX7Yo8rZDEjtG}{S1624}
\lookupPut{R_2YmiPfJ5XAmy1JA}{S1625}
\lookupPut{R_3Op6nRTpCVZYeVU}{S1626}
\lookupPut{R_2dSza1fbBig9MPL}{S1627}
\lookupPut{R_2zACgZU19dnSGNj}{S1628}
\lookupPut{R_2QzfFkbd1gkSlMZ}{S1629}
\lookupPut{R_dgwEplTeRH1OmQx}{S1630}
\lookupPut{R_vcAVPpRqdK3leBX}{S1631}
\lookupPut{R_2rClR25Na4E0aKc}{S1632}
\lookupPut{R_2E50OV1fYbPw5J6}{S1633}
\lookupPut{R_3DoI1nADnOIyfQu}{S1634}
\lookupPut{R_1dgjJiGqh9ycjt5}{S1635}
\lookupPut{R_tDjz9t0gznhwbyp}{S1636}
\lookupPut{R_24uZGhH0wYbbUhL}{S1637}
\lookupPut{R_2WGQvQHOhfSFw90}{S1638}
\lookupPut{R_C8kdf5ANuAJOGcx}{S1639}
\lookupPut{R_1CKIzHCqZAeap7I}{S1640}
\lookupPut{R_3PTxc3ASyWAInLn}{S1641}
\lookupPut{R_1CKm4V4sqiNSgAj}{S1642}
\lookupPut{R_3Oh485H8Cr10OLn}{S1643}
\lookupPut{R_25CqAG73RfoR1Id}{S1644}
\lookupPut{R_1lobVxup420M1Md}{S1645}
\lookupPut{R_2pVK8FQCW5j5Qpu}{S1646}
\lookupPut{R_2s5T9EKmP9SvA5m}{S1647}
\lookupPut{R_3mjDrgtxpoNtFDV}{S1648}
\lookupPut{R_2QAp7spHLgAr5Cl}{S1649}
\lookupPut{R_6rkWD1Ud2qvo265}{S1650}
\lookupPut{R_1EYheiPusDnqLbG}{S1651}
\lookupPut{R_3Li2pnZhRGCa6rK}{S1652}
\lookupPut{R_3OeT7k6uVGrCReZ}{S1653}
\lookupPut{R_1NbiAc7Io2ZeGUO}{S1654}
\lookupPut{R_2SjABmqohYIjkbb}{S1655}
\lookupPut{R_1OMGC8SeAxqbGcI}{S1656}
\lookupPut{R_3dSUDLAYI0yAm3J}{S1657}
\lookupPut{R_3nSI5nEOMWmokL4}{S1658}
\lookupPut{R_892P1TvhJxzE9Ed}{S1659}
\lookupPut{R_1mls0I6fpsAqnTz}{S1660}
\lookupPut{R_2ylbCpj80BeqTiA}{S1661}
\lookupPut{R_1gTTfe9XE3zTBcr}{S1662}
\lookupPut{R_riQNkUEUCqTV3TH}{S1663}
\lookupPut{R_1jHhimzOxyzKBZB}{S1664}
\lookupPut{R_rjoN5xdU1pPHOMh}{S1665}
\lookupPut{R_3fqIl3jU2GEb3c4}{S1666}
\lookupPut{R_2uWjzmBgo3O7Fhe}{S1667}
\lookupPut{R_1pVx7SSRqvREpxg}{S1668}
\lookupPut{R_cBI9KUr88QEUkEN}{S1669}
\lookupPut{R_qJfLlkh9GDDQLAd}{S1670}
\lookupPut{R_DAc8ZJ1EN5JrmmJ}{S1671}
\lookupPut{R_6GtJD5VVJe0JPoZ}{S1672}
\lookupPut{R_3oOaGFq5nFqxAHX}{S1673}
\lookupPut{R_3QKN6AE0ka8Wr2x}{S1674}
\lookupPut{R_1JDAc1S17z28nOw}{S1675}
\lookupPut{R_3oAEaPqTkfAql1W}{S1676}
\lookupPut{R_r3aGBuzNF0tTfqh}{S1677}
\lookupPut{R_1FDvmeNgHuH0n8q}{S1678}
\lookupPut{R_2TuwdGl95GpyL67}{S1679}
\lookupPut{R_3kGNB3oeI0oOaOA}{S1680}
\lookupPut{R_1jZMcnLAXsNW3Vf}{S1681}
\lookupPut{R_215mI1760SWJJDS}{S1682}
\lookupPut{R_1Hnr2m9xEkuV5m1}{S1683}
\lookupPut{R_3PvLCHSTChe9W34}{S1684}
\lookupPut{R_3OrHtvIe7X0mk39}{S1685}
\lookupPut{R_3R4MQQm7dSVN5fu}{S1686}
\lookupPut{R_1GVpCyeSUlRJaWC}{S1687}
\lookupPut{R_2WYBBmXunRkOgsc}{S1688}
\lookupPut{R_1IztEiRyr3aqoGw}{S1689}
\lookupPut{R_56VmeCG8nNY7DDX}{S1690}
\lookupPut{R_2QXO6e0xaGdfEg2}{S1691}
\lookupPut{R_3lRJcQbFrvBlBFo}{S1692}
\lookupPut{R_1pWE6LHM1LBrS2A}{S1693}
\lookupPut{R_2EfUUWdL32pINQe}{S1694}
\lookupPut{R_3EYngCYKfsP80ry}{S1695}
\lookupPut{R_3qlDD8eedwY2BeS}{S1696}
\lookupPut{R_3HRBIz6OvrK4ttO}{S1697}
\lookupPut{R_eJUJiJNO4lmNnk5}{S1698}
\lookupPut{R_27qljG8OYQOxxii}{S1699}
\lookupPut{R_Q5JXd7WlvJX14cN}{S1700}
\lookupPut{R_upRZWKqzZ8Kn4el}{S1701}
\lookupPut{R_1EhU2XNN1yECtaj}{S1702}
\lookupPut{R_3hbBFgOatLASJFF}{S1703}
\lookupPut{R_9Xfo0JBUr4B5HTb}{S1704}
\lookupPut{R_22qUZpLFEF8FjVG}{S1705}
\lookupPut{R_1Q3n78ru4E82c4H}{S1706}
\lookupPut{R_1joA83XOEbnPXX5}{S1707}
\lookupPut{R_tGUTKKH3ywaxNS1}{S1708}
\lookupPut{R_33r8L9lPyih5mok}{S1709}
\lookupPut{R_3ozJUeqTUbYQTYM}{S1710}
\lookupPut{R_24dS2ti4Jjwz1ue}{S1711}
\lookupPut{R_2fm61mhcY0ZunoH}{S1712}
\lookupPut{R_1nUY8JztZsmQiO5}{S1713}
\lookupPut{R_1PSkeUfLC6RMWfh}{S1714}
\lookupPut{R_1d4b7cB1YBqQah9}{S1715}
\lookupPut{R_WuI7W3t3MbGTwZz}{S1716}
\lookupPut{R_1Kd48jl8IV0OOlO}{S1717}
\lookupPut{R_3gT2F6FRBZRHNxz}{S1718}
\lookupPut{R_3g8GHmC76FYDKhP}{S1719}
\lookupPut{R_3dQcnXebDEKhHuH}{S1720}
\lookupPut{R_1l6UcUsKEKfgVAc}{S1721}
\lookupPut{R_ONkPkt4jhQSlbhL}{S1722}
\lookupPut{R_2EtvLqLzheYpR3k}{S1723}
\lookupPut{R_2dsuFkCWlTp8jj4}{S1724}
\lookupPut{R_2CHC9N00zaHMLKP}{S1725}
\lookupPut{R_1cUpnuPdhSa0ke4}{S1726}
\lookupPut{R_O1ZKUV41CwqxEWJ}{S1727}
\lookupPut{R_3Ej9ZFnye1Yf4Zs}{S1728}
\lookupPut{R_110lQn0BS8rgj3G}{S1729}
\lookupPut{R_1CIreXBuPUAkqaR}{S1730}
\lookupPut{R_1r86EjfcWW03yPJ}{S1731}
\lookupPut{R_1ptKzxvPZdMTiaI}{S1732}
\lookupPut{R_vIAXLvehGBRY7mN}{S1733}
\lookupPut{R_DcchxGIKgK5pbqN}{S1734}
\lookupPut{R_0qCcgXzXFsTff8d}{S1735}
\lookupPut{R_2EghYWNqSxVSFrj}{S1736}
\lookupPut{R_1hKz4vTyj9tZVjo}{S1737}
\lookupPut{R_2PCmXcicmdnPjuF}{S1738}
\lookupPut{R_2VropcEirhGKINz}{S1739}
\lookupPut{R_3KPWWfALcVhaGOx}{S1740}
\lookupPut{R_1E051YmYkaK2fJX}{S1741}
\lookupPut{R_3qHVYixJB2zQZ7u}{S1742}
\lookupPut{R_pKLFWY40yAdFaA9}{S1743}
\lookupPut{R_BLcKpxMyUzEGhP3}{S1744}
\lookupPut{R_2VrmJl8SLEmuhxf}{S1745}
\lookupPut{R_2frP51cu9Dhtq5p}{S1746}
\lookupPut{R_3EavxrPSiF4r1O5}{S1747}
\lookupPut{R_2aF7u4IvW6cPfaD}{S1748}
\lookupPut{R_pGHJ5h0eNof6KvT}{S1749}
\lookupPut{R_3kB6J8N22gLaDOS}{S1750}
\lookupPut{R_C8QGYNfZqGiWC7n}{S1751}
\lookupPut{R_3LZUn6OfMTKS3tY}{S1752}
\lookupPut{R_2ffIE3GMaTsrvWv}{S1753}
\lookupPut{R_2Pdur0CcOuBk8Of}{S1754}
\lookupPut{R_2tA86bGi8BUojj8}{S1755}
\lookupPut{R_w1QQVyfuW7WYPhD}{S1756}
\lookupPut{R_etxVat00eJukvIJ}{S1757}
\lookupPut{R_2R2NWRQ6v6PuRPI}{S1758}
\lookupPut{R_1P7rtnYdxUAdla8}{S1759}
\lookupPut{R_2cC0P7RFjInsKvx}{S1760}
\lookupPut{R_eKUhM8togb26TVT}{S1761}
\lookupPut{R_3qDm3kBlqKHK86i}{S1762}
\lookupPut{R_27P8iSHJ2l4G1He}{S1763}
\lookupPut{R_3L6DoeuUna40fOM}{S1764}
\lookupPut{R_1MLefKdPtY2HMX4}{S1765}
\lookupPut{R_CfwP0a3ky6ni5Il}{S1766}
\lookupPut{R_2TyGPeM4jvOB3vA}{S1767}
\lookupPut{R_1dpfaL6t1owQHjb}{S1768}
\lookupPut{R_2ypFyNp6eTaIorP}{S1769}
\lookupPut{R_behLeTisDHm7s1X}{S1770}
\lookupPut{R_2aQgfsExsJsHJ5d}{S1771}
\lookupPut{R_3n2Wj6s2XYoG7FE}{S1772}
\lookupPut{R_3Jt4Ou4rv5gBF4r}{S1773}
\lookupPut{R_3NI3lJNo2G2j4ur}{S1774}
\lookupPut{R_33qlFsO5W7h0Z2Y}{S1775}
\lookupPut{R_2X4BEY6gmyQsBjD}{S1776}
\lookupPut{R_2cbhXlQZMam5X3O}{S1777}
\lookupPut{R_27lipZvxZ4H0l4Z}{S1778}
\lookupPut{R_3dLWy7x6jiKGScR}{S1779}
\lookupPut{R_3qjPkeDlwc0SWy5}{S1780}
\lookupPut{R_2y7P4kkKXSR9Z15}{S1781}
\lookupPut{R_262Qkcrthyj6c2Z}{S1782}
\lookupPut{R_5gWeR3A4QTssXiF}{S1783}
\lookupPut{R_21HfBUKSJyQN6fo}{S1784}
\lookupPut{R_3ISs25uvnVI3gI3}{S1785}
\lookupPut{R_1MLDUqTn3WTDooh}{S1786}
\lookupPut{R_7TGIWcE4Se48i5P}{S1787}
\lookupPut{R_2qjBHxyRBJGBmGr}{S1788}
\lookupPut{R_UuUFa7uJ068DuF3}{S1789}
\lookupPut{R_30cNvkSjokmCTz4}{S1790}
\lookupPut{R_1E6bsQRtsAn036u}{S1791}
\lookupPut{R_rprxPl56P6Udr7X}{S1792}
\lookupPut{R_31ZCgqIYfyzgIjN}{S1793}
\lookupPut{R_2wzsiWcf0Zs8CFU}{S1794}
\lookupPut{R_2XiBnWDkIbiIEKQ}{S1795}
\lookupPut{R_1mIbkPQWTQ0QnFb}{S1796}
\lookupPut{R_baAnQOlN0M3mVmV}{S1797}
\lookupPut{R_3iPTDngCXdiDZeG}{S1798}
\lookupPut{R_2QmN8QhatJ3zVbz}{S1799}
\lookupPut{R_29brVKBzC1SHyfB}{S1800}
\lookupPut{R_2dfFNs2SUrossAW}{S1801}
\lookupPut{R_1dynn18K5cqNhLc}{S1802}
\lookupPut{R_3oTRJL25s5MreHp}{S1803}
\lookupPut{R_3lKMAYNthY9pcLW}{S1804}
\lookupPut{R_3M5L4JrRceKlMRJ}{S1805}
\lookupPut{R_28Gqu3tjMciT7gK}{S1806}
\lookupPut{R_3JmfUbiTjADGIb5}{S1807}
\lookupPut{R_3030UUEpqnpcY0N}{S1808}
\lookupPut{R_31QT41NDj6EnaeN}{S1809}
\lookupPut{R_2zC2wYWHdwR9Gvj}{S1810}
\lookupPut{R_22nT66B63Uq8eED}{S1811}
\lookupPut{R_2ykg15vv9CHcu5g}{S1812}
\lookupPut{R_8j0aowEa3EvlXix}{S1813}
\lookupPut{R_22YxfUI6C6pHsVq}{S1814}
\lookupPut{R_2fwVrv8tUVCNBrQ}{S1815}
\lookupPut{R_23epSrsp1TsSHba}{S1816}
\lookupPut{R_3Jn1mK1gHAbMEu4}{S1817}
\lookupPut{R_9HmmxLt4LZbcuDn}{S1818}
\lookupPut{R_1OQa3IGgldq9KbD}{S1819}
\lookupPut{R_2uBt9eVBBLVv9TE}{S1820}
\lookupPut{R_2SkTfY2pTCC3yH1}{S1821}
\lookupPut{R_10Ji8pamY8McQjk}{S1822}
\lookupPut{R_0JQRBkPmuc7JoBj}{S1823}
\lookupPut{R_2awrSFW92hAnKzF}{S1824}
\lookupPut{R_32Khe4uT2FvjxNF}{S1825}
\lookupPut{R_28AEGmkctYuUSqq}{S1826}
\lookupPut{R_3kuDAYuAINmuw58}{S1827}
\lookupPut{R_2y3ClHlT4COcqmZ}{S1828}
\lookupPut{R_2B4FevEwCISpdjY}{S1829}
\lookupPut{R_89a9uMdgjIONyFz}{S1830}
\lookupPut{R_2Yt6h331J8B1IWo}{S1831}
\lookupPut{R_27lNhA2yw0pQv0K}{S1832}
\lookupPut{R_1lbimxXMqOnnKkP}{S1833}
\lookupPut{R_3irIlfyCgBxNAyo}{S1834}
\lookupPut{R_XFzlUmYyf1y5zhL}{S1835}
\lookupPut{R_3NyNuK5PhhJ3ztb}{S1836}
\lookupPut{R_1HiWKnDqaVskMgf}{S1837}
\lookupPut{R_1g5wQ0fZFCDA494}{S1838}
\lookupPut{R_CmFOcXdiM2OmQXD}{S1839}
\lookupPut{R_pa4Azj57DrrARtD}{S1840}
\lookupPut{R_3n1rKRc8n56Wpih}{S1841}
\lookupPut{R_XNibxQtEmpXg2k1}{S1842}
\lookupPut{R_28GO5NgtVnaanu4}{S1843}
\lookupPut{R_zU2PACXU11klHRT}{S1844}
\lookupPut{R_32Xeri5zHxGCEXg}{S1845}
\lookupPut{R_3iqxuj81dAHSACa}{S1846}
\lookupPut{R_p9Tn4RfexNCOJIB}{S1847}
\lookupPut{R_3huOXVWJUPDMfY0}{S1848}
\lookupPut{R_9uxkfS9O9bVmfqF}{S1849}
\lookupPut{R_2VeEgV9BmmD1sVn}{S1850}
\lookupPut{R_2qE5sVIPw2BF5V5}{S1851}
\lookupPut{R_1rv5dPEzFl9ruqU}{S1852}
\lookupPut{R_Q6QSOfwpscbmu89}{S1853}
\lookupPut{R_2SGFOLMR1O5lGxL}{S1854}
\lookupPut{R_3HM5D6hIBH2dyRV}{S1855}
\lookupPut{R_1KvtMcWtwvVeCtY}{S1856}
\lookupPut{R_Cjq3zQUS30ofiKZ}{S1857}
\lookupPut{R_3nOfCHazyuX6Pv8}{S1858}
\lookupPut{R_1CdqfB5LT4G6PaG}{S1859}
\lookupPut{R_OGOcsYcqgrPaRBD}{S1860}
\lookupPut{R_2ayqYubfqbteTr3}{S1861}
\lookupPut{R_1qa3d7PzfSOQhBu}{S1862}
\lookupPut{R_1N2geZc53EWgSc6}{S1863}
\lookupPut{R_2B8OL8r7GR97pON}{S1864}
\lookupPut{R_RtenMoHZnjGRB8l}{S1865}
\lookupPut{R_3RpZxMqlz70Fowo}{S1866}
\lookupPut{R_242YkwKsDAnVE02}{S1867}
\lookupPut{R_2huKw1OL7ZirIk1}{S1868}
\lookupPut{R_2DZZ2aPYfC0GXYI}{S1869}
\lookupPut{R_pMfJI4decYay9dn}{S1870}
\lookupPut{R_2eakCO8BGHDdsXn}{S1871}
\lookupPut{R_yyiY94lMBnO8vE5}{S1872}
\lookupPut{R_1FwyE74xKu679jh}{S1873}
\lookupPut{R_1FzlsBh6K0ToiUL}{S1874}
\lookupPut{R_O3SG0DNoGlWqQ3T}{S1875}
\lookupPut{R_2aFwvkwQVCCMdKA}{S1876}
\lookupPut{R_1lv5nl5r6R8424A}{S1877}
\lookupPut{R_27QjqrwV0DkqBvh}{S1878}
\lookupPut{R_Au0H0bOpKuOGD3b}{S1879}
\lookupPut{R_eCBYudxy43nAg9j}{S1880}
\lookupPut{R_3pnajpLfFhWKhHA}{S1881}
\lookupPut{R_1QGBQ4seupTsb42}{S1882}
\lookupPut{R_0TXiRFuIz132EgN}{S1883}
\lookupPut{R_en8lO1TiLqljjot}{S1884}
\lookupPut{R_DUojphyGEmCzg2J}{S1885}
\lookupPut{R_tGazD15oWUePoWZ}{S1886}
\lookupPut{R_WuntqAxvCidOZuV}{S1887}
\lookupPut{R_2Qh7kAgkKIC9zVV}{S1888}
\lookupPut{R_1KlERrMltIBcuXQ}{S1889}
\lookupPut{R_3KxpOylNmQd8Dcn}{S1890}
\lookupPut{R_qWTI2UUx9zbofst}{S1891}
\lookupPut{R_12qXki55O859VEv}{S1892}
\lookupPut{R_21tcHsNWRmNdBN8}{S1893}
\lookupPut{R_12bsWs9WCLWxQIs}{S1894}
\lookupPut{R_1gI5Ki0JUBtiWWH}{S1895}
\lookupPut{R_5pyGnxA76vDYrKN}{S1896}
\lookupPut{R_3kgKCDU8QBBZBt7}{S1897}
\lookupPut{R_2Eo6uhO7xqhu5Oc}{S1898}
\lookupPut{R_2CjKyKvsD41FTPX}{S1899}
\lookupPut{R_2TAm2eu4XHdLzdM}{S1900}
\lookupPut{R_6MuI5gpXOtbNezT}{S1901}
\lookupPut{R_31Zsx3ARjAcWTn8}{S1902}
\lookupPut{R_3j39SThjumNQYcq}{S1903}
\lookupPut{R_237vrIqqzflKOV2}{S1904}
\lookupPut{R_41807FrOladsLMB}{S1905}
\lookupPut{R_2WxsHcVLGfVT66D}{S1906}
\lookupPut{R_3Ddw8u01DtXwegh}{S1907}
\lookupPut{R_2t9BSYyW64anbm9}{S1908}
\lookupPut{R_3OditCwLKWlL8df}{S1909}
\lookupPut{R_3GvNLn4CeG1Yghm}{S1910}
\lookupPut{R_2wGiIxtxrrJR8Hp}{S1911}
\lookupPut{R_25GzzGy29OIbbKk}{S1912}
\lookupPut{R_3j6vV3PoBtoVPOw}{S1913}
\lookupPut{R_2e5uMGw5IKFMUCO}{S1914}
\lookupPut{R_XtUJlHCFAH2KidP}{S1915}
\lookupPut{R_CfxWwlDGQyjQyrf}{S1916}
\lookupPut{R_ZsjxCaA5fVgwjoR}{S1917}
\lookupPut{R_3MokXRkknVzUlXu}{S1918}
\lookupPut{R_9YqR4LEV06Gom1X}{S1919}
\lookupPut{R_3nAvFVagieY5ud7}{S1920}
\lookupPut{R_2YQeWuviSc6T11A}{S1921}
\lookupPut{R_3nr3Ik0SHgNxfiG}{S1922}
\lookupPut{R_29hLwY0fh3rypyr}{S1923}
\lookupPut{R_2Y3QDbFDzYK1Fui}{S1924}
\lookupPut{R_OB9gb5zoj538ZPj}{S1925}
\lookupPut{R_3h6bPGPWbQBN57e}{S1926}
\lookupPut{R_3Rz3yzqc4EGEhEl}{S1927}
\lookupPut{R_32WMwePZYGChslK}{S1928}
\lookupPut{R_1oHody8VPRyYv2c}{S1929}
\lookupPut{R_1EgG57c4APFdYTc}{S1930}
\lookupPut{R_3futJoy4RbhHnkg}{S1931}
\lookupPut{R_3rNw8mHMZsEWZqQ}{S1932}
\lookupPut{R_1LogCAyX7veEZMv}{S1933}
\lookupPut{R_2fDuahVJjie7S95}{S1934}
\lookupPut{R_1SV1u9krVwp0Lux}{S1935}
\lookupPut{R_1hROtydqGpUAXDO}{S1936}
\lookupPut{R_3CCJnsRALNbX2gW}{S1937}
\lookupPut{R_2X4ceeSvIMmnISh}{S1938}
\lookupPut{R_3q7sHri5ndNi0Kc}{S1939}
\lookupPut{R_2VNLJ1OvelB3o45}{S1940}
\lookupPut{R_1dcWBo96r8IkyeV}{S1941}
\lookupPut{R_ysf9ql6P9UkBfdT}{S1942}
\lookupPut{R_PBrtipgcmun9Ish}{S1943}
\lookupPut{R_338Wvdc9OsvMDYa}{S1944}
\lookupPut{R_1HotBHafHu6sEaH}{S1945}
\lookupPut{R_uwgMt24OHvQRaQV}{S1946}
\lookupPut{R_3Jk3BgQus26ExZ5}{S1947}
\lookupPut{R_3RgBbMsROzia966}{S1948}
\lookupPut{R_WCEelVYpCgMvo4h}{S1949}
\lookupPut{R_riJcBPbFCYHC39T}{S1950}
\lookupPut{R_1kOjYAshpNY6Rxc}{S1951}
\lookupPut{R_1doAVRI6rTEFGI1}{S1952}
\lookupPut{R_3juiiRaIi17wGB3}{S1953}
\lookupPut{R_3gX2HpgFUd3YbNT}{S1954}
\lookupPut{R_1ocCZKTA99GgbkR}{S1955}
\lookupPut{R_ZK0ok5Qsl1sp01r}{S1956}
\lookupPut{R_2BzLrGAFn78iUOk}{S1957}
\lookupPut{R_1loy78sEUV0aECl}{S1958}
\lookupPut{R_eJA3JxqxtlVECyJ}{S1959}
\lookupPut{R_3PvkScviBcOwDZ6}{S1960}
\lookupPut{R_2EFu9FAJMbQF8sm}{S1961}
\lookupPut{R_31da3ya6GWRXZRN}{S1962}
\lookupPut{R_2xRCEpDqSjzP1g6}{S1963}
\lookupPut{R_1do8nfnoDBW0OyF}{S1964}
\lookupPut{R_1BWuPr4aZqYNG9N}{S1965}
\lookupPut{R_1ow0MY9dJyUkX2R}{S1966}
\lookupPut{R_1hG7mpLxasdFMdE}{S1967}
\lookupPut{R_yxYXPUVM2UivOsV}{S1968}
\lookupPut{R_2Vat0DGiJIt85IY}{S1969}
\lookupPut{R_0eypKiYrcAmjldL}{S1970}
\lookupPut{R_3P7GHHtvHQCXswX}{S1971}
\lookupPut{R_29hOVoMFDewbXbu}{S1972}
\lookupPut{R_3PtrRYkqADCEtu5}{S1973}
\lookupPut{R_1d4fLGuWIMloVIQ}{S1974}
\lookupPut{R_sRwkTblGXe1UPGF}{S1975}
\lookupPut{R_37VN841rJ0IYJFL}{S1976}
\lookupPut{R_2YgkAbCNX8rRwsi}{S1977}
\lookupPut{R_QhyRANbpqtAXDJD}{S1978}
\lookupPut{R_2QDJgOJzFrxucYg}{S1979}
\lookupPut{R_3e8oR55YU1bDyGY}{S1980}
\lookupPut{R_1NDKRz0A744q5iU}{S1981}
\lookupPut{R_2asfmLsXw5HtJD6}{S1982}
\lookupPut{R_23ZFp7E1KP6TOYH}{S1983}
\lookupPut{R_2y900dVHtzaElmy}{S1984}
\lookupPut{R_cPhrX2U7KeoE2kh}{S1985}
\lookupPut{R_2tgJNyxwbVPKRlD}{S1986}
\lookupPut{R_vldRpvbLjLRRGnL}{S1987}
\lookupPut{R_1Egq0RpOZaXHSev}{S1988}
\lookupPut{R_1LB9YdqS05xWdtQ}{S1989}
\lookupPut{R_1gbwC0Nzh6zr2y6}{S1990}
\lookupPut{R_OiDQixqHKXATdW9}{S1991}
\lookupPut{R_4Nmnke0TaFecAXT}{S1992}
\lookupPut{R_vc57Engrjy950kN}{S1993}
\lookupPut{R_tR7MAx2rxvyX453}{S1994}
\lookupPut{R_3oRPSgDAlaFeQHf}{S1995}
\lookupPut{R_xGCDWjX5SS39Yyt}{S1996}
\lookupPut{R_2bZSP3mnTsjcJyy}{S1997}
\lookupPut{R_wLiVP7MyoAQGGhX}{S1998}
\lookupPut{R_3Gvwojd5MClI4W8}{S1999}
\lookupPut{R_ABWaVza89MO44pj}{S2000}
\lookupPut{R_22LJyN6zBuOGsT0}{S2001}
\lookupPut{R_0PAMS5rjxiTmyzL}{S2002}
\lookupPut{R_2ruCaUJYnIaWX7S}{S2003}
\lookupPut{R_2f6tGLpnrnQlLSZ}{S2004}
\lookupPut{R_3npggdQZKSdYKMa}{S2005}
\lookupPut{R_3qDMgipuZ2MWsMF}{S2006}
\lookupPut{R_1CvvmjFGhRAywXp}{S2007}
\lookupPut{R_1j9fveIgQPx9XEq}{S2008}
\lookupPut{R_2CZXypmIkphyXaw}{S2009}
\lookupPut{R_2rw69jjW3GG3G0A}{S2010}
\lookupPut{R_qxboLqd67TpXIK5}{S2011}
\lookupPut{R_1FLfbgKGQmC1PWq}{S2012}
\lookupPut{R_2bZTgeLKnHeNn5D}{S2013}
\lookupPut{R_1BSpOSQHRYAIJmf}{S2014}
\lookupPut{R_2Cg4mqm2SB4KZKJ}{S2015}
\lookupPut{R_3qeNbrv6dDQFCR6}{S2016}
\lookupPut{R_1H5WH4Jv5NCMxEE}{S2017}
\lookupPut{R_29nTdECip5Pa6AF}{S2018}
\lookupPut{R_SYJKji8TQJOYogh}{S2019}
\lookupPut{R_41pc4dpmOIEMpW1}{S2020}
\lookupPut{R_3lWWAnWt3OrngDp}{S2021}
\lookupPut{R_1PS64xaayPiN23w}{S2022}
\lookupPut{R_25YI4nHUC73fs2Z}{S2023}
\lookupPut{R_21BW4UytaJA9KCJ}{S2024}
\lookupPut{R_3EbVDc52Fvkq6Ej}{S2025}
\lookupPut{R_xE2W31KnGsjyx1f}{S2026}
\lookupPut{R_XXvDrItbot0W1P3}{S2027}
\lookupPut{R_1N3rl1VcZtWOhbt}{S2028}
\lookupPut{R_3J2vHLogbd6hRZW}{S2029}
\lookupPut{R_3PavAx0KqwA8jnr}{S2030}
\lookupPut{R_2OVGSIHuefux8AF}{S2031}
\lookupPut{R_3FIQbhlZCrikqD3}{S2032}
\lookupPut{R_2Y38io5UTIWob04}{S2033}
\lookupPut{R_2c8CoARI4qXI3bm}{S2034}
\lookupPut{R_5sycSzYs4ihufuh}{S2035}
\lookupPut{R_1djg2tL1l3f2Zkz}{S2036}
\lookupPut{R_2YhNsVlERhduVnZ}{S2037}
\lookupPut{R_3qmtBA9F1Qik6tD}{S2038}
\lookupPut{R_UF9JXLizPCiD8Vr}{S2039}
\lookupPut{R_p6kjB3aGADsFqil}{S2040}
\lookupPut{R_22EKIe13nNzdCH3}{S2041}
\lookupPut{R_3qs1zpB9yPujxpg}{S2042}
\lookupPut{R_cvYcFL0QF2VZrTb}{S2043}
\lookupPut{R_3fVi82KsKJCLhRz}{S2044}
\lookupPut{R_3eqNo7C35E872Zn}{S2045}
\lookupPut{R_2BfcfSKiXwGtNVJ}{S2046}
\lookupPut{R_1lfWuG5jf8G2Tja}{S2047}
\lookupPut{R_W1LSGsyV7NmsBOx}{S2048}
\lookupPut{R_3Rsw5b4gmtj0JD1}{S2049}
\lookupPut{R_2CdG4eJ8pwLS0X2}{S2050}
\lookupPut{R_2SAz7KyQWaTdaIM}{S2051}
\lookupPut{R_1M3sSfBRQtWXexv}{S2052}
\lookupPut{R_tYtg9kRAhTiW8eZ}{S2053}
\lookupPut{R_1GBLMCATvfktoEZ}{S2054}
\lookupPut{R_2disXBkalQfkcN8}{S2055}
\lookupPut{R_1nUV71QgYDtS7Kq}{S2056}
\lookupPut{R_UzJCcfg3S9lXW4V}{S2057}
\lookupPut{R_dmAKN0FQ7ShTRdv}{S2058}
\lookupPut{R_1H1ri9ZaEfi0dnw}{S2059}
\lookupPut{R_1lv586TusvEuAfr}{S2060}
\lookupPut{R_27Qk2j2RXYCyThJ}{S2061}
\lookupPut{R_22Q5gaINUHCtVvE}{S2062}
\lookupPut{R_1EcwiyjtQOdW2VO}{S2063}
\lookupPut{R_1dfYLvM71zVbEIS}{S2064}
\lookupPut{R_cwK0RsBJWPpgIkV}{S2065}
\lookupPut{R_2AKZSt97kTtTif9}{S2066}
\lookupPut{R_247xLBTIMV3Bujk}{S2067}
\lookupPut{R_2f8unWpke7QBLec}{S2068}
\lookupPut{R_1ODmx1bDCan7XTG}{S2069}
\lookupPut{R_2rjAwQ5gfGULHt5}{S2070}
\lookupPut{R_1hEUUo46qMAMMdh}{S2071}
\lookupPut{R_3oC3J0EL6WEUV0t}{S2072}
\lookupPut{R_eeUqE2fuq8jfgFb}{S2073}
\lookupPut{R_qxWwuIo9akyTQMp}{S2074}
\lookupPut{R_zTQfR7cgEM02RDX}{S2075}
\lookupPut{R_3I6uiNS44JyHzH0}{S2076}
\lookupPut{R_3MfFVB4Zzm8lejS}{S2077}
\lookupPut{R_O1drAyWUjeLArC1}{S2078}
\lookupPut{R_2ePUkYwETZdKyC3}{S2079}
\lookupPut{R_3QQr1yzDVGBZcOk}{S2080}
\lookupPut{R_1Oy1wegk94LaozT}{S2081}
\lookupPut{R_3D1NMPkFr2zA6r1}{S2082}
\lookupPut{R_1mfnk1vnfhUI4ov}{S2083}
\lookupPut{R_e5Lcx04h1Mgr4Hf}{S2084}
\lookupPut{R_PASuEaZRDx4wlYR}{S2085}
\lookupPut{R_3lMIQt7NCLQZ6o9}{S2086}
\lookupPut{R_3g6p7BWsEDmnZHW}{S2087}
\lookupPut{R_8wT1WVa606zMJAl}{S2088}
\lookupPut{R_1OZTbLwbugZR2fY}{S2089}
\lookupPut{R_1f8HdfHBX6R8DlT}{S2090}
\lookupPut{R_1mXDYCcX7ueri58}{S2091}
\lookupPut{R_1C2NzSPdbnEx7Ev}{S2092}
\lookupPut{R_3et0SZYQPk9WKej}{S2093}
\lookupPut{R_C9zOWrHdVUE18pb}{S2094}
\lookupPut{R_3R4N4cMEaSNte26}{S2095}
\lookupPut{R_2QFfyfjTgaXpSUa}{S2096}
\lookupPut{R_pKpc7B3jQTMd88F}{S2097}
\lookupPut{R_27BZ1jUsis1hWWe}{S2098}
\lookupPut{R_3PsPx4SHODrVgRG}{S2099}
\lookupPut{R_2U3raYp8UTJkPEF}{S2100}
\lookupPut{R_3kjjhHuIyhMM1pb}{S2101}
\lookupPut{R_2riIxiiZe8Wv80J}{S2102}
\lookupPut{R_1NEqZxMTOE3oxTi}{S2103}
\lookupPut{R_XHRO7f95L8wgKVH}{S2104}
\lookupPut{R_bNq1UFEelvBUzZL}{S2105}
\lookupPut{R_1jiFwV9M1sKUnVs}{S2106}
\lookupPut{R_PUl3MMUH2sA61sl}{S2107}
\lookupPut{R_eo21RfshfLyb8K5}{S2108}
\lookupPut{R_24f1w9wCXx1rTEd}{S2109}
\lookupPut{R_2X4dZ9q5oJtQTqw}{S2110}
\lookupPut{R_3KphIBEhBjLl93c}{S2111}
\lookupPut{R_PY5o4m8cdIvYbPr}{S2112}
\lookupPut{R_1NfTEpxwyrubsSZ}{S2113}
\lookupPut{R_1mlxmmJsCPBVEJy}{S2114}
\lookupPut{R_paSLmyEEDpUIN3P}{S2115}
\lookupPut{R_2XgGI4EogESeAME}{S2116}
\lookupPut{R_3QLXnf2Ea0Ck74B}{S2117}
\lookupPut{R_3oHLy9VUPhvWduw}{S2118}
\lookupPut{R_O1KGuIzvBemB24V}{S2119}
\lookupPut{R_3IaxZKiwpJTwqOe}{S2120}
\lookupPut{R_1Kvv7jmLH91CYzw}{S2121}
\lookupPut{R_aWxKtc1T8ELm0rD}{S2122}
\lookupPut{R_DwpRGy19dGZehX3}{S2123}
\lookupPut{R_AGlT1mbvGY4dJyF}{S2124}
\lookupPut{R_3qHYegweUHHlRZe}{S2125}
\lookupPut{R_sb62wszmEgzv0dj}{S2126}
\lookupPut{R_2usxxeLvmhQEf6i}{S2127}
\lookupPut{R_3DnZMA4UT715mEk}{S2128}
\lookupPut{R_1n72VDgCmFlMhaK}{S2129}
\lookupPut{R_3Oq5UTBa2Qzvie0}{S2130}
\lookupPut{R_1pKimgRNHCIAmST}{S2131}
\lookupPut{R_1reKanbMdlpmiRB}{S2132}
\lookupPut{R_11bvcPN6vIVgej4}{S2133}
\lookupPut{R_1dHuDcTYL7yHHL5}{S2134}
\lookupPut{R_9SSNzKRGVD7Denf}{S2135}
\lookupPut{R_1jIKQQTQODezrV9}{S2136}
\lookupPut{R_3oN3C1Dbh4oI4Ue}{S2137}
\lookupPut{R_1jdYccV3E1HWO4C}{S2138}
\lookupPut{R_2AStXae55G5a4Ag}{S2139}
\lookupPut{R_BXG8YXPTSZx0R0J}{S2140}
\lookupPut{R_z1r1bJX3A6LA4MN}{S2141}
\lookupPut{R_2wEvuUEshHcdoGV}{S2142}
\lookupPut{R_3PhsHvO5WH2eAeB}{S2143}
\lookupPut{R_vJ1jhkafrLXPvDb}{S2144}
\lookupPut{R_1lAHLexytLNzIxF}{S2145}
\lookupPut{R_wZTF7panlY7c7VT}{S2146}
\lookupPut{R_3KHojydg8XIr5MR}{S2147}
\lookupPut{R_1GPjHBAooj7nAqZ}{S2148}
\lookupPut{R_1OP0izv2q6bOfJZ}{S2149}
\lookupPut{R_24GbjFFZcxubaDg}{S2150}
\lookupPut{R_ALP09fbFFHJltuN}{S2151}
\lookupPut{R_3soAxkqzzS9avA9}{S2152}
\lookupPut{R_1LNly6Eo3r68Cxg}{S2153}
\lookupPut{R_8IaMmIx334upQAh}{S2154}
\lookupPut{R_p5hnY9Z3a17mcM1}{S2155}
\lookupPut{R_pEQ2ccWNhcMFo0p}{S2156}
\lookupPut{R_1IsizO9esN504Mt}{S2157}
\lookupPut{R_2aJLHdEiyMQA2nT}{S2158}
\lookupPut{R_1gcf5C8Z0SU5xbX}{S2159}
\lookupPut{R_2OUQ73y2PNQOIpo}{S2160}
\lookupPut{R_2wGdKHLDYpCSTdt}{S2161}
\lookupPut{R_1gUOj7IsvvB3J5A}{S2162}
\lookupPut{R_Dpk8231eD8C82iJ}{S2163}
\lookupPut{R_2tG1VaGbxOtCW3a}{S2164}
\lookupPut{R_2RPUBy61foSbDFz}{S2165}
\lookupPut{R_24GJUwAB0HQ94fH}{S2166}
\lookupPut{R_cCLzuvXVfOvyXrr}{S2167}
\lookupPut{R_21tAs6Si7KTsGyC}{S2168}
\lookupPut{R_p0BttQXfBjj8vjb}{S2169}
\lookupPut{R_3nSng16zgjgoDQD}{S2170}
\lookupPut{R_1n9mh98Lii5OdhY}{S2171}
\lookupPut{R_2rGh6J45YJPRemX}{S2172}
\lookupPut{R_3mkFEyBvONnH2a5}{S2173}
\lookupPut{R_3m9R64yAXBU4m78}{S2174}
\lookupPut{R_1KkAkPLfZexD74h}{S2175}
\lookupPut{R_Ubv0umr8PyXqI1z}{S2176}
\lookupPut{R_2D2M6qrcWq71mjx}{S2177}
\lookupPut{R_2rSDDQsEL7MBS4X}{S2178}
\lookupPut{R_3lQzo3N4cMvPg8a}{S2179}
\lookupPut{R_eLqeklGz81lEnf3}{S2180}
\lookupPut{R_xsb8NtGYZvurUBP}{S2181}
\lookupPut{R_3m2HRMYbypYMnM2}{S2182}
\lookupPut{R_1E5KoO9pEXjo7BO}{S2183}
\lookupPut{R_2rVqLW2y1g7zCOJ}{S2184}
\lookupPut{R_6DaMOf10TYk0O0F}{S2185}
\lookupPut{R_24PfGieTEZs3sUa}{S2186}
\lookupPut{R_1GDJldAStSseowd}{S2187}
\lookupPut{R_umMPlWhYImwVVcJ}{S2188}
\lookupPut{R_3jxyk33KVDRoNfr}{S2189}
\lookupPut{R_1jxPMjHnrJqabSo}{S2190}
\lookupPut{R_x5ftNppZF8Q8TZf}{S2191}
\lookupPut{R_1mLgZE6VIBJYY4A}{S2192}
\lookupPut{R_1LuGtfYmX1XYiRc}{S2193}
\lookupPut{R_241uQDhbxp9qmnv}{S2194}
\lookupPut{R_xa5rHfEPjHvcd3j}{S2195}
\lookupPut{R_XOqrErfJEXDHgVH}{S2196}
\lookupPut{R_donHmUCqBZvWeHf}{S2197}
\lookupPut{R_10HJEsPUZZFh5Jh}{S2198}
\lookupPut{R_3kCcl3qRwwCvYMi}{S2199}
\lookupPut{R_22Dqd5ITMYZkpUY}{S2200}
\lookupPut{R_3hfKROMgbt6wyv0}{S2201}
\lookupPut{R_2cdRhhEUAJtjWBb}{S2202}
\lookupPut{R_pK8fiBvgHwXjutr}{S2203}
\lookupPut{R_9FeRwL3SNcPgBXz}{S2204}
\lookupPut{R_sROMewLPEkejL5T}{S2205}
\lookupPut{R_2e4kXukLvWmxib6}{S2206}
\lookupPut{R_1HqSFchuUtokBI5}{S2207}
\lookupPut{R_3e7XqV3j1oKjTKD}{S2208}
\lookupPut{R_3NxKLZ9iGgnt0Oj}{S2209}
\lookupPut{R_3fiWOrcu1s4LP5N}{S2210}
\lookupPut{R_2bZzLRapy1Obf88}{S2211}
\lookupPut{R_1FKgSgMVshHsOjx}{S2212}
\lookupPut{R_voBb7w5ScENa9mV}{S2213}
\lookupPut{R_2SxyZJC1Xj4pwuY}{S2214}
\lookupPut{R_2Wvd4qcxqJ5Fesp}{S2215}
\lookupPut{R_2tsscBcYx6B4qGk}{S2216}
\lookupPut{R_W7CJm0siVakI7Pr}{S2217}
\lookupPut{R_1Lb9fUOeHZT1uJM}{S2218}
\lookupPut{R_6lZoPEl7zK7rLYl}{S2219}
\lookupPut{R_210WUjQJDirbt5R}{S2220}
\lookupPut{R_3PEJOjQygZqxnaJ}{S2221}
\lookupPut{R_22JROakyyDqXNc2}{S2222}
\lookupPut{R_2dyeWN1DpzBHvH1}{S2223}
\lookupPut{R_agFS6EeonrRcfAJ}{S2224}
\lookupPut{R_1IKkTk4NUZfcteF}{S2225}
\lookupPut{R_3qwIeOcpTCYwyrn}{S2226}
\lookupPut{R_2pJKSv5fjwuOgPS}{S2227}
\lookupPut{R_2uqS1nskWLzhJ41}{S2228}
\lookupPut{R_2lszhhqK0WWeBJ7}{S2229}
\lookupPut{R_3kii8qboeCWPn8Q}{S2230}
\lookupPut{R_3kqcQevYrhiXqVW}{S2231}
\lookupPut{R_12hWkGcIPLGlzGz}{S2232}
\lookupPut{R_3iwNcywmX2KWhgw}{S2233}
\lookupPut{R_10TyWpwOGiWloQV}{S2234}
\lookupPut{R_3gMmptKKjdeBy2p}{S2235}
\lookupPut{R_1gvwLAqE6Mr9PmD}{S2236}
\lookupPut{R_2ydRvWAxxiYkrtt}{S2237}
\lookupPut{R_3Ppo1SsMGdrlUzm}{S2238}
\lookupPut{R_RfXwMb2enxqa91v}{S2239}
\lookupPut{R_1dGOOrNoqyukc2Q}{S2240}
\lookupPut{R_cBV2nzCXyxKijkt}{S2241}
\lookupPut{R_aW4jTm59FRYnJn3}{S2242}
\lookupPut{R_2azR3amTXG6v2tZ}{S2243}
\lookupPut{R_3MaNQ0wrIaByW68}{S2244}
\lookupPut{R_1DCIICt7YxJqHVH}{S2245}
\lookupPut{R_1NE4wo5XJip6OUg}{S2246}
\lookupPut{R_22VQbalXxFNCpwM}{S2247}
\lookupPut{R_aWZ6kKQi7Dx5So1}{S2248}
\lookupPut{R_2ysdfm8VTqor9cw}{S2249}
\lookupPut{R_NWRH2OCh5rCmkW5}{S2250}
\lookupPut{R_2ziiMdveJZWAMf8}{S2251}
\lookupPut{R_3NKLoi1OXJe3opr}{S2252}
\lookupPut{R_AAuwmUB1gAUzco9}{S2253}
\lookupPut{R_2BbRCvwI4cFU0r2}{S2254}
\lookupPut{R_3ptNkcY03AaDhk8}{S2255}
\lookupPut{R_uekasyIq0AG8XHX}{S2256}
\lookupPut{R_3RpVCDk9xukQv02}{S2257}
\lookupPut{R_3qNWZ7H07zEn5AX}{S2258}
\lookupPut{R_2QPukYfyWuWkmdP}{S2259}
\lookupPut{R_28zEhL7tLCUIxoq}{S2260}
\lookupPut{R_3FQA4P8rQt89kwE}{S2261}
\lookupPut{R_1ghwQKVQucCSOq8}{S2262}
\lookupPut{R_1g0jlnFltKVsCcE}{S2263}
\lookupPut{R_31tQxK2Mwoaidln}{S2264}
\lookupPut{R_1KRSOENclbC0S5z}{S2265}
\lookupPut{R_yw63tXZ1ZSA63T3}{S2266}
\lookupPut{R_XB5KJRVGSx62Dpn}{S2267}
\lookupPut{R_ZEl1b4oxZeHVs0p}{S2268}
\lookupPut{R_22StgdXay4Bj8Xh}{S2269}
\lookupPut{R_1NyJH6jH2QLe4Tj}{S2270}
\lookupPut{R_1onbzQOva04md8k}{S2271}

\lookupPut{R_2QhYxxrOzIJ8xcE}{P1}
\lookupPut{R_2yec50up2YE3MFg}{P2}
\lookupPut{R_2SdgeMfsYXuiNez}{P3}
\lookupPut{R_2uqyJTbPN1G8uNL}{P4}
\lookupPut{R_BEZYRo3WhyMopQR}{P5}
\lookupPut{R_1d0uJyznXvNvmDa}{P6}
\lookupPut{R_1OPFk7XW80g52V1}{P7}
\lookupPut{R_w5jdQfEOrGZQF0t}{P8}
\lookupPut{R_31ASeSO1VkXlXJ8}{P9}
\lookupPut{R_2dGc6Iztrx5kgGg}{P10}
\lookupPut{R_2uybGocFlttRSZb}{P11}
\lookupPut{R_3rUHyHzC9I0E5PZ}{P12}
\lookupPut{R_1LTjhYsr6YlisbY}{P13}
\lookupPut{R_1E061czKYYqisgp}{P14}
\lookupPut{R_3p8ZBsVqroAOPRp}{P15}
\lookupPut{R_ypPI0WNCipWRVxT}{P16}
\lookupPut{R_2zeEUFZaCZ1YyrW}{P17}
\lookupPut{R_qwQ0z1lnV6QvlQZ}{P18}
\lookupPut{R_24ut71AH3jmzjjB}{P19}
\lookupPut{R_24iLqbAJOv4OT2Q}{P20}
\lookupPut{R_263kXSgt11Q6XME}{P21}
\lookupPut{R_3KZuZTqH2xrRWrz}{P22}
\lookupPut{R_3kOezPysJPNV1FF}{P23}
\lookupPut{R_2SCCa8OoLuJgqDp}{P24}
\lookupPut{R_1NCTFXkeDgDc6og}{P25}
\lookupPut{R_2qlqurlcghoHphk}{P26}
\lookupPut{R_PCfMq75Hw44ENzP}{P27}
\lookupPut{R_27U06WlU6igY2Bn}{P28}
\lookupPut{R_Qht6lHTRma8Jly9}{P29}
\lookupPut{R_2X69xzk9boloIKb}{P30}
\lookupPut{R_2QxkOqAVCsSIiE5}{P31}
\lookupPut{R_2az6235kOVuP6Nh}{P32}
\lookupPut{R_2ClDHYit6qdzUJm}{P33}
\lookupPut{R_3HgX0AYGgmRWl2e}{P34}
\lookupPut{R_cSL3ZlNXD8jNW2B}{P35}
\lookupPut{R_vwVLqA7RiwCMD3X}{P36}
\lookupPut{R_2y2RYhzd2Ogysvl}{P37}
\lookupPut{R_ZyLjhxur9bd029b}{P38}
\lookupPut{R_10GcerD7kXaLzrv}{P39}
\lookupPut{R_Q6yJTfPnCRgtBCN}{P40}
\lookupPut{R_1gvVhR1eeEpZDir}{P41}
\lookupPut{R_246FLoQPhWvkxgM}{P42}
\lookupPut{R_2t5uYNqejqNNlqw}{P43}
\lookupPut{R_27qHZG3G4O8kBzq}{P44}
\lookupPut{R_yJwmzVifabHq0sF}{P45}
\lookupPut{R_roNDzIsneX6BIAN}{P46}
\lookupPut{R_1F4WnzNfVTC6ZQz}{P47}
\lookupPut{R_2dsTV5Z8nG9DmW7}{P48}
\lookupPut{R_2zpX2l8aEkGMTMv}{P49}
\lookupPut{R_3psfVQOzvvQ0pKI}{P50}
\lookupPut{R_3lG5IMc8mO1q82f}{P51}
\lookupPut{R_1NlfW9AzD8Rttxp}{P52}
\lookupPut{R_1Lev62noFWmf2N2}{P53}
\lookupPut{R_9MqHB6OX0ft0bbH}{P54}
\lookupPut{R_3p54sLnFp1FC3m4}{P55}
\lookupPut{R_1KiG8EEU5Bl6oSV}{P56}
\lookupPut{R_3OjlX4UtMe4SNwJ}{P57}
\lookupPut{R_8p44FZdAgP4Xmxj}{P58}
\lookupPut{R_vwQbPDsRR4O3y6J}{P59}
\lookupPut{R_1dMNkrzmrhNuLmx}{P60}
\lookupPut{R_2uQE8iGBPhrAsBc}{P61}
\lookupPut{R_3MlNczDKngkAUhr}{P62}
\lookupPut{R_1Qt1rrOQPEmjPkE}{P63}
\lookupPut{R_sMqFVMumSKyRAnn}{P64}
\lookupPut{R_07DnymPaSvLvpHb}{P65}
\lookupPut{R_1HhQdS7M4ZuOZ3v}{P66}
\lookupPut{R_2fx4zGNUB5aAzT3}{P67}
\lookupPut{R_zcZ8yEzaYsTLJhD}{P68}
\lookupPut{R_1mK4roVUn5bDrEZ}{P69}
\lookupPut{R_wTAZ1uQxo5xyXYt}{P70}
\lookupPut{R_ritNjFhf7eAVDwt}{P71}
\lookupPut{R_1K2yfUfcFB1wjLn}{P72}
\lookupPut{R_1LzFvjfrSUNcEnU}{P73}
\lookupPut{R_86sKEzl1KJHhdvP}{P74}
\lookupPut{R_2eW6bK9FYsGSmyH}{P75}
\lookupPut{R_2dhyHPfnESToWJe}{P76}
\lookupPut{R_1hHmcHnftEat9NR}{P77}
\lookupPut{R_2as27CDxgGZAJ6Q}{P78}
\lookupPut{R_3oLL90vAb5X36TW}{P79}
\lookupPut{R_1gOvu7NBkFhTkrc}{P80}
\lookupPut{R_BtsrgMRSw2nyAfv}{P81}
\lookupPut{R_3sgoxm1m0OXjXXR}{P82}
\lookupPut{R_3Kw5QGeoVdwKE28}{P83}
\lookupPut{R_2zciC6BTPcFhSwB}{P84}
\lookupPut{R_28ZArZkX9sfWKci}{P85}
\lookupPut{R_3qTKULFIQXyuz2F}{P86}
\lookupPut{R_3HoZnh3BUg8oOTY}{P87}
\lookupPut{R_3fuWSF2zZovfH4B}{P88}
\lookupPut{R_C8QFvMHRgRpm3aV}{P89}
\lookupPut{R_2XmMf7DWwDvTOMV}{P90}
\lookupPut{R_qWxdknK126JbTMZ}{P91}
\lookupPut{R_1hXPR5DTWSHiTli}{P92}
\lookupPut{R_eQzl5OJltP01b3P}{P93}
\lookupPut{R_cIqEmqSaj7UuFup}{P94}
\lookupPut{R_VPtaYKAcrxcWONX}{P95}
\lookupPut{R_2duc4sOpGoyWwQt}{P96}
\lookupPut{R_URsvwVym57gXPBD}{P97}
\lookupPut{R_3kh7dAbvTFvUODs}{P98}
\lookupPut{R_1NFJfRWRpVQqrjB}{P99}
\lookupPut{R_33mAgQTiIeeKI3p}{P100}
\lookupPut{R_38n8etKfdB1gf29}{P101}
\lookupPut{R_3plu6LTn9ufuBLR}{P102}
\lookupPut{R_PMMjc41KvBY2Pwl}{P103}
\lookupPut{R_1FEcpXZJ8McnSJm}{P104}
\lookupPut{R_3KB1QpJWvHc9S9b}{P105}
\lookupPut{R_2VyDRYfns44pNL0}{P106}
\lookupPut{R_UGimlI0xLi5LFwR}{P107}
\lookupPut{R_zcfYOjt4DKP33q1}{P108}
\lookupPut{R_3F3DvSuwbpgocYe}{P109}
\lookupPut{R_1i86kuKSZCwm3A2}{P110}
\lookupPut{R_2PcJpH0O6NEOSmy}{P111}
\lookupPut{R_3hA6hMdbJyUEt0f}{P112}
\lookupPut{R_zTHtRgiaTeG883T}{P113}
\lookupPut{R_XNA2xtrluHD1InD}{P114}
\lookupPut{R_1odQgDXWbeeM7wg}{P115}
\lookupPut{R_2wuI155wwAOdoPo}{P116}
\lookupPut{R_VLsxnwAeN8N5uwh}{P117}
\lookupPut{R_3Rfnac86j1HufbD}{P118}
\lookupPut{R_2bOu5DBEyMjKk52}{P119}
\lookupPut{R_2qFbI0PVK7OT9pe}{P120}
\lookupPut{R_cMggqn03fhHsZ2h}{P121}
\lookupPut{R_2ztGzJ9hjrThn2v}{P122}
\lookupPut{R_C7uYO4QDc4jCAsp}{P123}
\lookupPut{R_2PjvFMzOL8SJfZm}{P124}
\lookupPut{R_1kZeD32GoRPLf93}{P125}
\lookupPut{R_1LG7J5tuVFMfQI2}{P126}
\lookupPut{R_W7LyE4vCEPjrpNT}{P127}
\lookupPut{R_pc8zC0D72BdCRvr}{P128}
\lookupPut{R_2Yb5PREKlyLwCkd}{P129}
\lookupPut{R_cMFeXkjzNsCU2LD}{P130}
\lookupPut{R_2QXnFU2xaIjATzA}{P131}
\lookupPut{R_2VR4F4LfZK5oeT8}{P132}
\lookupPut{R_1OIFFmE5ZbP8QKg}{P133}
\lookupPut{R_3COEJEGjrN27Akd}{P134}
\lookupPut{R_1nNLMm7Tpwl7uRl}{P135}
\lookupPut{R_2AEjD9TC6oY8jse}{P136}
\lookupPut{R_RlQ40sZ1vWqgqNb}{P137}
\lookupPut{R_2CDGro15oOlzY7W}{P138}
\lookupPut{R_3Gf7Up40MGWE6qq}{P139}
\lookupPut{R_23W9feuiPcOPyDm}{P140}
\lookupPut{R_1mJCxiNxrZj0GCZ}{P141}
\lookupPut{R_2bQ5hpsf0cLwttA}{P142}
\lookupPut{R_2fvODXiyftq6nis}{P143}
\lookupPut{R_263dtopdzmNfW7T}{P144}
\lookupPut{R_Q4D6syPn394SeqZ}{P145}
\lookupPut{R_3DERDufB5YtfC5b}{P146}
\lookupPut{R_1Netav5ATYNhgLY}{P147}
\lookupPut{R_aeHsQVP2zxLZthT}{P148}
\lookupPut{R_2UhAKSESSyjrx26}{P149}
\lookupPut{R_3lJrmIQhD1TBlIx}{P150}
\lookupPut{R_saRI451J7Kn7uKt}{P151}
\lookupPut{R_xAFUa3TqDIPflSh}{P152}
\lookupPut{R_bHNiOEbhnEhMtG1}{P153}
\lookupPut{R_22qRRvdY7X057Cu}{P154}
\lookupPut{R_3h53A1oNOlvt0uI}{P155}
\lookupPut{R_XA11UlWMxsiPHTH}{P156}
\lookupPut{R_2tL0BH0t9M8yTUI}{P157}
\lookupPut{R_yOqMom5Jmq75LI5}{P158}
\lookupPut{R_32VEa92xmxnxu4E}{P159}
\lookupPut{R_1CDcs7hNkfBAzGL}{P160}
\lookupPut{R_3iQgCcULw8KZOS8}{P161}
\lookupPut{R_8dg16ECJ4p2to1X}{P162}
\lookupPut{R_snAKw4628EsDy9P}{P163}
\lookupPut{R_VJxIwLovTsVTFlv}{P164}
\lookupPut{R_3D5AB6024euQ4sl}{P165}
\lookupPut{R_3iEjaee3U9a2tps}{P166}
\lookupPut{R_1mEO1siQSQxO2JE}{P167}
\lookupPut{R_1r6tIJdLDXZ53NZ}{P168}
\lookupPut{R_3kp8Jdv3xF8sLlW}{P169}
\lookupPut{R_1pseHJa8wXOGLGJ}{P170}
\lookupPut{R_2nLRFGLrjsPmbgB}{P171}
\lookupPut{R_25zpZRbo2i1dXGP}{P172}
\lookupPut{R_2zSF7deUqILgPxP}{P173}
\lookupPut{R_urzu4fDjMutW8aB}{P174}
\lookupPut{R_abmG1f5sUFDzGU1}{P175}
\lookupPut{R_1rJTavlyGEPu1wP}{P176}
\lookupPut{R_1jrmaS4lZtnG8Uq}{P177}
\lookupPut{R_uk890vApAdg3Wbn}{P178}
\lookupPut{R_2y8yHQ7WbRTmxnb}{P179}
\lookupPut{R_3qPvBTddlFdezQ7}{P180}
\lookupPut{R_1CC7G8Szz27ZMqY}{P181}
\lookupPut{R_24CYwbutnarT7Mb}{P182}
\lookupPut{R_b7NB23FuGVMWQBr}{P183}
\lookupPut{R_2VEYc6ssookniGu}{P184}
\lookupPut{R_3ffcuPqYR1gyBtl}{P185}
\lookupPut{R_2uQyLbyDO80ltib}{P186}
\lookupPut{R_1dt2EiEzrA69Cvw}{P187}
\lookupPut{R_2SCzNJqz4gfxD1P}{P188}
\lookupPut{R_3D6kOSpOvEo4QKe}{P189}
\lookupPut{R_1ePOSZQEwSbLm5R}{P190}
\lookupPut{R_1rjP5XiStS6mKMD}{P191}
\lookupPut{R_2uVEQgDMhAGfAel}{P192}
\lookupPut{R_3DjWbJXzzMzFUK7}{P193}
\lookupPut{R_3D8eyLr1ZcdxPTt}{P194}
\lookupPut{R_2Y9XvduDVpracQo}{P195}
\lookupPut{R_3CPWNBTQLXAUVTl}{P196}
\lookupPut{R_1LcFQJSZFfpj5zn}{P197}
\lookupPut{R_1mxRzhVR2bS4EAI}{P198}
\lookupPut{R_2QWWPK7EE69CpLv}{P199}
\lookupPut{R_3ewrdYhCPyOH4NT}{P200}
\lookupPut{R_3NDoq1PIRR2fMMf}{P201}
\lookupPut{R_10NlJL7TJWl4o6y}{P202}
\lookupPut{R_1mUGOvVroKygr3F}{P203}
\lookupPut{R_2qC1FpXxbN9yKOq}{P204}
\lookupPut{R_24CMItWb8oUY30U}{P205}
\lookupPut{R_1Dp94PMtYUKwoT8}{P206}
\lookupPut{R_2Wv8wMxg2jSRYnr}{P207}
\lookupPut{R_2SvpTpItziTfm0R}{P208}
\lookupPut{R_3qU8Ant97kZDa6m}{P209}
\lookupPut{R_1Qs94WBWvxJyLaT}{P210}
\lookupPut{R_27amzn4O1xlgTjv}{P211}
\lookupPut{R_1joYf29ApLyGaAY}{P212}
\lookupPut{R_6A5Y6DfVPAU8Vzj}{P213}
\lookupPut{R_8q2mDiTKVvoMf3r}{P214}
\lookupPut{R_2dKZf5IG6crztdq}{P215}
\lookupPut{R_ripeYoeh8wOfQ1X}{P216}
\lookupPut{R_1DY0mn6VxiO6Qxu}{P217}
\lookupPut{R_UH39zSVd1du1Dk5}{P218}
\lookupPut{R_3QQ9OVCzTb7TKRd}{P219}
\lookupPut{R_31nhmZaOJG5E3C4}{P220}
\lookupPut{R_3PnPgxkqeEEggfI}{P221}
\lookupPut{R_2bJI5GxKT31IaMF}{P222}
\lookupPut{R_1rD0XIqTLYEcnpq}{P223}
\lookupPut{R_27eQD9fY67lBkdT}{P224}
\lookupPut{R_31MC5a99v7NIc4s}{P225}
\lookupPut{R_2PdiQpYWl09Ly9J}{P226}
\lookupPut{R_117liAwRyqWYeSR}{P227}
\lookupPut{R_23eoNVYdPRKpkQv}{P228}
\lookupPut{R_3iRociSQUC7WFJr}{P229}
\lookupPut{R_2vY6mPne6j1mRmn}{P230}
\lookupPut{R_0lypTF3WmPQHpct}{P231}
\lookupPut{R_1qdPBS18m8ON2Rk}{P232}
\lookupPut{R_22JhjLQglqxy3Fo}{P233}
\lookupPut{R_3gRTjONOe4ip6ho}{P234}
\lookupPut{R_1pzmVREaSpUixK1}{P235}
\lookupPut{R_33kEEBhA73F3vBX}{P236}
\lookupPut{R_27TdJ9DU6himlly}{P237}
\lookupPut{R_sHluAWJNuB3M8Gl}{P238}
\lookupPut{R_1JUCX5aT3ibnk9w}{P239}
\lookupPut{R_2DSvmG6QmMfudS8}{P240}
\lookupPut{R_1K2YKX8oIIat5mU}{P241}
\lookupPut{R_1JL14inAYypcPCh}{P242}
\lookupPut{R_2tnGSQkXgy4aU6P}{P243}
\lookupPut{R_esnxM2V2NtAMa3L}{P244}
\lookupPut{R_10VqcFDroE9voYy}{P245}
\lookupPut{R_3qeyY0MxNVK7ZkU}{P246}
\lookupPut{R_2D1YtNFY6xENPG4}{P247}
\lookupPut{R_1cTgKieFHUT8x9h}{P248}
\lookupPut{R_3Dno8nq0CAdhZOc}{P249}
\lookupPut{R_1hA2rimN3kIyFpG}{P250}
\lookupPut{R_3g5HUAJP4eZInTz}{P251}
\lookupPut{R_wY08ghCZL7Xpk7D}{P252}
\lookupPut{R_plOJWNvkhExTXJ7}{P253}
\lookupPut{R_bCn11CoJmM9yQlb}{P254}
\lookupPut{R_1N47ZKlLj0dFsu9}{P255}
\lookupPut{R_1gwHQkfD7tSs5zh}{P256}
\lookupPut{R_2VyVa2bn4dKQ8G9}{P257}
\lookupPut{R_2SernBYVQAWUIR7}{P258}
\lookupPut{R_1oneuCw6zSnOeaH}{P259}
\lookupPut{R_AuIvVAaEq7A1HEZ}{P260}
\lookupPut{R_1d3MbxEeoXzteNA}{P261}
\lookupPut{R_9pN21mNpaVJy0jT}{P262}
\lookupPut{R_3rG5j3xXL8nsguN}{P263}
\lookupPut{R_1hSFppfk9kmnQBC}{P264}
\lookupPut{R_2z7rJaNfrOsDzoS}{P265}
\lookupPut{R_3LiOoDmpDLw0RPx}{P266}
\lookupPut{R_1rqdpRwbgJ6nsYT}{P267}
\lookupPut{R_3KpQYBDJqZRrjam}{P268}
\lookupPut{R_3sAo493oY3jcnSW}{P269}
\lookupPut{R_XmI3LDGR4iT1KcF}{P270}
\lookupPut{R_1dKPKfAvVKnHNsj}{P271}
\lookupPut{R_0qwl9lBJlBNAxod}{P272}
\lookupPut{R_2dhV70dWvH4rRei}{P273}
\lookupPut{R_2wtqI7xtRZPWLCl}{P274}
\lookupPut{R_2di2r4Vya2dhBzp}{P275}
\lookupPut{R_1LOkOCtqEX4WVlo}{P276}
\lookupPut{R_1hBhy4czqX2tmHK}{P277}
\lookupPut{R_qURqxoG0p0OlC0x}{P278}
\lookupPut{R_0D6fzJNzkTvZz0Z}{P279}
\lookupPut{R_24rfY33H5RuwygC}{P280}
\lookupPut{R_1jjPiaJLD0OPsq1}{P281}
\lookupPut{R_25BMmykC4FUHiE1}{P282}
\lookupPut{R_1ocoJj2znV7h3z0}{P283}
\lookupPut{R_cSnE1iGUypXdgqZ}{P284}
\lookupPut{R_2OUNc64O6VypHc8}{P285}
\lookupPut{R_2thwMO7G0FfTLVC}{P286}
\lookupPut{R_Rf3SeSMg8oyHeCZ}{P287}
\lookupPut{R_xhCWE8qDaMq6smZ}{P288}
\lookupPut{R_31sBkX3q0PZV6o6}{P289}
\lookupPut{R_BFL28hyc3IPT8iZ}{P290}
\lookupPut{R_2SfdTJgydwoyvSJ}{P291}
\lookupPut{R_2vjlRAIjMyFgrA1}{P292}
\lookupPut{R_1gHMbhc7GQNGjK7}{P293}
\lookupPut{R_ehV75JV12f7S1eF}{P294}
\lookupPut{R_yjB6Nru2HbLXwmB}{P295}
\lookupPut{R_1CpXdz7QHGGTe72}{P296}
\lookupPut{R_2BeOzQUO9eaeiso}{P297}
\lookupPut{R_2BhWAVBiG0l5D3Z}{P298}
\lookupPut{R_AgH6tvlbjnzAgw1}{P299}
\lookupPut{R_Okj81j1io5kZ62B}{P300}
\lookupPut{R_3lFjuW6VwdRmBfG}{P301}
\lookupPut{R_sLJWp3Ym37BGcGR}{P302}
\lookupPut{R_2cbUihVC3Dv6kiV}{P303}
\lookupPut{R_2CoZISB8FFoInKV}{P304}
\lookupPut{R_3ELOJOiP5IdYfY5}{P305}
\lookupPut{R_25NMc04EQLBYTQL}{P306}
\lookupPut{R_29c3of0cEW9r2AZ}{P307}
\lookupPut{R_2V9JI8O05IDgI9F}{P308}
\lookupPut{R_3HHb7BNkil4pwJH}{P309}
\lookupPut{R_3fCqI2NKXllCB4t}{P310}
\lookupPut{R_2gCCb1gEzSxlumZ}{P311}
\lookupPut{R_2trPSUFUj2NvcId}{P312}
\lookupPut{R_1dAzjdfk48upEn2}{P313}
\lookupPut{R_2SGEFmgobQjQ8PX}{P314}
\lookupPut{R_3ELxbvWYi1uRZzJ}{P315}
\lookupPut{R_1BVs4ilMQOANnUO}{P316}
\lookupPut{R_3nvBZSH99z6v7xf}{P317}
\lookupPut{R_XO1hFRbuXuv53DH}{P318}
\lookupPut{R_25Tk6AIykepP9VW}{P319}
\lookupPut{R_3MEZiVXADnefQf5}{P320}
\lookupPut{R_6QBhGWrTysGFt0B}{P321}
\lookupPut{R_2v8oZBfQGrDgkrf}{P322}
\lookupPut{R_2upSs5ctw4atEMX}{P323}
\lookupPut{R_tFisJqI5ZXUxsBP}{P324}
\lookupPut{R_2aFDcTv5D89AvCd}{P325}
\lookupPut{R_32KsnV2i9rU3P0L}{P326}
\lookupPut{R_9mKzu6ZXWKPwbQZ}{P327}
\lookupPut{R_2Pbc5xMxN89j19f}{P328}
\lookupPut{R_1oAtyXs5Lzd91a8}{P329}
\lookupPut{R_31QJb6Uk5LeRIi2}{P330}
\lookupPut{R_2tJzdjKNaKDDItC}{P331}
\lookupPut{R_2tL22NE58as1wUP}{P332}
\lookupPut{R_1immlUzAEXPS5Yc}{P333}
\lookupPut{R_p9mBUYkp3nGna8h}{P334}
\lookupPut{R_2B4oog2pDyJsuB6}{P335}
\lookupPut{R_1QgSTIUrqkTFkYm}{P336}
\lookupPut{R_390LxvoAGXyiLgB}{P337}
\lookupPut{R_C9QURDTfT96z7AR}{P338}
\lookupPut{R_1pRKZdKywGjyVGy}{P339}
\lookupPut{R_1ihbXeOSH7WNXAI}{P340}
\lookupPut{R_3JFdVZtVHPd0jsZ}{P341}
\lookupPut{R_2fDHmT8sYn0PHxq}{P342}
\lookupPut{R_7aE5kUKe5As4mQx}{P343}
\lookupPut{R_sTetq07mzpLTm9j}{P344}
\lookupPut{R_UDygtpcxmqF95yp}{P345}
\lookupPut{R_31bdGbOoTC0O36O}{P346}
\lookupPut{R_2dnnA0Oy9SC0ldj}{P347}
\lookupPut{R_3MhdlJnJaDddr8S}{P348}
\lookupPut{R_30cSbqs2aT7B585}{P349}
\lookupPut{R_2DS6eqxDIwR8KhM}{P350}
\lookupPut{R_zTIMb1o6jhiMvE5}{P351}
\lookupPut{R_2uxMzYIyWhQpSaU}{P352}
\lookupPut{R_3EiKrDrB3aHWr2g}{P353}
\lookupPut{R_Z1XP7NY2CpQCjvP}{P354}
\lookupPut{R_DMKQ7QWeojQeqrf}{P355}
\lookupPut{R_2cdfsLIOtUVF6km}{P356}
\lookupPut{R_yW9ld2e4jJsUpEZ}{P357}
\lookupPut{R_1QnKULWzGkYEPL5}{P358}
\lookupPut{R_3KIRGZHKFOt5wmG}{P359}
\lookupPut{R_0GMVjj8AGUu8ZKV}{P360}
\lookupPut{R_3g1MuqjxxoePvbl}{P361}
\lookupPut{R_sIukv6LxHBaYNlT}{P362}
\lookupPut{R_2YkT5joiClRR52a}{P363}
\lookupPut{R_2Cyu7JUuv4c6hly}{P364}
\lookupPut{R_zdpoNPsLURodyNj}{P365}
\lookupPut{R_yWUss6P9aGUfLsB}{P366}
\lookupPut{R_1i5tt0NzsJZ9hyh}{P367}
\lookupPut{R_30wdqRilTw1dRSF}{P368}
\lookupPut{R_2eW1wJTRFgZMbBG}{P369}
\lookupPut{R_VOVEaLSI52TKE3n}{P370}
\lookupPut{R_6hxPRq3I1dVhE8V}{P371}
\lookupPut{R_2BleNqqDd1VQ3ME}{P372}
\lookupPut{R_2ifCltwPGa4wy3L}{P373}
\lookupPut{R_3ltBuJgHw4FZYfT}{P374}
\lookupPut{R_2QVQOFa3uD1gYJw}{P375}
\lookupPut{R_3JDazXkeTrY34A8}{P376}
\lookupPut{R_2ZZyxQC7L7C0HaF}{P377}
\lookupPut{R_qDCQSQgSWmYj65X}{P378}
\lookupPut{R_3rIktuQ5WhtENnX}{P379}
\lookupPut{R_12D4dW7ANxTnIwB}{P380}
\lookupPut{R_22nzYgD0TdzHk1F}{P381}
\lookupPut{R_db6hKQyZRBCBunf}{P382}
\lookupPut{R_2OVbH4VVXhVkdX0}{P383}
\lookupPut{R_32PHtzXJggT42hZ}{P384}
\lookupPut{R_1LhTAsbJihdf8Vd}{P385}
\lookupPut{R_1o0v0z3KgrKGlFb}{P386}
\lookupPut{R_1QmRfcbf3QQgcI4}{P387}
\lookupPut{R_1Nea8HJ46usu0b9}{P388}
\lookupPut{R_3R2FBCbw96hICaV}{P389}
\lookupPut{R_1ovdkDTXpl7CgEB}{P390}
\lookupPut{R_1d51TyOa75B4UBK}{P391}
\lookupPut{R_1Ka4gqNe3nnRF4V}{P392}
\lookupPut{R_3hAO6dmmrSjTU7V}{P393}
\lookupPut{R_2qlr2coKJQddzwy}{P394}
\lookupPut{R_3hychKhlHBYfWuD}{P395}
\lookupPut{R_3GClcYwGt18Kk7X}{P396}
\lookupPut{R_by2a2nQVhN8g6bf}{P397}
\lookupPut{R_2rBpaJY8Oc2mSvv}{P398}
\lookupPut{R_DTrwG3cL9X9Y1CF}{P399}
\lookupPut{R_27yzjDTh6KYXvJR}{P400}
\lookupPut{R_30l608tVczGNLEQ}{P401}
\lookupPut{R_1FtBWeUuzNsFO0h}{P402}
\lookupPut{R_1OPIgbkj0lHb05E}{P403}
\lookupPut{R_3Hi6Cr1zpHvejOk}{P404}
\lookupPut{R_3J5jX4zozcJJEt6}{P405}
\lookupPut{R_23W8Z1OuAUi14Nk}{P406}
\lookupPut{R_2U4oU45gD68VZl6}{P407}
\lookupPut{R_2OPjeQIw1PTNISY}{P408}
\lookupPut{R_UR8gjYMf9bd1hND}{P409}
\lookupPut{R_1FmlPRJC38FQ8uL}{P410}
\lookupPut{R_12LuLJOKoc5ewdR}{P411}
\lookupPut{R_3QFpsxUR3RCA99h}{P412}
\lookupPut{R_tGuPlwwTEDiExuF}{P413}
\lookupPut{R_3RyJmgDyS7R4emh}{P414}
\lookupPut{R_eR1j35RyvigWBOh}{P415}
\lookupPut{R_3psDMrJfJB9IAVh}{P416}
\lookupPut{R_1hVV1zWldPVMHZ7}{P417}
\lookupPut{R_3ELpQVFXlFPLVXg}{P418}
\lookupPut{R_3Pzd3hVZK6iyan0}{P419}
\lookupPut{R_3kbxn47kHr9A0m7}{P420}
\lookupPut{R_2q98lUdyWJi4Pfe}{P421}
\lookupPut{R_yx3TcRRZLpkJyF3}{P422}
\lookupPut{R_27ythWmrfg5QY40}{P423}
\lookupPut{R_3nSGOKTqj7r6YzL}{P424}
\lookupPut{R_3ESvPsAjM1QijCv}{P425}
\lookupPut{R_28NcKPlPyu60TNH}{P426}
\lookupPut{R_1dmhM2tvePMfU2q}{P427}
\lookupPut{R_2zGeOMbxSS9sTC1}{P428}
\lookupPut{R_2m1h93SIYXFZNux}{P429}
\lookupPut{R_2czrcGXNa8Riwhf}{P430}
\lookupPut{R_rl2U8G4Ouqgncfn}{P431}
\lookupPut{R_3PXBfwPjK3UqSei}{P432}
\lookupPut{R_3rOQE9XhwoaBCIk}{P433}
\lookupPut{R_cYgLpNB5pGtQOwV}{P434}
\lookupPut{R_3QDY0gllmhb8kDN}{P435}
\lookupPut{R_1MJWfIh3Hl2QRwZ}{P436}
\lookupPut{R_BPPRrqHqMI8DSQV}{P437}
\lookupPut{R_2fAKy9qtcFqI6kD}{P438}
\lookupPut{R_2OU1NYjO9YPXxcH}{P439}
\lookupPut{R_3L29joBsVOssQ9r}{P440}
\lookupPut{R_2bZf6qYybIbEo7L}{P441}
\lookupPut{R_WvCqhVGQNqXOFnr}{P442}
\lookupPut{R_2usaiaXJioBiZJK}{P443}
\lookupPut{R_1M5pigAOBAlOjTB}{P444}
\lookupPut{R_2dvkxLJ9ceA2JrC}{P445}
\lookupPut{R_W2LfQ6gsxz5vxYJ}{P446}
\lookupPut{R_3k4VwA2uuxfgiiX}{P447}
\lookupPut{R_2ZErMz5vmgeLzqs}{P448}
\lookupPut{R_5i4oa0BUB1lyhjP}{P449}
\lookupPut{R_ueS14VAQ9I5EHZL}{P450}
\lookupPut{R_qUSOqaaWDOTB8at}{P451}
\lookupPut{R_3novrSIFGwJtNTA}{P452}
\lookupPut{R_8CFrvjTc9SI03Wp}{P453}
\lookupPut{R_2qypJ4BWECDUyOi}{P454}
\lookupPut{R_279yYMIOwv2eyw2}{P455}
\lookupPut{R_2TRb2wmnMQ9yoQu}{P456}
\lookupPut{R_23nluYZnNDgx5mN}{P457}
\lookupPut{R_1MSsJ24Y3fxZvAo}{P458}
\lookupPut{R_2ymFG1W8hhxg0Xo}{P459}
\lookupPut{R_2dzuw8l4ku46xJe}{P460}
\lookupPut{R_24D9BbdJt6w7PzR}{P461}
\lookupPut{R_2fJ44PventcqXgB}{P462}
\lookupPut{R_1pVwxXd1A5lON3d}{P463}
\lookupPut{R_3iWpT4gkfZbR1L8}{P464}
\lookupPut{R_4IvLlGbO0vet2Ct}{P465}
\lookupPut{R_T73ey7QX76wk3PH}{P466}
\lookupPut{R_1BQkUcvubr8kBya}{P467}
\lookupPut{R_2rJybzBo8CjAzDL}{P468}
\lookupPut{R_27JBYI8bI0oSHNM}{P469}
\lookupPut{R_2AYXkgg7oW0zHcA}{P470}
\lookupPut{R_w18XWZrl5QwGqFH}{P471}
\lookupPut{R_2v05hMtz83iP8QS}{P472}
\lookupPut{R_2TCaIVIlsesERcC}{P473}
\lookupPut{R_a4TDGiMtUrm1SSd}{P474}
\lookupPut{R_2E00uAwyx4RVwY4}{P475}
\lookupPut{R_2z6JgYxbVcxQ9DD}{P476}
\lookupPut{R_vYOHiHIPqHbM9z3}{P477}
\lookupPut{R_1roALdYLC1cY7hE}{P478}
\lookupPut{R_2b1uGMmm4vFGi3Q}{P479}
\lookupPut{R_30o9E3jls9Fmxmj}{P480}
\lookupPut{R_2DY6iyOxdTzvKQS}{P481}
\lookupPut{R_2PveDuIW1xSXPnB}{P482}
\lookupPut{R_1kMoGk1UPGYuybk}{P483}
\lookupPut{R_3QWcrQqPwCbt7s9}{P484}
\lookupPut{R_3qk6MIjwqVni3Cu}{P485}
\lookupPut{R_3e7X9ukI4Chv84l}{P486}
\lookupPut{R_3kG17vE4fk6zAy0}{P487}
\lookupPut{R_1PTb9SE9PLMm1EW}{P488}
\lookupPut{R_22GhCSvmLzw3Vkm}{P489}
\lookupPut{R_20VjqtNTNbQWlBp}{P490}
\lookupPut{R_10ZB31w9IIczbJC}{P491}
\lookupPut{R_1IQbI4y4sucQj42}{P492}
\lookupPut{R_20SeMyFnr02Baic}{P493}
\lookupPut{R_Pv96O5pA1gVsld7}{P494}
\lookupPut{R_2Vl46TaYa5qutt0}{P495}
\lookupPut{R_0xlwfvBocPIgdKV}{P496}
\lookupPut{R_12b4dVX9K4oUdXH}{P497}
\lookupPut{R_2954wfjgbmeP5kV}{P498}
\lookupPut{R_tFpL1pXbTs76YuZ}{P499}
\lookupPut{R_3k0adONtvYQNx11}{P500}
\lookupPut{R_XnccZZlni4vmmdP}{P501}
\lookupPut{R_Zh3CADAyXQpnVTz}{P502}
\lookupPut{R_1FQSYPHqUV6czsE}{P503}
\lookupPut{R_C9yAmQlgmCwlD33}{P504}
\lookupPut{R_1Nh3t5RH93w6hnZ}{P505}
\lookupPut{R_bQSXpPnY0BoFfnb}{P506}
\lookupPut{R_eY8UoNT1qNZBA5z}{P507}
\lookupPut{R_ZltABV8DO4sv5kZ}{P508}
\lookupPut{R_3CBnFzCj0T3LunY}{P509}
\lookupPut{R_3hfGUUrmwT0TsUK}{P510}
\lookupPut{R_1HqpWVJhn1KXx3f}{P511}
\lookupPut{R_2oFb6G4rTsGu6Vr}{P512}
\lookupPut{R_xG7x0O0JKY5Njwt}{P513}
\lookupPut{R_1Itm4JHlk5EcVtU}{P514}
\lookupPut{R_3HZz8PM9kTSqJ6l}{P515}
\lookupPut{R_1dd9t1VhdMxIpEH}{P516}
\lookupPut{R_3lSlo8kkjR53pS0}{P517}
\lookupPut{R_D03QcjHQF89BkD7}{P518}
\lookupPut{R_2tsC0zXifP0l6aR}{P519}
\lookupPut{R_vCbIR7P0fKcRsZz}{P520}
\lookupPut{R_2RR4MwdJApLo3eq}{P521}
\lookupPut{R_2EsLiW5WkbTOOoH}{P522}
\lookupPut{R_2AZI7Pz6IeC9Ig1}{P523}
\lookupPut{R_1d7grWka0tUzeLz}{P524}
\lookupPut{R_0x5pZHzhB2hEvUl}{P525}
\lookupPut{R_10C9CfcR2DABuYI}{P526}
\lookupPut{R_aUZaozmIx5mChCF}{P527}
\lookupPut{R_BtUfLnBV2ZsCTzH}{P528}
\lookupPut{R_3kfQjXS8yJog7uN}{P529}
\lookupPut{R_1I9Hf4cgnFtOttj}{P530}
\lookupPut{R_2tgmuEtLpdUmeDt}{P531}
\lookupPut{R_8jZjruF7JSWQyEV}{P532}
\lookupPut{R_1NhpvHlFOunAzIs}{P533}
\lookupPut{R_2PjJNjWJ7JJSj2e}{P534}
\lookupPut{R_2wEK7iouSP9ii6S}{P535}
\lookupPut{R_zeAklvOWtIwUD7z}{P536}
\lookupPut{R_TvYYjR6xRs82CWt}{P537}
\lookupPut{R_3EKbN4XMLRdhOPJ}{P538}
\lookupPut{R_3CNJDMVRDlaDUeF}{P539}
\lookupPut{R_djnYn3UndlAx0l3}{P540}
\lookupPut{R_3PHMr03sWpGJc6V}{P541}
\lookupPut{R_3HFqJ0eMtFKqSml}{P542}
\lookupPut{R_1qXvHDYa6663MYB}{P543}
\lookupPut{R_3qF1J5lGenS4LSV}{P544}
\lookupPut{R_1dnrWm47wl7UWhF}{P545}
\lookupPut{R_2VBFh65XflPD1k9}{P546}
\lookupPut{R_1DOWjFQHYpsaacG}{P547}
\lookupPut{R_3KDB8lQ1O4XSOgY}{P548}
\lookupPut{R_1gbGR1fHLOmuEHZ}{P549}
\lookupPut{R_2tsQTevU9GehGd5}{P550}
\lookupPut{R_3GqPVd9urP2FPAc}{P551}
\lookupPut{R_3EbfzmvK96BSYJI}{P552}
\lookupPut{R_2ZTxbVHtqY7pq71}{P553}
\lookupPut{R_3Ras30IhMXftfun}{P554}
\lookupPut{R_6PtHdi4udsXn9PX}{P555}
\lookupPut{R_OkgRGtotY66sBLX}{P556}
\lookupPut{R_AMrqGQ7Rc4gUFax}{P557}
\lookupPut{R_3PACVRYskz9rhsW}{P558}
\lookupPut{R_bNNgV8AOSoKIHMB}{P559}
\lookupPut{R_p5Y1rroZXBMMOHv}{P560}
\lookupPut{R_yJTEq6Te1BjRmgh}{P561}
\lookupPut{R_1N5rvPo072Pay7K}{P562}
\lookupPut{R_VOkExUuSzre79Kx}{P563}
\lookupPut{R_2qggkHKxll4vnUW}{P564}
\lookupPut{R_cUP7DcTQ4UqUiu5}{P565}
\lookupPut{R_0celH9oEiGnAkaB}{P566}
\lookupPut{R_1yP7oNZIui88rzb}{P567}
\lookupPut{R_PCb1vmXM8gPXhMR}{P568}
\lookupPut{R_1gcMXNZ2PJNvmh9}{P569}
\lookupPut{R_2rG6ZTZJicVoDgR}{P570}
\lookupPut{R_1MLgmRQGuAwBfIA}{P571}
\lookupPut{R_9mnvmUbGrQr47Lz}{P572}
\lookupPut{R_2Wv4y4xXTOnhGsc}{P573}
\lookupPut{R_OpUFK5jQArToB6p}{P574}
\lookupPut{R_Anx7w6jf34JV3vr}{P575}
\lookupPut{R_1itk2BsKI53Ied0}{P576}
\lookupPut{R_3MGKnTxdsKRzxWe}{P577}
\lookupPut{R_1CfPmmlU1Wiz6Vb}{P578}
\lookupPut{R_3HUEXT8us3Ow24o}{P579}
\lookupPut{R_1C1LsfTu7JrBT2y}{P580}
\lookupPut{R_bwJykqUw1JolJGp}{P581}
\lookupPut{R_31WpfXR6Br9gyqR}{P582}
\lookupPut{R_2CTk1AI0oGJSp7R}{P583}
\lookupPut{R_2dHjDshgz276hVn}{P584}
\lookupPut{R_ZBJL4gKzoj1f92p}{P585}
\lookupPut{R_2wNAIR7xxH4S4O6}{P586}
\lookupPut{R_2ANo3EuRJzyWGes}{P587}
\lookupPut{R_Rl5836USj8kHA9b}{P588}
\lookupPut{R_24AF5iBmzuFpeS6}{P589}
\lookupPut{R_1LJvUaJJceq0Hmh}{P590}
\lookupPut{R_2WI1ZXxBaFo7hgN}{P591}
\lookupPut{R_3RruZEU2ofciHO2}{P592}
\lookupPut{R_2RUQZMv2QoUGbdA}{P593}
\lookupPut{R_2rTMAeNF13wtOiz}{P594}
\lookupPut{R_3suETDPEnjdEaIF}{P595}
\lookupPut{R_22LhsPdaRlZlbmB}{P596}
\lookupPut{R_2teSTCglf0l7Dfg}{P597}
\lookupPut{R_stVhqrceEm9Oisx}{P598}
\lookupPut{R_3I3SPVGtMpCnRQq}{P599}
\lookupPut{R_oZfbVsTy2X9tDBn}{P600}
\lookupPut{R_3fJXgb2TQfqnCV3}{P601}
\lookupPut{R_OBYcRjLsj4qfmjn}{P602}
\lookupPut{R_1LZ0JToA3n6Cstg}{P603}
\lookupPut{R_2zBZBQTmUF7zCgM}{P604}
\lookupPut{R_265CSKGVMAK5k69}{P605}
\lookupPut{R_1f9gTgSJVj93ZtL}{P606}
\lookupPut{R_1P4MtlTAEWu3AOV}{P607}
\lookupPut{R_1dpkctpR4nkX4FB}{P608}
\lookupPut{R_1H6k5xVBSigRTqk}{P609}
\lookupPut{R_b7QBbRFVgPJLNG9}{P610}
\lookupPut{R_3CT4zha0L6MiTEu}{P611}
\lookupPut{R_2fqQXLolqkGceg8}{P612}
\lookupPut{R_Bzdi7g4YUGn4p45}{P613}
\lookupPut{R_1M3w0xqhVMBQniI}{P614}
\lookupPut{R_29tf6mNgwpyIbbw}{P615}
\lookupPut{R_3EEJnW4aldQE9Rc}{P616}
\lookupPut{R_2YRhr7ZndVNEfRZ}{P617}
\lookupPut{R_wXFz8W4saUbpvr3}{P618}
\lookupPut{R_31XI2rebAnejIqC}{P619}
\lookupPut{R_8Aq7wBTnJCxRpDP}{P620}
\lookupPut{R_2AGd1HKvERqpOmc}{P621}
\lookupPut{R_2cchrVAI7b1kYtq}{P622}
\lookupPut{R_2tDaYKYTdaB5kHj}{P623}
\lookupPut{R_sjy4SJRi9AoMYfv}{P624}
\lookupPut{R_1DI8MxIDOikKOoj}{P625}
\lookupPut{R_sRwqVdLeUGMTsch}{P626}
\lookupPut{R_3ivTpEsnqkL3bg9}{P627}
\lookupPut{R_2ygbzRXLwFhkcex}{P628}
\lookupPut{R_3FRJMuk8CVduA34}{P629}
\lookupPut{R_1Q9HWVFMUTZ8VKg}{P630}
\lookupPut{R_QgBoz895c2Z1FQJ}{P631}
\lookupPut{R_XGHiobDwzEXtWcp}{P632}
\lookupPut{R_ym9DdCldebQ9TNv}{P633}
\lookupPut{R_6Rm45Q9J4hpqZi1}{P634}
\lookupPut{R_2VdxwUcL9kFfEn4}{P635}
\lookupPut{R_AjO6ZXY7JoF969z}{P636}
\lookupPut{R_2To9VkwldrkuuvC}{P637}
\lookupPut{R_PM08lxtxOQ4tY2J}{P638}
\lookupPut{R_1riTvwf3vhSsE2s}{P639}
\lookupPut{R_32OPCUw1zw3cRZK}{P640}
\lookupPut{R_30iXEwFp9FvYy3o}{P641}
\lookupPut{R_2Eb32x980Tk1Bis}{P642}
\lookupPut{R_3NCX9VpQpluY3OF}{P643}
\lookupPut{R_29nt19EDs6laNRH}{P644}
\lookupPut{R_1pEjzYeozgisP63}{P645}
\lookupPut{R_xu3KJDkdlK1w49z}{P646}
\lookupPut{R_30jf8xiVF9Od4Cz}{P647}
\lookupPut{R_22ukI9pBecuZhm2}{P648}
\lookupPut{R_1ls21923Py8PC51}{P649}
\lookupPut{R_33DuzxVMDP8VQTX}{P650}
\lookupPut{R_3qPRDBGtfCWnCEo}{P651}
\lookupPut{R_1Kd9mUZsLGiXSLw}{P652}
\lookupPut{R_2YaJrj6qqXfEm0Y}{P653}
\lookupPut{R_1LAqSANA3wHNpPW}{P654}
\lookupPut{R_3wOcURxYquERoit}{P655}
\lookupPut{R_3isqFXM5NEOrlsv}{P656}
\lookupPut{R_3CHMNXUCoopMKov}{P657}
\lookupPut{R_25AQfv50oY9o8UX}{P658}
\lookupPut{R_3j7fQGfQfnUnlgd}{P659}
\lookupPut{R_3QQX0HzGgPEtz7o}{P660}
\lookupPut{R_3kHB9YjdWWE7V0d}{P661}
\lookupPut{R_11ZauEIdsVy91br}{P662}
\lookupPut{R_1PbLrOBFq83uor9}{P663}
\lookupPut{R_2Ce5xqGXzXTuRKl}{P664}
\lookupPut{R_3nOfIpD4vkMBygu}{P665}
\lookupPut{R_2S0ozbTYQMJp37a}{P666}
\lookupPut{R_3hfHJULgw1QhVE7}{P667}
\lookupPut{R_3ixlwGuIC1Ftkcc}{P668}
\lookupPut{R_1mruQc8Ny3qH6Wv}{P669}
\lookupPut{R_3D6DKjjxvjiVj2P}{P670}
\lookupPut{R_vc9rfLMi6ljfm25}{P671}
\lookupPut{R_1gUcbBH4CIjt2lE}{P672}
\lookupPut{R_UYozoyzobSGAyml}{P673}
\lookupPut{R_2wvn1VIOvuxufXW}{P674}
\lookupPut{R_1pAKVLN8WASgam0}{P675}
\lookupPut{R_3nAFJsJthnZ24Sk}{P676}
\lookupPut{R_1Dv6pjFbJ5UFWp6}{P677}
\lookupPut{R_10vmKitSIhXK10f}{P678}
\lookupPut{R_1CHGAHXk5Sh8DKa}{P679}
\lookupPut{R_vvIXB1U0UPTv0dP}{P680}
\lookupPut{R_1meDUhvlmb9zhrA}{P681}
\lookupPut{R_2EmAVDs6AGudmp0}{P682}
\lookupPut{R_eMbfltCT1IZzr0d}{P683}
\lookupPut{R_3kb6OpljnzUJIWA}{P684}
\lookupPut{R_3O1YsXOlNzDVumS}{P685}
\lookupPut{R_u9rxVmoI6DVGOZ3}{P686}
\lookupPut{R_22xAQxZn7ghoCJt}{P687}
\lookupPut{R_1dBt8B8QK76rTZm}{P688}
\lookupPut{R_3EA94mBy6gEvjIT}{P689}
\lookupPut{R_WpQLOp0JeJnERJD}{P690}
\lookupPut{R_sROzubcZSHDdNgB}{P691}
\lookupPut{R_279hEXAtMvGXOdD}{P692}
\lookupPut{R_1KvZCo8pKMAuWvu}{P693}
\lookupPut{R_1dKX9WvYqTNhWz4}{P694}
\lookupPut{R_3VMZl7YbsTQ9OcF}{P695}
\lookupPut{R_3nqMMmi9S24lDpF}{P696}
\lookupPut{R_1hAqSYdfMd6cdsS}{P697}
\lookupPut{R_2P7trrWAf7UeFk9}{P698}
\lookupPut{R_1N4xItp0k1iiZl3}{P699}
\lookupPut{R_2fWkB3lMgqUTb33}{P700}
\lookupPut{R_1EYxqerKjeH39Ht}{P701}
\lookupPut{R_3ZRFbS2BGS3HkBj}{P702}
\lookupPut{R_2c6kYGOjLOI5WI3}{P703}
\lookupPut{R_1P6j8GV4gXcOi2H}{P704}
\lookupPut{R_1rC7iDCXfliHR9b}{P705}
\lookupPut{R_3qfzVV2HgKXiQD0}{P706}
\lookupPut{R_3hEpkb0s53zVpfL}{P707}
\lookupPut{R_3iqjjrsS0ofKDhy}{P708}
\lookupPut{R_11i3KWzXnEkpQQr}{P709}
\lookupPut{R_2YlE8QdS0LgrDA2}{P710}
\lookupPut{R_eL02pn4ZqHcNWYF}{P711}
\lookupPut{R_2TpYINBFZU1T9TF}{P712}
\lookupPut{R_3Ok9gqVoixzAdWM}{P713}
\lookupPut{R_1IDUxwQL4YSLkOW}{P714}
\lookupPut{R_3EYXB8IoDxsgG3P}{P715}
\lookupPut{R_2Vl7cYM8yNA1LcE}{P716}
\lookupPut{R_6XaFRih7Wzii5d7}{P717}
\lookupPut{R_UMjz2EJabwubyzD}{P718}
\lookupPut{R_3CI7GaqTtyHL5cs}{P719}
\lookupPut{R_1Dv6qZ298lJDcXf}{P720}
\lookupPut{R_2AFgmQiaBzye2F8}{P721}
\lookupPut{R_3h3JwEjG6SdElaz}{P722}
\lookupPut{R_214W2Dimh6Xtk5i}{P723}
\lookupPut{R_3HoVC1bKzUkIY3z}{P724}
\lookupPut{R_2B4ZWxkbCrHkiyE}{P725}
\lookupPut{R_W3sV5veBz8G2gql}{P726}
\lookupPut{R_3NOEQoQZ7KCWF9M}{P727}
\lookupPut{R_9oeSQ7FL9HuuGad}{P728}
\lookupPut{R_1Tif7259koaHj1v}{P729}
\lookupPut{R_1H06SamhMBcgVsk}{P730}
\lookupPut{R_30eKhsUaNflxmQD}{P731}
\lookupPut{R_bPMpTuA5COBvgl3}{P732}
\lookupPut{R_3g1kpqO0k1V3kFp}{P733}
\lookupPut{R_9uhMIz5Vxgs4POF}{P734}
\lookupPut{R_3iWplonXLKSmLto}{P735}
\lookupPut{R_2rvgJ2FaoUXOCB9}{P736}
\lookupPut{R_2VK31rvFCJ87ejI}{P737}
\lookupPut{R_pgCImwaEmRHJlgl}{P738}
\lookupPut{R_24HOtELfrTovMpG}{P739}
\lookupPut{R_3nfxpAmi1KJVNgM}{P740}
\lookupPut{R_YVOfflZMPJ0GjOV}{P741}
\lookupPut{R_2b2EYDHYZmVIBF9}{P742}
\lookupPut{R_3MPzZn0jzGTgeO8}{P743}
\lookupPut{R_21pmD25bm66eQtc}{P744}
\lookupPut{R_xzxVQlHpLbQiMh3}{P745}
\lookupPut{R_11j9HNj4cOM1pfa}{P746}
\lookupPut{R_2CTxBxhmOcrUnzz}{P747}
\lookupPut{R_3gUdJSj5SeM68gr}{P748}
\lookupPut{R_2zYez9aW5eSco6F}{P749}
\lookupPut{R_DnNCgyRXmr2fTyh}{P750}
\lookupPut{R_2bN0UPLkr72H2bQ}{P751}
\lookupPut{R_pF7z7MUGyFrllYJ}{P752}
\lookupPut{R_AdFkpWIkoo73snn}{P753}
\lookupPut{R_xr83MWoPYpZ5gd3}{P754}
\lookupPut{R_1gjD9RHUhGdPlCF}{P755}
\lookupPut{R_2rO6yEIAqeliRj3}{P756}
\lookupPut{R_2rFDSXbAzg1epUt}{P757}
\lookupPut{R_Ry2Dr9E4Y4AfNL3}{P758}
\lookupPut{R_z8AQfywjMIDnTbP}{P759}
\lookupPut{R_3G0tgIfLYOM9qdV}{P760}
\lookupPut{R_1CrooKGLrpGP8i2}{P761}
\lookupPut{R_9tLid2Mh2VSXtn3}{P762}
\lookupPut{R_2ZWuDgnd1WeUZaq}{P763}
\lookupPut{R_3FUS1dupyrbU06D}{P764}
\lookupPut{R_1dayzupM8DdFI30}{P765}
\lookupPut{R_2QtW7EMSYQ4ibGL}{P766}
\lookupPut{R_22of40DU48EyRwL}{P767}
\lookupPut{R_2q7wfOyKCTzOWky}{P768}
\lookupPut{R_3qeULwkpRaXDGnG}{P769}
\lookupPut{R_2D5j0gKRMp6J7JY}{P770}
\lookupPut{R_27dKpru7hX2y0XL}{P771}
\lookupPut{R_yl8ThKcllgBUVdn}{P772}
\lookupPut{R_0ofCH705i6Xpxfz}{P773}
\lookupPut{R_3PZCztraJWdpA9H}{P774}
\lookupPut{R_2fB84x9S8vYsBpl}{P775}
\lookupPut{R_308EEvJcekurO9J}{P776}
\lookupPut{R_2rNmmPZI9A7bQMZ}{P777}
\lookupPut{R_p4s2OrAM4DGjz7b}{P778}
\lookupPut{R_C1aAI5PduYIEjtv}{P779}
\lookupPut{R_1JCjQayokmOtVXW}{P780}
\lookupPut{R_BPOp1yK5vgc1VLP}{P781}
\lookupPut{R_1rlhphEJQ8jgnzc}{P782}
\lookupPut{R_2f1Kke10utIIOy8}{P783}
\lookupPut{R_25Z9lDGd7W0dD7b}{P784}
\lookupPut{R_31a9vJePRUEHxaO}{P785}
\lookupPut{R_1eJhh93KSbcnHoj}{P786}
\lookupPut{R_00nUjvZT9bdnDTb}{P787}
\lookupPut{R_3EuIQBWzQYIOs1z}{P788}
\lookupPut{R_1mRFyEPsICaOYGh}{P789}
\lookupPut{R_1oAWUcUoESLHnF9}{P790}
\lookupPut{R_2EsqYDXocagFMUo}{P791}
\lookupPut{R_2vhQtvNHaX9dzhs}{P792}
\lookupPut{R_2ZP4GXQenGFgIJC}{P793}
\lookupPut{R_30hFG7OJVno5faG}{P794}
\lookupPut{R_325eCwTchvw6y7A}{P795}
\lookupPut{R_1Cl7f3O11lkx4DU}{P796}
\lookupPut{R_33evIFp16rMb7Ty}{P797}
\lookupPut{R_3CHjAFDxUSQ0heE}{P798}
\lookupPut{R_2aRzWxnhQBuYBLa}{P799}
\lookupPut{R_3h9qPAqOssMWcbJ}{P800}
\lookupPut{R_73656UZsWujNJ1n}{P801}
\lookupPut{R_2bJGOzf3cfzFmBa}{P802}
\lookupPut{R_3CBduVNtLmNJI94}{P803}
\lookupPut{R_1NbmBLtwVa6K1Ma}{P804}
\lookupPut{R_3HF7zegKAqZURZQ}{P805}
\lookupPut{R_2uDnIM9YywmBW4v}{P806}
\lookupPut{R_2EFSL5evu6lRlu5}{P807}
\lookupPut{R_3CCPlSLF7DRwxMS}{P808}
\lookupPut{R_3PQSY6t7mZtMMPv}{P809}
\lookupPut{R_3MfDhkAuJbgweYv}{P810}
\lookupPut{R_31nq8JSqBtKaUkh}{P811}
\lookupPut{R_3gUhfc8xUgQUD0Y}{P812}
\lookupPut{R_2AMbOOYs9nmeH1X}{P813}
\lookupPut{R_1CdLbIhO3FgeWqW}{P814}
\lookupPut{R_1fdZWNkMnIwK6vb}{P815}
\lookupPut{R_3OlFnpbYfdNeH9S}{P816}
\lookupPut{R_sO25ymu0VtLjXr3}{P817}
\lookupPut{R_2QJRXiVowNdo5bq}{P818}
\lookupPut{R_1jrEbMkTMU00RNq}{P819}
\lookupPut{R_10Bl1NUr8hk0wiX}{P820}
\lookupPut{R_2qBG1wCJcHR86K1}{P821}
\lookupPut{R_1oaMZOj8jJvNKZR}{P822}
\lookupPut{R_1IWd8SOlwM8sNAQ}{P823}
\lookupPut{R_2lgLgldc55XB453}{P824}
\lookupPut{R_3rNxF4zVzGuXVHU}{P825}
\lookupPut{R_vdBS1cD3Ok8MlHP}{P826}
\lookupPut{R_9F7rPJ5ceoT2Os1}{P827}
\lookupPut{R_1hBkA6wna6XAaJQ}{P828}
\lookupPut{R_1LYHv2CkDgADXXR}{P829}
\lookupPut{R_yqDOUYlkpCfZFWV}{P830}
\lookupPut{R_1CrP7DLZkETQ989}{P831}
\lookupPut{R_3MgPUxMfIOdkwkx}{P832}
\lookupPut{R_03AB2ycRjm2WuFX}{P833}
\lookupPut{R_p9ICYhP39q9KMxj}{P834}
\lookupPut{R_3nBiWCBhA2MlFnP}{P835}
\lookupPut{R_2SwVKX0OkfJwlWc}{P836}
\lookupPut{R_3QYT0GXf77MNduc}{P837}
\lookupPut{R_2xXB9ZgYNer5aFt}{P838}
\lookupPut{R_3s4v0Ih4O8myhWj}{P839}
\lookupPut{R_8CCfJWho4K4URVf}{P840}
\lookupPut{R_PvxSoqC0YSJfkHv}{P841}
\lookupPut{R_2Tpea7vT7HIUXJd}{P842}
\lookupPut{R_1HjnFH1gpmrJWyu}{P843}
\lookupPut{R_06AFofuZpPTYKIx}{P844}
\lookupPut{R_1hycnKfGlFjRB8E}{P845}
\lookupPut{R_1mUv5KSuazAjQrp}{P846}
\lookupPut{R_2aXmIkfEWO8AgQC}{P847}
\lookupPut{R_A1lJxujHc5K13DX}{P848}
\lookupPut{R_qQGKrAjwYPyw8LL}{P849}
\lookupPut{R_1LSDuxXsz3ItZXw}{P850}
\lookupPut{R_2ckF4Lb6YvnLAGl}{P851}
\lookupPut{R_2WCKxekSq7zgYrT}{P852}
\lookupPut{R_3fTbZ67wbiLkyL5}{P853}
\lookupPut{R_3qJvzd2BwIQULx1}{P854}
\lookupPut{R_2bTZb1MtpMn7USl}{P855}
\lookupPut{R_3PX3jlGQ7vv6mCM}{P856}
\lookupPut{R_2wvllC2KZio8vUV}{P857}
\lookupPut{R_3kHVGnvsJTXBXcR}{P858}
\lookupPut{R_e4YmPjYMzuydUwp}{P859}
\lookupPut{R_2PvzVDfPcpMmJxr}{P860}
\lookupPut{R_1lhEbvOpFRZFpIA}{P861}
\lookupPut{R_3g5SaT7rK8JdSve}{P862}
\lookupPut{R_27W0pCuewtJaCEY}{P863}
\lookupPut{R_3iHVUHGVMCdqPOf}{P864}
\lookupPut{R_2AZO0AA9C6aZvHQ}{P865}
\lookupPut{R_12SEcdADTOtbl1o}{P866}
\lookupPut{R_Uo0MGDk1CyXVmqB}{P867}
\lookupPut{R_9prBDqLI2xVhlUB}{P868}
\lookupPut{R_0ibDpwiEdtjvrxL}{P869}
\lookupPut{R_1daFOHN8ZqSoKFe}{P870}
\lookupPut{R_RlzreebAE4ieWyt}{P871}
\lookupPut{R_1jJa2eohV8tLhSC}{P872}
\lookupPut{R_3ewcxFO2dtFamUN}{P873}
\lookupPut{R_BrGiA2RqPnLhCY9}{P874}
\lookupPut{R_2y7QbtKpFUbqe5R}{P875}
\lookupPut{R_1PTUejux9q4aciL}{P876}
\lookupPut{R_acaZEVAatRpWg3n}{P877}
\lookupPut{R_2dWmL00PEiCDj7n}{P878}
\lookupPut{R_22rWPmwmShQjRqI}{P879}
\lookupPut{R_27lLGeBkmhUdBGZ}{P880}
\lookupPut{R_3PUog910Ut2BOgP}{P881}
\lookupPut{R_1Fm7FlagiFqN5a9}{P882}
\lookupPut{R_3DoPgbYBMtH8QG0}{P883}
\lookupPut{R_3GdJzjY9LwY90ri}{P884}
\lookupPut{R_2wSR3WGvZOQhtHB}{P885}
\lookupPut{R_1rrxiSVJVOySSfj}{P886}
\lookupPut{R_1CJbeOXrszX43zT}{P887}
\lookupPut{R_DxeiSyylib98vgB}{P888}
\lookupPut{R_C2AHSKDkYIJAbJL}{P889}
\lookupPut{R_1QGGzsOxCK255va}{P890}
\lookupPut{R_1Om0SQTG1ltL99a}{P891}
\lookupPut{R_1FKGfD04ju5u8fO}{P892}
\lookupPut{R_2THLgTy4dKgggh6}{P893}
\lookupPut{R_1rxKrSPFfJZFtMN}{P894}
\lookupPut{R_29mNGcorrGQqAGM}{P895}
\lookupPut{R_2VlOSW7nf9Hm2nT}{P896}
\lookupPut{R_2A0VrvzPiSRxOMX}{P897}
\lookupPut{R_2YfwW8XAwVdZmAx}{P898}
\lookupPut{R_1OV2K377879LWL8}{P899}
\lookupPut{R_1CHfiFfZDOrkWVq}{P900}
\lookupPut{R_1OUL5JwXWqySvEh}{P901}
\lookupPut{R_1LUJHwZiKmqmomG}{P902}
\lookupPut{R_2TvpPwm2uzNyQvQ}{P903}
\lookupPut{R_2eUMuYhNF80MUIo}{P904}
\lookupPut{R_2RRlZFlcJdaFR3F}{P905}
\lookupPut{R_1Iza3yCtjvUmQS6}{P906}
\lookupPut{R_2c7rcZZZTiJpKWB}{P907}
\lookupPut{R_25Eez8sloLGliJq}{P908}
\lookupPut{R_3DoxMUjrnZFCa96}{P909}
\lookupPut{R_uylkgEmw2EEuFMt}{P910}
\lookupPut{R_2zO8MVD1mg8mZcs}{P911}
\lookupPut{R_5w4csxELm8S4EmZ}{P912}
\lookupPut{R_2UW7fAsvCPfjecr}{P913}
\lookupPut{R_31Mdq2LEbt9ClE1}{P914}
\lookupPut{R_qCMMTsmqWMrNN05}{P915}
\lookupPut{R_3O1ZTQqIbTgEEbM}{P916}
\lookupPut{R_XNffiSnpZWQPq3D}{P917}
\lookupPut{R_1ojYhV8Q9elDGre}{P918}
\lookupPut{R_1jZdw49Ixa8u2ZD}{P919}
\lookupPut{R_3HFCoGE7YOmbjhT}{P920}
\lookupPut{R_1mfnyVFAzMiyxvj}{P921}
\lookupPut{R_2uEe4youONZBwAk}{P922}
\lookupPut{R_3dDDqBMmq1fl3iz}{P923}
\lookupPut{R_2XhxkyVRGok3PiB}{P924}
\lookupPut{R_Cmhs6uTjdEuxnLr}{P925}
\lookupPut{R_2CkBmbzn06g2JvQ}{P926}
\lookupPut{R_3J3qu9GzYM2KAaf}{P927}
\lookupPut{R_3stVxOaPuXcUgJP}{P928}
\lookupPut{R_2aP5gbH6HHk4pDU}{P929}
\lookupPut{R_Y4bK9fRAQNF03kt}{P930}
\lookupPut{R_1OPsqHkxnIfwxtG}{P931}
\lookupPut{R_1IRAcPopyNfYbpG}{P932}
\lookupPut{R_YXOo2vYnBi4zGxz}{P933}
\lookupPut{R_1dKpzADDEmGGrep}{P934}
\lookupPut{R_2eRnnSGsUWUznTj}{P935}
\lookupPut{R_vkpwL4X8TgyLMKB}{P936}
\lookupPut{R_1dcXHNN6ZbeURQy}{P937}
\lookupPut{R_125U78s7Qs21PMi}{P938}
\lookupPut{R_2S948KHQlApGBai}{P939}
\lookupPut{R_3dMPaDDcZNYexYy}{P940}
\lookupPut{R_2WD802oitAQCLSE}{P941}
\lookupPut{R_1rBO7bUPL86joek}{P942}
\lookupPut{R_2QgtTEeUw4U0zdu}{P943}
\lookupPut{R_2P5mTY1vgoCWLm2}{P944}
\lookupPut{R_3DvVjEBhum8lDsG}{P945}
\lookupPut{R_1Hq5rXTEcu3UGBD}{P946}
\lookupPut{R_3EnTrkN4cjb06S1}{P947}
\lookupPut{R_2znALRsVHIsAAxj}{P948}
\lookupPut{R_2ZE1ZhDaoMoRS3b}{P949}
\lookupPut{R_3m4QuhahdssTPpx}{P950}
\lookupPut{R_3dX1qJKCXmFwzCt}{P951}
\lookupPut{R_r1oTo1NyKMJoEGl}{P952}
\lookupPut{R_dajYFyh6llyjUyJ}{P953}
\lookupPut{R_2a8j5yT3nJXhquX}{P954}
\lookupPut{R_3isPdLN0ZH46ytE}{P955}
\lookupPut{R_3PzQP5QSlICIVtd}{P956}
\lookupPut{R_yl5sCr6JfzyDSbD}{P957}
\lookupPut{R_1Du814bCaPzaf53}{P958}
\lookupPut{R_WpVF1tErtbCFxqF}{P959}
\lookupPut{R_sjbhGX7JoEWMkqR}{P960}
\lookupPut{R_1Ow3iuowq6TEvx3}{P961}
\lookupPut{R_1IiSh180CAtWfQ4}{P962}
\lookupPut{R_2ZOlP7NrqBgKc4r}{P963}
\lookupPut{R_AtcPVzUEo7JCyK5}{P964}
\lookupPut{R_3JFyq5b9TgpaNon}{P965}
\lookupPut{R_1GWjw2XfeenFibj}{P966}
\lookupPut{R_sbaSPj0rplBFuEx}{P967}
\lookupPut{R_3dHa6AMNobCkre7}{P968}
\lookupPut{R_3j1FR6g7peYe0o2}{P969}
\lookupPut{R_3D54pnaA09wS4le}{P970}
\lookupPut{R_3QM3nE4RRPKuphL}{P971}
\lookupPut{R_XjGWRl15lL3Nmzn}{P972}
\lookupPut{R_XN6uTfZ8XrMgMZb}{P973}
\lookupPut{R_Dcs5AAgxCwSYX05}{P974}
\lookupPut{R_1gT15NbPosujoPF}{P975}
\lookupPut{R_vo67G4xGmCndzpf}{P976}
\lookupPut{R_spPHcXhsOy1Mjuh}{P977}
\lookupPut{R_215qJXDdeQvfWwE}{P978}
\lookupPut{R_2wQt0njZeExIB0F}{P979}
\lookupPut{R_2cAQTy06OWQ0twk}{P980}
\lookupPut{R_25NOPP7l7kmc7Zl}{P981}
\lookupPut{R_1hEw0gPKS9XHVHc}{P982}
\lookupPut{R_22Y2FYymUpOe1PA}{P983}
\lookupPut{R_eD8TNDaNxjWTnZ7}{P984}
\lookupPut{R_2cqNsESluNYmRHS}{P985}
\lookupPut{R_07nwQMFy55bAFzz}{P986}
\lookupPut{R_2COpBSBUW8g5sQt}{P987}
\lookupPut{R_1LtewUMRC81lrHC}{P988}
\lookupPut{R_sc9VqLt101o1E41}{P989}
\lookupPut{R_3ERmga8Qvu4YGdC}{P990}
\lookupPut{R_1GU4gcwUVWoh0kg}{P991}
\lookupPut{R_xAEeaa7dxfGjz45}{P992}
\lookupPut{R_7TVMPUENwN8LrUd}{P993}
\lookupPut{R_1E703c3ifvaeSOm}{P994}
\lookupPut{R_3Hv72SmHJGFU9Nz}{P995}
\lookupPut{R_11au5A1mvqoSp1f}{P996}
\lookupPut{R_1IQ6bkf28KENtGF}{P997}
\lookupPut{R_2TNv9thEkDTvur4}{P998}
\lookupPut{R_3CGUVw40PSbFYZ4}{P999}
\lookupPut{R_OMXiedx7N52TqLv}{P1000}
\lookupPut{R_32Q39KOhYUBFvj0}{P1001}
\lookupPut{R_QoBdiCebl8YO0uZ}{P1002}
\lookupPut{R_3M3tovcZxhRMBtF}{P1003}
\lookupPut{R_1g6COFm1UUnO98q}{P1004}
\lookupPut{R_DB7ayVrJll93dHX}{P1005}
\lookupPut{R_3PTbl1r7Q7pjgYS}{P1006}
\lookupPut{R_2VsY7Tr2NIQZMNG}{P1007}
\lookupPut{R_vlwS2SfhXvhMP73}{P1008}
\lookupPut{R_1d1fYmo9dQUtODx}{P1009}
\lookupPut{R_8evIyAQLpzWipnb}{P1010}
\lookupPut{R_278MmyB3x80zUol}{P1011}
\lookupPut{R_t0tbPiiZ4b2vs2Z}{P1012}
\lookupPut{R_3ER3V3aSCaZkbS7}{P1013}
\lookupPut{R_RfXwdogbLGztrUJ}{P1014}
\lookupPut{R_3qwYFY2BLU36w7Q}{P1015}
\lookupPut{R_25BdiNhowgFFQZI}{P1016}
\lookupPut{R_pTaEUWW7Uxid2GB}{P1017}
\lookupPut{R_2CZKhGtLbLOOz3K}{P1018}
\lookupPut{R_2vYMMoRPTuq7hD8}{P1019}
\lookupPut{R_OuNkRKRIS5g7jmF}{P1020}
\lookupPut{R_237icOE3srAbSmV}{P1021}
\lookupPut{R_22xL2NBKSRrlyqA}{P1022}
\lookupPut{R_2uIEjOVsQjWeFQe}{P1023}
\lookupPut{R_3KNrcROUuhkIrUE}{P1024}
\lookupPut{R_zfoLqmNA06rlfRn}{P1025}
\lookupPut{R_3HIg7WV8IBGXrTY}{P1026}
\lookupPut{R_DwKI2SEYW5TpPl7}{P1027}
\lookupPut{R_27BSf7htpAju5bm}{P1028}
\lookupPut{R_3luBXRbJkZKzqfT}{P1029}
\lookupPut{R_1GQhT9rEoQW1jow}{P1030}
\lookupPut{R_11jypE5Uu1ZgdLe}{P1031}
\lookupPut{R_1IGyjeZo7vOOMwG}{P1032}
\lookupPut{R_3OjGku31S2j5qIh}{P1033}
\lookupPut{R_232eLNU8osGoLh9}{P1034}
\lookupPut{R_2Ph6FXaGooj1yRF}{P1035}
\lookupPut{R_1MXIQNHEhW4DBIw}{P1036}
\lookupPut{R_1KlkhWK0C4CZtOH}{P1037}
\lookupPut{R_2zSDg1HkPCpBb1e}{P1038}
\lookupPut{R_bE4u95QdGvA6VGN}{P1039}
\lookupPut{R_1o1kgjika901VJ9}{P1040}
\lookupPut{R_3ho8UpPxEyL8Mq5}{P1041}
\lookupPut{R_OBAvpKnEIGe023f}{P1042}
\lookupPut{R_eaE0qC0VdjHAayt}{P1043}
\lookupPut{R_27Vhr4hTjwhmlnF}{P1044}
\lookupPut{R_1LSE9QiwnuxNphN}{P1045}
\lookupPut{R_1rvMKsakew7AE0K}{P1046}
\lookupPut{R_33wdKDr6FFzrAIV}{P1047}
\lookupPut{R_1DTfsPqdppaBXoB}{P1048}
\lookupPut{R_30izw8NJ9m0plFk}{P1049}
\lookupPut{R_dcA1qgi7sBDCowN}{P1050}
\lookupPut{R_33lL7GSZRZXmpiO}{P1051}
\lookupPut{R_3iDWN85DTkLJhWv}{P1052}
\lookupPut{R_3hGCF4qvkplcpaY}{P1053}
\lookupPut{R_24794aHozvAp59l}{P1054}
\lookupPut{R_2dLakyJtpq0ovgU}{P1055}
\lookupPut{R_3qInQEtj4h6UrWd}{P1056}
\lookupPut{R_2PqbnBg6MNzPj8s}{P1057}
\lookupPut{R_1i20KKdEZG7KLMh}{P1058}
\lookupPut{R_3HoALxxI9ppwRTY}{P1059}
\lookupPut{R_3POfjbvEkP9SAIE}{P1060}
\lookupPut{R_2QRGqWTlj85FnVZ}{P1061}
\lookupPut{R_1ISCza5V0SUHhz0}{P1062}
\lookupPut{R_3Mb8woQNQQFnDF2}{P1063}
\lookupPut{R_sMOfMHNWNyEmEVP}{P1064}
\lookupPut{R_1j6UA413Mhtskfi}{P1065}
\lookupPut{R_9AcIthtG4mDt26J}{P1066}
\lookupPut{R_85FbcOOvK1Hrpyp}{P1067}
\lookupPut{R_2s6AAoPAG1u0Dos}{P1068}
\lookupPut{R_3KNMZhUSPPeFtFQ}{P1069}
\lookupPut{R_R37fylNX4ZnzYdz}{P1070}
\lookupPut{R_w767RZQ6ysat4nn}{P1071}
\lookupPut{R_pDATl3Oa6FYxSPD}{P1072}
\lookupPut{R_1dnb63XxxO8BjxO}{P1073}
\lookupPut{R_Od6BiCbo1Op7OX7}{P1074}
\lookupPut{R_3qg1ct2CxJZYEer}{P1075}
\lookupPut{R_3KSD4QQOpJuKLFT}{P1076}
\lookupPut{R_1H2AWFyQ8rZrM0l}{P1077}
\lookupPut{R_3s4TSpZcHdTFATy}{P1078}
\lookupPut{R_3O6ctJ6mTQnngI9}{P1079}
\lookupPut{R_1QlRLgdtVrACOp6}{P1080}
\lookupPut{R_25RQ3QNofD16QAk}{P1081}
\lookupPut{R_AcFLsTD2Fe0NwVH}{P1082}
\lookupPut{R_3Jk1t04cgx7oUTu}{P1083}
\lookupPut{R_aaAjjJ9U6YL0vct}{P1084}
\lookupPut{R_Z90e2aI6pnKieDn}{P1085}
\lookupPut{R_1QxO9gb1AcnOUGd}{P1086}
\lookupPut{R_10VtCJEiSRXXTHZ}{P1087}
\lookupPut{R_2ckoU49UxJYp8db}{P1088}
\lookupPut{R_10U7CbliunQVx5e}{P1089}
\lookupPut{R_2f24DLJcqApEojq}{P1090}
\lookupPut{R_1mRedoti2uxOZGK}{P1091}
\lookupPut{R_2ztGhD6P1PqR1hs}{P1092}
\lookupPut{R_1F8tfWmdRoSDK3a}{P1093}
\lookupPut{R_3HjFjcDmvHb4EbX}{P1094}
\lookupPut{R_BPT9CLM5sUJIUHT}{P1095}
\lookupPut{R_Z1MFatR1amuYRXj}{P1096}
\lookupPut{R_eQ0bgOv0y1KCDfz}{P1097}
\lookupPut{R_0CAoCrYym8qY1Oh}{P1098}
\lookupPut{R_0lhyPUVyc2I4zmN}{P1099}
\lookupPut{R_2RV8khuYOMHKKAI}{P1100}
\lookupPut{R_3HjAuuTMZpn3aUs}{P1101}
\lookupPut{R_3Dba1Ft8BNecltg}{P1102}
\lookupPut{R_yF27pRRnYo7yZvr}{P1103}
\lookupPut{R_2akpGm0i1o1xDfV}{P1104}
\lookupPut{R_3m2eGxdUehv9F1w}{P1105}
\lookupPut{R_daSrbIbQ902Lt8B}{P1106}
\lookupPut{R_22J0iRoU6jcN5rk}{P1107}
\lookupPut{R_3g19rwnBRLFuNvP}{P1108}
\lookupPut{R_4UVKPDWnianYNTb}{P1109}
\lookupPut{R_ZxEamq6b0E7sxod}{P1110}
\lookupPut{R_9Z7pu5IM2YMF02Z}{P1111}
\lookupPut{R_1n9NSitdQcdlpwo}{P1112}
\lookupPut{R_1kFa0iWYgJnHXU5}{P1113}
\lookupPut{R_3MDqqENMnT0nlww}{P1114}
\lookupPut{R_3JsgTkafrR6RNBW}{P1115}
\lookupPut{R_3JqhRhaacnYZv6h}{P1116}
\lookupPut{R_1mqjd5iLTfct2e2}{P1117}
\lookupPut{R_3dDJJcGQIZWo4Pd}{P1118}
\lookupPut{R_1NF8WUXoZJ5JZMU}{P1119}
\lookupPut{R_2zo9xcxBg3NXdk0}{P1120}
\lookupPut{R_BrIlP2cBr8ngxA5}{P1121}
\lookupPut{R_yZQgmsRfiXjv9Tz}{P1122}
\lookupPut{R_2uPAGayQlnljsz8}{P1123}
\lookupPut{R_2PyVe6erzujwqhK}{P1124}
\lookupPut{R_25Sdt6WlC2xKuwn}{P1125}
\lookupPut{R_Av2V4qgyOBi82I1}{P1126}
\lookupPut{R_dg0qwkWIWkdqEEx}{P1127}
\lookupPut{R_1nW5blHGmrQaG0V}{P1128}
\lookupPut{R_3HSIFsxtpZutfWB}{P1129}
\lookupPut{R_1IbFKrn4rDifPDa}{P1130}
\lookupPut{R_x6rcs5dZfu0Xd6x}{P1131}
\lookupPut{R_3Mbtz7U6UNq2ae5}{P1132}
\lookupPut{R_1eXlrTVSBmuQ7Lu}{P1133}
\lookupPut{R_1FwvDXvIPeJ9rng}{P1134}
\lookupPut{R_2S5ZF8UDlrSU8hC}{P1135}
\lookupPut{R_25zJ5TgUKqcODi5}{P1136}
\lookupPut{R_3G6Q1zcjLRosqdP}{P1137}
\lookupPut{R_2xS53tZ0OgurcqL}{P1138}
\lookupPut{R_6ogDR4YGwQ1Dxzr}{P1139}
\lookupPut{R_3EQe6C8WJjKPFvv}{P1140}
\lookupPut{R_3F4G1gQRnV4Y4bi}{P1141}
\lookupPut{R_xyjGVJQnyWY3vr3}{P1142}
\lookupPut{R_29dm1ksA2UL5b5l}{P1143}
\lookupPut{R_216OZ4ukfvXLxop}{P1144}
\lookupPut{R_2VPbxV0kQKTYRBS}{P1145}
\lookupPut{R_eJ83LD2l42o4iY1}{P1146}
\lookupPut{R_27scbayPakhTjeD}{P1147}
\lookupPut{R_PXnHpM4Y910fVg5}{P1148}
\lookupPut{R_XmONWtdQM5MdXzj}{P1149}
\lookupPut{R_2uqxKlvJLuxdQWe}{P1150}
\lookupPut{R_3oYZg03kinfXGbV}{P1151}
\lookupPut{R_38WbMe4aPfUNU2Z}{P1152}
\lookupPut{R_123ATmAv8zG1zaM}{P1153}
\lookupPut{R_1k0Oxced3Nuobgo}{P1154}
\lookupPut{R_3kzDZnAK1NA1oSk}{P1155}
\lookupPut{R_3gO4i4OvQ8nAoIp}{P1156}
\lookupPut{R_2Slsk34Di5tizNg}{P1157}
\lookupPut{R_3qIlD9GA1gkkTpv}{P1158}
\lookupPut{R_1H1VHPvqx9mFIqo}{P1159}
\lookupPut{R_yE1PugW7KWHidGN}{P1160}
\lookupPut{R_31H6S5VxBfooa14}{P1161}
\lookupPut{R_O7FevUmoaWD1PNv}{P1162}
\lookupPut{R_1gvVWO1a5zbHjKE}{P1163}
\lookupPut{R_3PFXRGAKnRO5X5i}{P1164}
\lookupPut{R_2uxo27Zu61Lmwnd}{P1165}
\lookupPut{R_1K259nev257f9PQ}{P1166}
\lookupPut{R_3oN6yjVE4NJ4MG1}{P1167}
\lookupPut{R_2xIOuZAdDWuGodJ}{P1168}
\lookupPut{R_2wM2trNgmLd0u7B}{P1169}
\lookupPut{R_zUNu6MmpxOVa9iN}{P1170}
\lookupPut{R_71UXNx084wrtqtX}{P1171}
\lookupPut{R_A7nDCT5URu76tpL}{P1172}
\lookupPut{R_1irJ2NoRN14onFb}{P1173}
\lookupPut{R_xsYCYMjGFuT2g9z}{P1174}
\lookupPut{R_00uLDyyfjsXnztD}{P1175}
\lookupPut{R_W1Wi9I6tkE6K9R7}{P1176}
\lookupPut{R_cLPUGqbsDdCvNlf}{P1177}
\lookupPut{R_2P1F8fAnrfPr4va}{P1178}
\lookupPut{R_pc56WeJlyQCRBjb}{P1179}
\lookupPut{R_1SRUcxQ4WoNwn6h}{P1180}
\lookupPut{R_pmauUrmnJZjJUQh}{P1181}
\lookupPut{R_0vqU2zhDJ3xuxuF}{P1182}
\lookupPut{R_2ydNFrgPm6ErV4B}{P1183}
\lookupPut{R_vP3t4EORrS7pwnT}{P1184}
\lookupPut{R_3Li0IXHM6bq05hw}{P1185}
\lookupPut{R_1jx0ynuEnwSBQ6N}{P1186}
\lookupPut{R_p4UBAO9QATka3dL}{P1187}
\lookupPut{R_1d9V7P68DE1EW6S}{P1188}
\lookupPut{R_3QXnbOChER8ptUI}{P1189}
\lookupPut{R_VX5SfXugYSk8EaB}{P1190}
\lookupPut{R_yD3CFKO2Ig21MHL}{P1191}
\lookupPut{R_2ZTwjFqZ1phuRiq}{P1192}
\lookupPut{R_3sztjdKmtQuvYjX}{P1193}
\lookupPut{R_vDI2a6huvHjrgHf}{P1194}
\lookupPut{R_Q6Vhg6dIKaGZ0el}{P1195}
\lookupPut{R_1CjUKHHPaZUS8SW}{P1196}
\lookupPut{R_3QMqxgj94rvkx7O}{P1197}
\lookupPut{R_zSHmXjVCcghSU4V}{P1198}
\lookupPut{R_Zy0TSyIG8K2irWF}{P1199}
\lookupPut{R_308hjlT84H4ZYoN}{P1200}
\lookupPut{R_1Eh3AhviiirxIIc}{P1201}
\lookupPut{R_24rer2jiUHhbo7I}{P1202}
\lookupPut{R_1iaOAgsLf8pTvX2}{P1203}
\lookupPut{R_3O8ksVDFsvqsm8w}{P1204}
\lookupPut{R_SE877JOEPuo8EmZ}{P1205}
\lookupPut{R_dg97eFqGI4QenQJ}{P1206}
\lookupPut{R_1OZDhDZ6AozybVF}{P1207}
\lookupPut{R_3HFDYqfP0n7pZSB}{P1208}
\lookupPut{R_1EchiP3COAVjAWQ}{P1209}
\lookupPut{R_DMMNmzfioVDuJqh}{P1210}
\lookupPut{R_sAlzID58RV5vOxP}{P1211}
\lookupPut{R_26alTYnj7XMD1dU}{P1212}
\lookupPut{R_3g7KK0mCSrsRnCg}{P1213}
\lookupPut{R_2SxxP74N4g2AgMt}{P1214}
\lookupPut{R_1Q9KkDAJLmuGl0O}{P1215}
\lookupPut{R_Rsqi57297Zhs3n3}{P1216}
\lookupPut{R_3ew42vwSIh8zkHo}{P1217}
\lookupPut{R_1IhgJenn0zsd2il}{P1218}
\lookupPut{R_2YD2vzL2ohBOZ4P}{P1219}
\lookupPut{R_3CPi3XvrpfTqcYA}{P1220}
\lookupPut{R_1Ol4qvvOLzyslMH}{P1221}
\lookupPut{R_3EYC7CTfe2Fz2BK}{P1222}
\lookupPut{R_ZvIPh3BviqkWdJD}{P1223}
\lookupPut{R_1rwgesezsRPpCOL}{P1224}
\lookupPut{R_2VBJ7vt7ZngF7xK}{P1225}
\lookupPut{R_T0cKOyvmcrzdrrj}{P1226}
\lookupPut{R_vDNwBlb0PaIdRbr}{P1227}
\lookupPut{R_2Cm0ZJ6pobg03lX}{P1228}
\lookupPut{R_5haHLV3DPGzKqpX}{P1229}
\lookupPut{R_3e9r8HwZ53nS5z1}{P1230}
\lookupPut{R_3lYsnuKMwIePh5a}{P1231}
\lookupPut{R_vqmRBkp8RJYSIcF}{P1232}
\lookupPut{R_1i2U3Oa5z0qd3D3}{P1233}
\lookupPut{R_3EYzeYnE7FfBTZg}{P1234}
\lookupPut{R_1mxefQYMgrb1Gjk}{P1235}
\lookupPut{R_3rYv78QDkbNsAiw}{P1236}
\lookupPut{R_Yb5tCHeyTfTEMYp}{P1237}
\lookupPut{R_1P8jQb3srOJyMsN}{P1238}
\lookupPut{R_22P17f4CFJrl357}{P1239}
\lookupPut{R_2VeIm4ho5bvrsis}{P1240}
\lookupPut{R_3PjXIewmbtzOxAu}{P1241}
\lookupPut{R_3s0aBWbysqTPwEg}{P1242}
\lookupPut{R_1dHxoAZI3wWGowQ}{P1243}
\lookupPut{R_1K7PPERvh7lb2Xh}{P1244}
\lookupPut{R_3OqxUEbpl72qeVK}{P1245}
\lookupPut{R_2yab0yIXYvctOMW}{P1246}
\lookupPut{R_3qD5koRJkcNDyH6}{P1247}
\lookupPut{R_10Bgt4uqLZTpKS2}{P1248}
\lookupPut{R_1oaO7FOSOtbkg0A}{P1249}
\lookupPut{R_2rYrCJb1dO9GDCv}{P1250}
\lookupPut{R_20ZIQoVrjk54eFt}{P1251}
\lookupPut{R_9yLXk2wX0MOThGp}{P1252}
\lookupPut{R_1Eg2jXOcPfEhXLl}{P1253}
\lookupPut{R_PTe8vfVjbZ7shB7}{P1254}
\lookupPut{R_3ktZ9MPIJvkJnju}{P1255}
\lookupPut{R_PG1pPcLqdH0uwZX}{P1256}
\lookupPut{R_3kkbxbuLur0evUN}{P1257}
\lookupPut{R_BQEMf9RBECk4yJz}{P1258}
\lookupPut{R_sLKskQPFiHwpg9H}{P1259}
\lookupPut{R_3irieQeNpjMJsT5}{P1260}
\lookupPut{R_AguPFuHfef4kwdb}{P1261}
\lookupPut{R_11jtLfSDYeRftaC}{P1262}
\lookupPut{R_1o23gvJUzZhuRnW}{P1263}
\lookupPut{R_PwioOeZ7vwHWYKZ}{P1264}
\lookupPut{R_3RscSpCHH7ci6Cl}{P1265}
\lookupPut{R_pmHG0JZ9Su8UE3n}{P1266}
\lookupPut{R_1LnzXFUNwgWkXWu}{P1267}
\lookupPut{R_ezFRpAsq9GLJzGN}{P1268}
\lookupPut{R_1f6NoOYKjXY9Nm1}{P1269}
\lookupPut{R_2YVBfNzPWagHmgv}{P1270}
\lookupPut{R_2y7pW492voDuy9k}{P1271}
\lookupPut{R_cuLmdqDEaNexMD7}{P1272}
\lookupPut{R_BKPkODZ1FAQ6rE5}{P1273}
\lookupPut{R_1rCbqFwFtUiSYNA}{P1274}
\lookupPut{R_YW8K14lf2MrgDDP}{P1275}
\lookupPut{R_1P5UOJ6id6W61sc}{P1276}
\lookupPut{R_232kf2XMtgFVNVS}{P1277}
\lookupPut{R_2c2B8gdDkKilYPc}{P1278}
\lookupPut{R_2VaaXdILYcIUEAx}{P1279}
\lookupPut{R_3kgEaEWOwjNxZg5}{P1280}
\lookupPut{R_2s0awqEYTZM7wQL}{P1281}
\lookupPut{R_25ZXcPJpMZvJNhL}{P1282}
\lookupPut{R_28T0txiwRQwsKpC}{P1283}
\lookupPut{R_2Cp5u5KjPJTWkMZ}{P1284}
\lookupPut{R_28zwrmS68vVCYZX}{P1285}
\lookupPut{R_2bPf7ozaUvsWv41}{P1286}
\lookupPut{R_1QZWNDMeBPuGL9n}{P1287}
\lookupPut{R_1hy8XuJRwe5tPXy}{P1288}
\lookupPut{R_sRu5g2TlKbcQLVD}{P1289}
\lookupPut{R_p6hiac6AwwmNkGt}{P1290}
\lookupPut{R_WfD6zaKupTmEq2d}{P1291}
\lookupPut{R_2q93BeGDTftj4So}{P1292}
\lookupPut{R_2dBenQkCdhtFFV4}{P1293}
\lookupPut{R_1ihbQfyJNTZIU5r}{P1294}
\lookupPut{R_1rNTkHVKd2Qoykd}{P1295}
\lookupPut{R_2OTLXjuksCOQOUr}{P1296}
\lookupPut{R_wO8hPFy88l0S7jb}{P1297}
\lookupPut{R_AgOmVN8nRkHR3cR}{P1298}
\lookupPut{R_p3tec5dAHQcXWmd}{P1299}
\lookupPut{R_2f2TcE01QVuNBfU}{P1300}
\lookupPut{R_2qCB4D8GEiWRRSr}{P1301}
\lookupPut{R_1OO9ne1JO7bzDkA}{P1302}
\lookupPut{R_2Ej9QwC4drqTtM8}{P1303}
\lookupPut{R_XvT3nypUD7SpBqp}{P1304}
\lookupPut{R_2CwCpPrITUNA2wX}{P1305}
\lookupPut{R_z7og8ya636z4CnT}{P1306}
\lookupPut{R_3HuxNkhagsvwqco}{P1307}
\lookupPut{R_2VeDNTgNcJEDrCs}{P1308}
\lookupPut{R_p48npNWz6Q3MvMB}{P1309}
\lookupPut{R_3ijXkf2ZsNDsSd7}{P1310}
\lookupPut{R_tWIux8zQH0y0Ns5}{P1311}
\lookupPut{R_SKb4WW7ELTJlmp3}{P1312}
\lookupPut{R_2V4w3c89vCZG4uB}{P1313}
\lookupPut{R_31t4wZcQkDPP4K9}{P1314}
\lookupPut{R_21AqtrOFvXBtOyR}{P1315}
\lookupPut{R_PZmoY1jxkoXJB8B}{P1316}
\lookupPut{R_2f1jpqiSJ1ZKUIL}{P1317}
\lookupPut{R_1rP3pIIs06EtOMF}{P1318}
\lookupPut{R_1nUf920iu7yE44X}{P1319}
\lookupPut{R_doqPhmABnWFcuAh}{P1320}
\lookupPut{R_qz5Ylui0T2dMkX7}{P1321}
\lookupPut{R_24GXPznQcG0ycvM}{P1322}
\lookupPut{R_24cJCA3wKQkvSJo}{P1323}
\lookupPut{R_3s0PYik3LkWGEb6}{P1324}
\lookupPut{R_3p8QAlp1lhpPXE3}{P1325}
\lookupPut{R_Ox9SGpyL2UWSf9n}{P1326}
\lookupPut{R_1hGvqlVhNsPLnob}{P1327}
\lookupPut{R_2bZUiWpv4Sq2R8r}{P1328}
\lookupPut{R_6xISrKB429R9SG5}{P1329}
\lookupPut{R_2WGR9ZsdxumegbY}{P1330}
\lookupPut{R_2rx2GerDxaGhG33}{P1331}
\lookupPut{R_sZOqmUSyXNklMuR}{P1332}
\lookupPut{R_SPGXlXcaTkMMcet}{P1333}
\lookupPut{R_sBwzwhxT1S13bX3}{P1334}
\lookupPut{R_3DhFxWmt8nP1MNb}{P1335}
\lookupPut{R_RLCCM6AquxKYzPH}{P1336}
\lookupPut{R_2R8QL40WAHVBGHS}{P1337}
\lookupPut{R_3FVuK8u8mltYdll}{P1338}
\lookupPut{R_TvdFf0hXPAUQhiN}{P1339}
\lookupPut{R_u1vCd5L2MVzvvAB}{P1340}
\lookupPut{R_1jHQkkShMUoeADg}{P1341}
\lookupPut{R_1Nb6JhWgFuV6cSI}{P1342}
\lookupPut{R_2Cm0KQeRGuoW8Xv}{P1343}
\lookupPut{R_2VpwyhDCqMqXhTD}{P1344}
\lookupPut{R_2t5RnMyc7nNuowv}{P1345}
\lookupPut{R_12DV3RkiL3yOrHm}{P1346}
\lookupPut{R_29fSCXZEyfvDJ56}{P1347}
\lookupPut{R_bOh1lOyXVdI0TvP}{P1348}
\lookupPut{R_1mrVP9EO7fHPJ52}{P1349}
\lookupPut{R_3qQGP8Mmo4awoNY}{P1350}
\lookupPut{R_2DTuj0F3FYdRaih}{P1351}
\lookupPut{R_3MrPRL5VmZjiQPu}{P1352}
\lookupPut{R_1o88RXWarZoFqtT}{P1353}
\lookupPut{R_x4w7HqaG9qddI5z}{P1354}
\lookupPut{R_1E6y5r069quaTeJ}{P1355}
\lookupPut{R_2CDniMkpmXX1A5F}{P1356}
\lookupPut{R_3qIhyK2HV1DPdVV}{P1357}
\lookupPut{R_3rIApURAVnu1z8E}{P1358}
\lookupPut{R_d41aLgqUKselzb3}{P1359}
\lookupPut{R_RXMtU3UsO0w4xLH}{P1360}
\lookupPut{R_9uFmEFZ1foENHPP}{P1361}
\lookupPut{R_2uX320lu95Og3oQ}{P1362}
\lookupPut{R_ernTUoYLS39CnBv}{P1363}
\lookupPut{R_3lQR3xKIzgcyPE9}{P1364}
\lookupPut{R_OPPQTwVr2oAfj1v}{P1365}
\lookupPut{R_27OuMxGU3IOrqDU}{P1366}
\lookupPut{R_269nIeTeWhTVxSj}{P1367}
\lookupPut{R_3NE3qPGMXc6I9MX}{P1368}
\lookupPut{R_3NCiPFla3egA1KT}{P1369}

\newif\ifIncludeAuthorInfo
\IncludeAuthorInfotrue

\usepackage{mybchart}
\usepackage{enumitem}
\usepackage{balance}
\usepackage{hyperref}

\begin{document}
\title{``How Was Your Weekend?''  Software Development Teams Working From Home During COVID-19}
\ifIncludeAuthorInfo

\author{\IEEEauthorblockN{Courtney Miller\IEEEauthorrefmark{1},
Paige Rodeghero\IEEEauthorrefmark{3},
Margaret-Anne Storey\IEEEauthorrefmark{4}, 
Denae Ford\IEEEauthorrefmark{5} and
Thomas Zimmermann\IEEEauthorrefmark{6}}
\medskip\IEEEauthorblockA{\IEEEauthorrefmark{1} New College of Florida, FL, USA. Email: courtney.miller17@ncf.edu}
\IEEEauthorblockA{\IEEEauthorrefmark{3} Clemson University, SC, USA. 
Email: prodegh@clemson.edu}
\IEEEauthorblockA{\IEEEauthorrefmark{4} University of Victoria, BC, Canada. Email: mstorey@uvic.ca}
\IEEEauthorblockA{\IEEEauthorrefmark{5} Microsoft Research, WA, USA. Email: denae@microsoft.com}
\IEEEauthorblockA{\IEEEauthorrefmark{6} Microsoft Research, WA, USA. Email: tzimmer@microsoft.com}}
\else
    \author{\IEEEauthorblockN{Anonymous Author(s)}}
\fi

\definecolor{darkGreenAccentColor}{HTML}{808064}
\definecolor{mediumGreenAccentColor}{HTML}{A6A68A}
\definecolor{lightGreenAccentColor}{HTML}{B7B7A4}
\definecolor{neutralAccentColor}{HTML}{EDC9B9}
\definecolor{lightBrownAccentColor}{HTML}{9E6952}
\definecolor{mediumBrownAccentColor}{HTML}{784F31}
\definecolor{darkBrownAccentColor}{HTML}{59371F}

\newcommand\crule[3][black]{\textcolor{#1}{\rule{#2}{#3}}}

\maketitle

\thispagestyle{plain}
\pagestyle{plain}

\begin{abstract}

The mass shift to working at home during the COVID-19 pandemic radically changed the way many software development teams collaborate and communicate.
To investigate how team culture and team productivity may also have been affected, we conducted two surveys at a large software company. The first, an exploratory survey during the early months of the pandemic with 2,265 developer responses, revealed that many developers faced challenges reaching milestones and that their team productivity had changed.  We also found through qualitative analysis that important team culture factors such as communication and social connection had been affected. For example, the simple phrase ``How was your weekend?'' had become a subtle way to show peer support.

In our second survey, we conducted a quantitative analysis of the team cultural factors that emerged from our first survey to understand the prevalence of the reported changes. 
From 608 developer responses, we found that
74\% of these respondents missed social interactions with colleagues and 51\% reported a decrease in their communication ease with colleagues.
We used data from the second survey to build a regression model to identify important team culture factors for modeling team productivity.  We found that the ability to brainstorm with colleagues, difficulty communicating with colleagues, and satisfaction with interactions from social activities are important factors that are associated with how developers report their software development team's productivity.  
Our findings inform how managers and leaders in large software companies can support sustained team productivity during times of crisis and beyond. 

\end{abstract}

\IEEEpeerreviewmaketitle

\section{Introduction}

As COVID-19 spread globally, many companies, including Google, Microsoft, Twitter, Amazon, and Facebook, instructed their software developers to go home and work remotely~\cite{duffy20}.  
Entire software development teams that used to work predominantly in-person suddenly had to pivot their work and quickly establish effective remote collaboration and communication.

\begin{figure}
  \centering
  \includegraphics[scale=0.5]{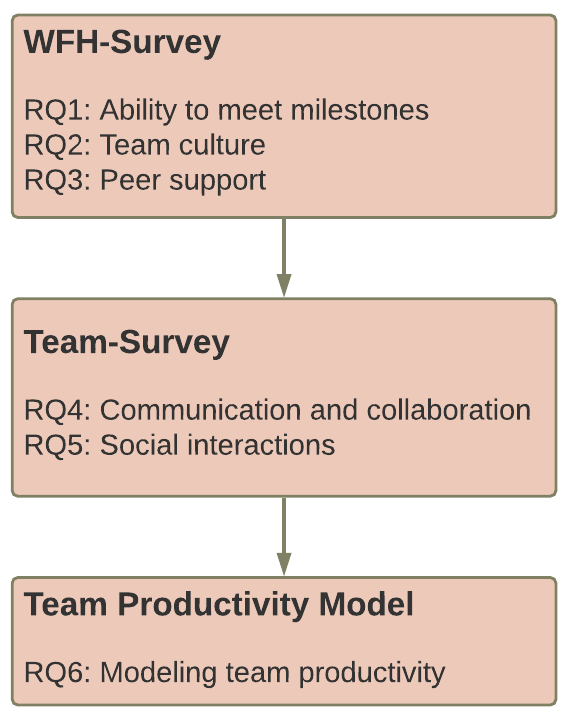}
  \caption{Methodology Flow Chart}
  \label{fig:methodology}
\end{figure}

Prior research has studied how working from home (WFH) affects productivity~\cite{baker07, neufeld05}. 
While regular WFH is not the same as WFH during a pandemic,
the COVID-19 pandemic has created a natural experiment for researchers to study WFH on a much larger scale than previously possible. %
For instance, it has helped reduce selection bias in studies of WFH since almost everyone has to work from home, and it has helped researchers identify and understand the concrete challenges faced by developers while working remotely. There are recent papers that already began investigating the impacts of this unique work setting. Bao \textit{et al.}\ performed a case study using automated trace data, along with other metrics, to determine how productivity has been affected. They found that productivity was affected in various ways depending on the productivity metrics used~\cite{bao20}.  Ralph \textit{et al.}\ performed an international large-scale questionnaire survey of developer well-being and productivity and found that productivity and well-being are closely related, and both are currently suffering~\cite{ralph20}. As insightful as these works are at providing empirical evidence of factors affecting individual developer productivity, they lack a deeper understanding of a major responsibility of industrial software developers---collaborating with a team.
In our work, we identify factors that affect software development team productivity such as team culture factors, including communication, camaraderie, and team cohesion~\cite{wagner19, storey19}. We hypothesize these factors are at particular risk of being disrupted by this unexpected shift to WFH.  Thus, we investigated the effects of WFH on teams and answered the following research questions (RQs):
\begin{description}[leftmargin=2.5em]
    \item[\textbf{RQ1}] How has the ability for teams to meet milestones changed during WFH?  %
    \item[\textbf{RQ2}] How has team culture changed during WFH? %
    \item[\textbf{RQ3}] How have teams supported their members during WFH? %
    \item[\textbf{RQ4}] How has team communication and collaboration changed during WFH? %
    \item[\textbf{RQ5}] How have social interactions within teams changed during WFH?
    \item[\textbf{RQ6}] Which factors are associated with a change in team productivity during WFH? 
\end{description}

To answer our research questions, we conducted two surveys.  The first survey was an exploratory survey to understand how team productivity, ability to meet milestones, culture, and support structures have changed. 
We refer to this survey throughout the paper as the ``WFH-Survey.''  
We analyze data collected from the exploratory survey that was aimed to better understand team culture factors and changes in communication and interaction practices. We also achieve early insights about team productivity, specifically how some believe it has been reduced and the reasons why teams may not meet milestones. 

We found a common narrative among developers about how \emph{ad-hoc} in-person communication has been replaced with online communication that, despite the often times more meaningful intent, has more friction.  
For example, as one of our respondents put it,  formerly shallow questions from team members like ``How was your weekend?'' are now a deep inquiry of well-being and genuine concern. Many developers also discussed how their teams had trouble finding satisfying replacements for the low-effort in-person social activities like lunch or coffee breaks that were used to help maintain camaraderie and social connection on their team.

Consequently, we designed an additional quantitative survey, which we refer to throughout the paper as the ``Team-Survey'' to focus on team-specific factors and their prevalence.
From our Team-Survey, we found that 66\% of respondents reported a decrease in social connection with their team members, and 78\% and 65\% cited a decrease in impromptu and scheduled social activities, respectively.  Across the board, we saw a dramatic decrease in feelings of social connection and team cohesion.  We also found that 57\% of respondents felt a decrease in their ability to brainstorm with their colleagues.  Likewise, the quality and ease of communication were commonly cited challenges.  
To understand the refined impact of these factors, we built a model of team productivity where we found that factors including the ability to brainstorm with team members, the frequency of scheduled social activities, and experiencing difficulties communicating with colleagues all have a significant relationship with change in team productivity.

As remote work will continue beyond the pandemic, it is important for us to know how to evolve our approaches to support remote work.  
Our work can help improve the experience of remote teams by identifying crucial factors that they should focus on strengthening, like the ability to effectively communicate and brainstorm, and factors they should work on maintaining like social connection and team cohesion.

\section{Methodology}\label{sec:methodology}

To answer our research questions, we distributed two surveys to full-time software engineers at a large software company in the United States. The company started working from home in March 2020 around the same time when Amazon, Facebook, Google, and Microsoft started remote work due to the COVID-19 pandemic. 

Figure~\ref{fig:methodology} shows a methodology flowchart.  
The first survey (WFH-Survey, Section~\ref{subsec:WFHSurvey}) collected qualitative data and was used to answer RQ1, RQ2, and RQ3. The second survey (Team-Survey, Section~\ref{subsec:TeamProductivitySurvey}) builds on the first survey and collected quantitative data to answer RQ4, RQ5, and RQ6. 

Both surveys were anonymous, and no personally identifiable information was collected.  We emphasized the anonymity of the survey to help organically create a more-candid space for respondents~\cite{hewson96}.  Respondents were invited to reach out to us via separate email if they had any questions or concerns.  Both surveys included a filter question, in the beginning, asking respondents whether they were currently working from home to ensure they qualified to participate. The complete surveys can be found in the replication package~\cite{courtney_miller_2021_4456041}. 

In both surveys, we asked participants about the change in perceived team productivity:
\begin{itemize}
\item Compared to working in office, how has your team productivity changed? (\QteamproductivityWFH, \Qteamproductivity) \footnote{We renumbered the questions for this paper. The prefix indicates whether the question belongs to the \textbf{WFH-Survey} or the \textbf{Team-Survey}.}
\end{itemize}
The responses to this question were on a five-point scale from \textit{significantly decreased} to \textit{significantly increased}.

\subsection{WFH-Survey}\label{subsec:WFHSurvey}
In April 2020, we ran a large-scale three-week survey with the goal of understanding how developers are being affected by WFH at the large software company. 

\smallskip \textbf{Survey Instrument.}
We asked participants a wide array of questions (42 in total) about their experiences working from home. Most questions were focused on the respondents' individual experiences. For this paper, we analyzed the subset of questions that were specifically related to teams. 
We used three open-ended questions for the qualitative analysis of how teams are affected as they work from home:
\begin{itemize}
    \item How has your team culture changed?~(\Qculture)
    \item How has your team supported you during this crisis? (\Qsupport)
    \item Compared to working in office, do you think the ability of the team to reach milestones has changed? (No/Yes).\newline If yes, please explain.~(\Qmilestones)~\footnote{The question \Qmilestones was asked only in Weeks 2 and 3.}
\end{itemize}

To identify common ways teams are being affected by WFH, we triangulated the responses from these three questions.  This helps create a more holistic perspective on the effects of WFH by considering team culture, team support, and challenges in meeting milestones.

\smallskip \textbf{Participants.}
The WFH-Survey was deployed using the Qualtrics survey tool and participants were invited via personalized emails. We sent email invitations over a period of three weeks to a random sample of 9,000 software engineers (3,000 per week). This survey received 2,265 responses (response rate of 25\%, which is comparable to other surveys in software engineering~\cite{smith13}). After completion of the WFH-Survey, participants could enter a sweepstakes to win USD \$100 Amazon.com gift certificates.  

\smallskip \textbf{Analysis.}
We used \emph{qualitative content analysis coding}~\cite{schreier12} (sometimes also known as card sorting), which provides a framework to ascertain the meaning and summarize overarching themes of responses within the context of the research question. In the analysis, we grouped similar responses to the open-ended questions into codes. The first iteration of coding was performed by one author, and any responses the first author was unsure of were decided by discussion with another author. 
All the codes were then reviewed by the other authors.

For the questions about team culture (\Qculture) and team provided support (\Qsupport), we coded a random sample of 600 responses. 
We selected 
this sample based on (1) previous experience with similar surveys and (2) a sample size determination for a confidence interval of 95\% and a margin of error of 5\%, which required 385 or more responses. 
For both questions we reached saturation, defined as coding 100 responses without new codes (for WFH-Q2 saturation was after 358 responses, for WFH-Q3 after 511 responses). %

For the question about whether the team's ability to meet milestones has changed (\Qmilestones, shown only in Weeks 2 and 3), we used all 416 explanations provided when participants responded ``Yes.'' Of the 1,477 responses, 446 responded ``Yes'' (30\%) but not all provided an explanation.

In Section~\ref{sec:results}, we discuss the most common responses (codes) in more detail and include examples from the WFH-Survey. The codes are discussed in order from most to least frequently cited in the responses. The codes were selected until there was a natural dip in frequency.  
The full list of codes can be seen in our replication package~\cite{courtney_miller_2021_4456041}.

\subsection{Team-Survey}\label{subsec:TeamProductivitySurvey}

From the analysis of the WFH-Survey, we learned that communication and collaboration with team members were among the most commonly cited challenges impacting the ability to reach milestones. We also found that social connectedness and communication were the most commonly cited team culture factors affected by WFH. 
Because the first survey was primarily open-ended, we were not able to quantify how widespread these experiences were and how they are associated with team productivity. We, therefore, designed a second survey (``Team-Survey''). To identify specific common developer experiences that relate to \emph{social connectedness} and \emph{quality of communication} to include in the Team-Survey, we referred to the responses to the WFH-Survey that cited social connectedness and communication factors.

\smallskip \textbf{Survey Instrument.}
The Team-Survey contained 19 questions in total, including questions about demographics, team productivity, team culture, communication and collaboration, social connectedness, as well as three open-ended questions at the end to provide respondents with an opportunity to share additional thoughts and feedback. The majority of the questions were Likert response questions to allow for more quantitative analysis.

We asked two key questions exploring team culture factors:
\begin{itemize}
    \item How has communication with your team changed since working from home?~(\Qcommunication)
    \item How has social interaction with your team members changed since working from home?~(\Qsocialinteraction)
\end{itemize}
The items and response formats for \Qcommunication and \Qsocialinteraction   can be found in Tables~\ref{Q12LikertTable} and~\ref{Q13LikertTable}.  The items for \Qcommunication and \Qsocialinteraction were derived from a combination of the WFH-Survey and previous literature~\cite{murphy19}.

We also asked about how respondents stayed socially connected to colleagues~(\Qsocialconnection), about work-related challenges while working from home, and how impactful these challenges are~(\Qchallenges).

To quantify and model which factors are associated with a change in team productivity, we included team culture factors as items in the Team-Survey because team culture factors have been established as important factors of productivity~\cite{wagner18}. %
Related work has shown that there are additional team-related factors that can predict productivity including confidence and supportiveness of a team~\cite{murphy19}. Therefore, we added a question to the survey with agreement about:
(1) my manager is highly capable; %
(2) my team members are highly capable; %
(3) my team members are supportive of new ideas; and %
(4) I feel positively about other people on my team. %

\smallskip \textbf{Participants.}
The Team-Survey was implemented with the Microsoft Forms Pro survey tool and participants were invited via personalized emails. 
Invitations were sent to a random sample of 3,500 software engineers in July 2020.  We received 178 out of office responses giving us a pool of 3,322 potential participants, of whom 608 responded yielding about an 18\% response rate, which is comparable to other software engineering surveys~\cite{smith13}. 
There was no overlap between the invitees of the WFH-Survey and the Team-Survey.
The gender identification breakdown of respondents is as follows, 455 identified as men, 90 as women, 2 as non-binary/gender diverse, and 17 preferred not to answer. 
While our survey collected the gender identity of respondents, we did not ask about the gender composition of teams as our scope was on team productivity. Thus, we did not analyze gender composition.

Survey respondents did not receive compensation, but for each response USD \$2  were donated to the Black Lives Matter Global Network\cite{blm} for a total of up to USD \$400.  

\smallskip \textbf{Analysis.}
To understand the overall trends of the responses and to quantify the themes that emerged in the Team-Survey we used visualizations and descriptive analyses.

To model the change in perceived team productivity with the factors in the survey (RQ6), we ran a regression analysis with three steps described as follows. We first built a model using all factors, then we used AIC backwards stepwise regression for variable selection, followed by standard model diagnostics.   

\smallskip
\emph{Step 1: Build Regression Model.}
We initially built a multiple linear regression model using all the factors included in the Team-Survey~\cite{fox15}.  
Multiple linear regression analysis is a form of regression modeling used to model a quantitative response variable using multiple explanatory variables that can be both quantitative and categorical~\cite{montgomery12}. The response variable was the perceived change in team productivity~(\Qteamproductivity), which was formatted as a 5-point Likert response item.  Since we used a 5-point scale, it is appropriate to treat it as an ordinal version of a continuous variable and to code it as a numerical variable for the model~\cite{norman10, sullivan13}.  However, the explanatory variables remained categorical because some of them are Likert response items with only three levels rather than five, so they could not be coded numerically.  All explanatory variables were kept categorical to help preserve the consistency of interpretation.  

\smallskip \emph{Step 2: Perform Variable Selection.}
Once we had the full model built with all the factors from the Team-Survey, we reduced the model using variable selection.
The goal of variable selection is to reduce the model down to only the most important variables, for this we use Akaike Information Criterion (AIC)~\cite{akaike74}.    
AIC considers two key components in its formula.  First, it considers the quality of fit for the model given by the model's sum of squared error (SSE).  Second, it considers the complexity of the model, based on the number of parameters.  Removing a variable from the model will lower the complexity of the model, but it will also possibly increase the error of the SSE, this balance is what the criterion is based on.
We used backward stepwise selection, which starts with the full model and removes variables one at a time according to the AIC~\cite{yamashita07} with the goal of minimizing the AIC.  

\smallskip \emph{Step 3: Model diagnostics.}
We first examined the Variance Inflation Factor (VIF) for each factor, which signals whether multicolinearity is an issue~\cite{dodge08}.  We removed any variables with a VIF score greater than 5, which is a standard benchmark within the statistics community~\cite{fox15}.  
We also performed the standard diagnostic plots to assure the linear regression assumptions of linearity, constant variance, and normality were upheld~\cite{fox15}.
We then used ANOVA testing to identify which variables in the model had a statistically significant relationship with perceived change in team productivity~\cite{scheffe99}.

\section{Results}
\label{sec:results}

\newcommand{\inlinequestion}[2]{\textbf{``#1''} (#2)}

\subsection{How has the ability for teams to meet milestones changed during WFH?~(RQ1)}\label{sec:RQ1Results}

We discuss the responses to the question \inlinequestion{Compared to working in office, do you think the ability of the team to reach milestones has changed?}{\Qmilestones}.
30\% of respondents said ``Yes'' the ability to reach milestones is affected.  
When people responded that their ability was affected, they were asked to elaborate. 
Although this question was phrased neutrally, the vast majority of respondents who elaborated cited challenges that \textit{hindered} their ability to reach milestones.  
Beyond general reduced productivity and current events, participants commonly cited collaboration challenges, and challenges surrounding effective communication.  These communication-related factors were frequently raised.

\newcommand{\codetitle}[2]{\smallskip\textbf{#1.}~($#2\times$)\xspace}

\codetitle{Reduced Productivity -- General}{72}
General reduced productivity was characterized by an overall decrease in general efficiency/productivity among team members, it could also manifest as the velocity of work decreasing.  
\myquote{We  have lost somewhere between 20\%-40\% effectiveness in use of time. In order to keep up, people are working longer hours. We are starting to see burnout.}{R_UEoCJWvIHNWojvz}
\myquote{We definitely are not as efficient.}{R_2QzfFkbd1gkSlMZ}

\codetitle{Reduced Productivity -- Current Events/Kids}{71}  
This code encompasses reduced productivity/efficiency due to WFH challenges specifically related to current events including COVID-19 and childcare responsibilities.
\myquote{The time pressure due to child care expectations and more-frequent scheduled meetings (vs. quick hallway chats) has definitely slowed the team down.  We have explicitly and officially postponed some work, citing Coronavirus-rooted challenges as reasons.}{R_2dWpecDRqxBOmm4}

\codetitle{Communication Challenges}{46}  
This code captured issues related to brainstorming sessions, discussions, and meetings, or issues with miscommunications. 
\myquote{It is more difficult to land nuanced discussions.  1:1 is OK, but in a multi-person discussion it is very, very hard to keep focus.}{R_2cbhXlQZMam5X3O}
\myquote{Difficulties in communication lead to additional time necessary to connect and sync. Similar communication difficulties lead to wasted work, re-work on at least a couple occasions.}{R_3oC3J0EL6WEUV0t}

\codetitle{Collaboration Challenges}{43}  
This code included issues with collaborative brainstorming, effective discussions, and other coordination tasks.
\myquote{I think we spend a lot more time trying to coordinate with each other and driving for clarity and shared understanding is harder. This means more time solidifying what we need to be doing and less time doing/building it.}{R_1rojnV1ntpeqYnm}
\myquote{Brainstorming activities that need whiteboard and involve more passionate conversation progresses slower than before and can be frustrating.}{R_1lbimxXMqOnnKkP}

\subsection{How has team culture changed during WFH?~(RQ2)}

Here, we discuss the responses to \inlinequestion{How has your team culture changed?}{\Qculture}. %
The most common answer to this question was actually \textit{No Change}.
After \textit{No Change}, the most commonly cited changes were related to social connection and communication.  There appears to be a general push for increased conscious interactions, especially focused on social connectedness and communication. 
 
\codetitle{No Change}{109}  
 No change constituted not noticing or experiencing any significant change in team culture.  
 \myquote{I do not think there was any drastic changes}{R_1CKm4V4sqiNSgAj}
 \myquote{I feel no difference}{R_2dfFNs2SUrossAW}
 
\codetitle{Social Interaction Emphasized}{49}
 Respondents reported teams were investing a conscious effort into promoting, reinforcing, and improving social interactions and connections within a team, or a focus on implementing social events.  
 \myquote{We have had to setup dedicated socializing meetings/channels/chats to fill the void of hallway conversations.}{R_p4tFScS1BtYeYxj}

Teams were putting an active effort into ensuring they were able to engage socially during WFH. Participants often reported that this active effort was necessary due to the loss of \textit{ad-hoc} social events that they engaged in when in the office.  
 \myquote{More explicit efforts to facilitate social events and foster team relationships.}{R_3oAEaPqTkfAql1W}
 
\codetitle{More Meetings}{39}  
 Respondents described an increased number of either formal or informal meetings, including sync up meetings, collaboration meetings, and social meetings.  
 \myquote{We are trying to adapt to daily sync ups and face time between all team members. which helps with communication but adds into the million meetings}{R_ONkPkt4jhQSlbhL}
\myquote{We have more meetings in general, but we also have had some team meetings to discuss how people are feeling/coping with the situation}{R_1Egq0RpOZaXHSev}

\codetitle{Missing General Interaction}{39}   Respondents reported a general lack of communication, interaction, and/or connection with team members.  
\myquote{I don't really interact/talk to them very much anymore}{R_zTCKFxGSn0nrM1r}
\myquote{fewer interactions overall. More interactions with fewer people.}{R_3fleMBQQo46z4lf}

\codetitle{Missing Social Interaction}{36} Respondents reported experiencing a lack of casual interpersonal communication and socialization.  Pre-WFH, these informal interactions often used to be manifested as lunch, quick chats, and other team bonding experiences.  Respondents commonly cited not having found equivalent remote substitutions for these interactions.  
\myquote{Fewer informal sessions of ``just chatting'' before/after meetings makes things feel lonely}{R_r7QWDc95UDMHlfP}
\myquote{We used to have lunches together, and had good comeraderie. No more lunches together, less comeraderie, and less incidental `water cooler' information exchange.}{R_3Rsw5b4gmtj0JD1}

\codetitle{Increased Empathy}{36} Respondents reported that their team has an increased understanding, consideration, or empathy for the situations of others on the team.
\myquote{We are definitely acting bit more caring and respectful of each others struggles during this time. We are going out of our way to make everyone comfortable. At the same time, fatigue is seeping in and folks are gradually giving up hope on this whole situation.}{R_22StgdXay4Bj8Xh}

\subsection{How have teams supported members during WFH?~(RQ3)}

We next discuss the responses to \inlinequestion{How has your team supported you during this crisis?}{WFH-Q3}
After \textit{Peer Support}, the most commonly cited methods of support focused on  communication, social connection, and empathy/understanding.  
Many participants described a focus on supporting social connection and an empathetic culture.  

\codetitle{Peer Support}{56}  The most commonly cited method of support was peer support.  This constituted respondents reporting general shows of support from team members.  
\myquote{Folks are always there to help each other out.}{R_bvzddb7l9RFyaDT}
\myquote{Everyone is supportive of each other.  It is easy to talk with someone if needed.  Feel supported.}{R_2eQYoRRjHxYTTgG}

\codetitle{Social Engagements}{42}  Social engagements were defined as meetings organized with the purpose of fostering social interaction, including lunch calls and virtual happy hour. 
\myquote{We try to organize online events such as Friday night chats and virtual lunches.}{R_1mDHqlhVOHEYxgf}
\myquote{We have weekly social hours, which helps with some of the social isolation I've been feeling.}{R_25RUhPXZbCQzRQi}

\codetitle{Communication Tools}{41}  Respondents reported the use of communication tools such as Slack and Microsoft Teams to connect with other employees across the company. Respondents reported the use of such tools to help facilitate meetings, discussions, and other forms of communication.
\myquote{Lots of online meetings.}{R_2zGbBEbu6wasQqd}
\myquote{A lot of interactions, face time via communication tools to help us feel less socially isolated.}{R_riQNkUEUCqTV3TH}

\codetitle{Understanding Personal WFH Situations}{40}  
The respondents described team members being understanding and supportive of child care responsibilities, limited schedule availability, and various WFH challenges such as needing personal breaks between meetings.  
\myquote{We've more intentionally focused on learning about people's personal lives \& situations. \textbf{``How was your weekend''} has transitioned from a perfunctory pleasantry into real personal concern/care for others.}{R_pa4Azj57DrrARtD}
\myquote{Agreement that meetings should start at 5- or 10- after the hour to allow for personal breaks.}{R_PASuEaZRDx4wlYR}
\myquote{Everyone is in the same position and therefore very understanding of each person's individual circumstances, so there is empathy and flexibility.}{R_3EB7FF1nJXrVKfG}
\myquote{Everyone has been very supportive when I needed flexibility to help my kids get started with their home schooling}{R_WiKER04AR4imEeZ}

\codetitle{Personal Check Ins}{39}   
Personal check-ins were defined as check-in meetings with a small number of team members, often with the objective of exploring personal well-being or work progress.   
\myquote{More meetings with our direct manager's team just to talk and check in.}{R_tDjz9t0gznhwbyp}
\myquote{Open conversation and checking in during 1:1's.}{R_3dRVJJUbvbeshu9}

 \begin{table*}[ht]
\renewcommand{\arraystretch}{1.2}
\caption{How has communication with your team members changed since working from home?~(\Qcommunication, 608 responses)}
\vspace{-0.5\baselineskip}
\label{Q12LikertTable}
\centering
\setlength\tabcolsep{.2 cm}
\begin{tabular}{r l c}
    \toprule
    & Communication Method & \\
    \midrule
    CM1 & Frequency of scheduled meetings &  \includegraphics[width = 6.5cm]{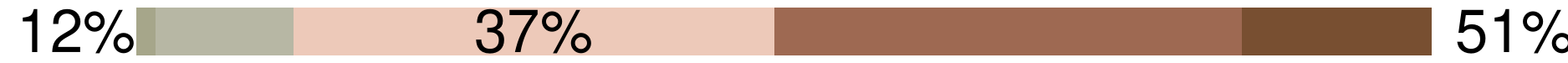}\\
    CM2 & Frequency of team member notifications & \includegraphics[width = 6.5cm]{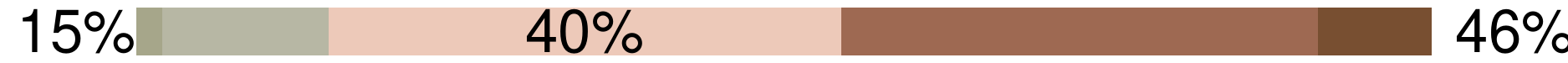} \\
    CM3 & Frequency of ad-hoc meetings & \includegraphics[width = 6.5cm]{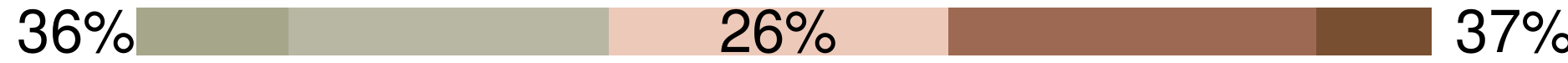}\\
    CM4 & Impactful contributions I make to team & \includegraphics[width = 6.5cm]{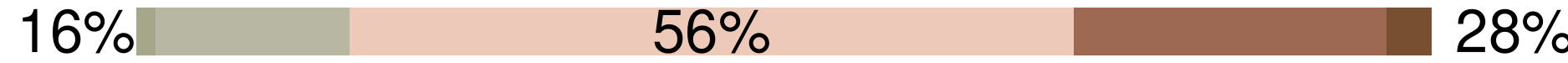}\\
    CM5 & Quality of scheduled meetings & \includegraphics[width = 6.5cm]{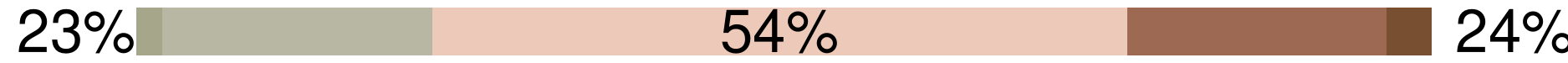}\\
    CM6 & Positive interactions with my team & \includegraphics[width = 6.5cm]{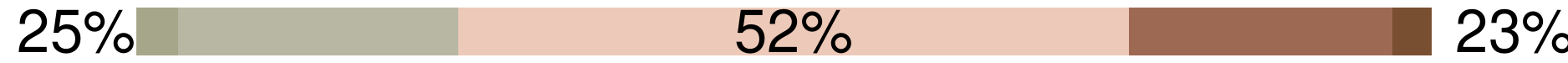} \\
    CM7 & Ability to collaborate with colleagues 1:1 & \includegraphics[width = 6.5cm]{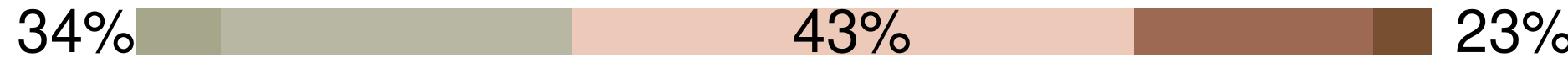} \\
    CM8 & Positive interactions with my manager & \includegraphics[width = 6.5cm]{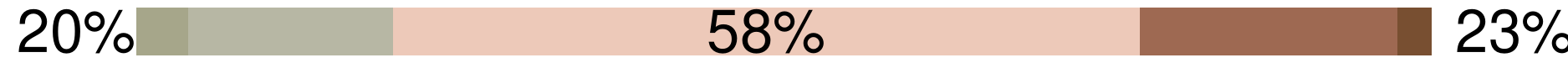}\\
    CM9 & Satisfaction with communication with team & \includegraphics[width = 6.5cm]{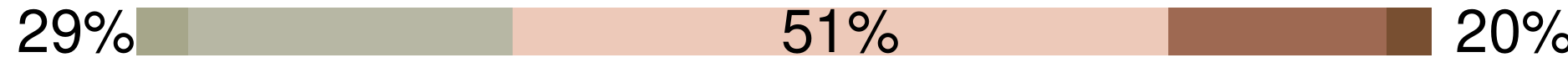}\\
    CM10 & Ability to ask questions in group meetings & \includegraphics[width = 6.5cm]{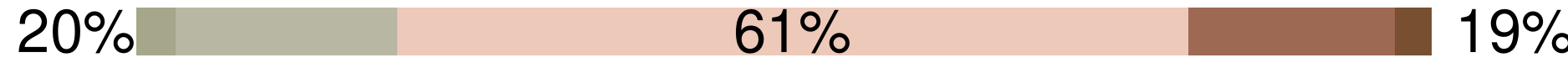} \\
    CM11 & Effectiveness of communication with colleagues &\includegraphics[width = 6.5cm]{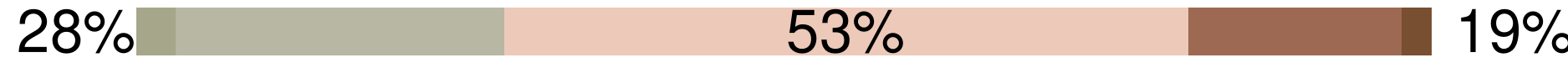}\\
    CM12 & Ability to share thoughts in group meetings & \includegraphics[width = 6.5cm]{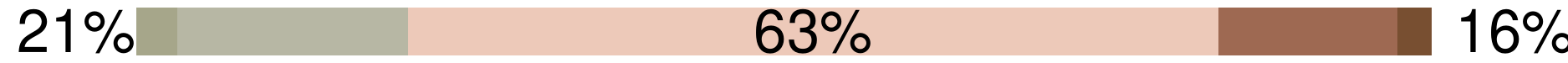} \\
    CM13 & Communication ease with colleagues &  \includegraphics[width = 6.5cm]{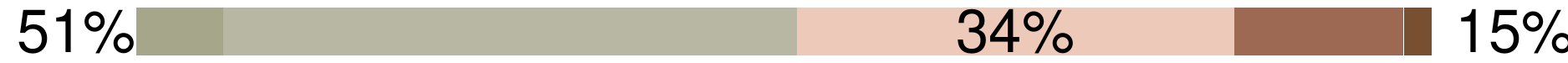} \\
    CM14 & Communication breakdowns w/in my team & \includegraphics[width = 6.5cm]{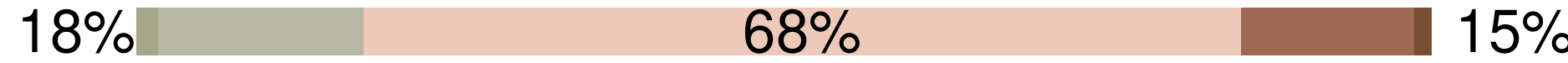}\\
    CM15 & Awareness of colleague's work & \includegraphics[width=6.5cm]{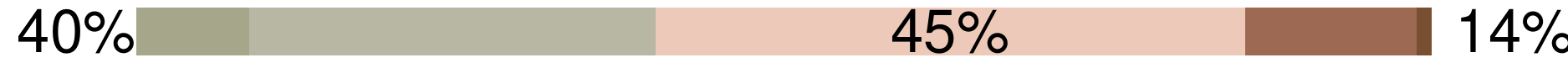}\\
    CM16 & Speed of decision making on my team & \includegraphics[width = 6.5cm]{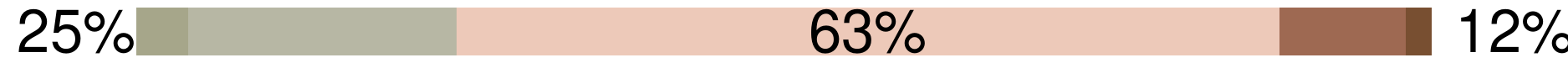}\\
    CM17 & Ability to brainstorm with colleagues & \includegraphics[width = 6.5cm]{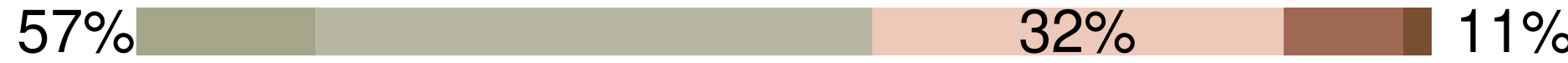}\\
    CM18 & Knowledge flow within my team &  \includegraphics[width = 6.5cm]{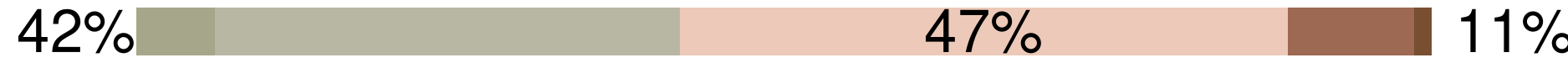}\\
    CM19 & Ability to make decisions as a team & \includegraphics[width = 6.5cm]{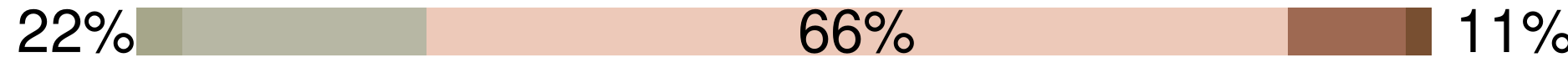}\\
    CM20 & Feeling socially connected to my team & \includegraphics[width = 6.5cm]{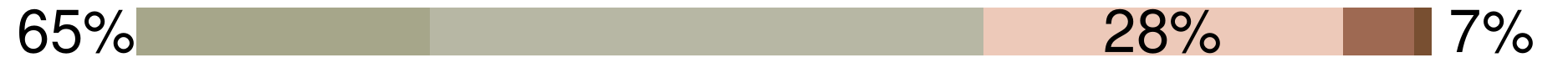} \\
    \midrule
    \multicolumn{2}{l}{\scriptsize \crule[darkGreenAccentColor]{.2cm}{.2cm} Significant Decrease \crule[lightGreenAccentColor]{.2cm}{.2cm} Decrease \crule[neutralAccentColor]{.2cm}{.2cm} About the Same \crule[lightBrownAccentColor]{.2cm}{.2cm} Increase \crule[mediumBrownAccentColor]{.2cm}{.2cm} Significant Increase}
    & {\scriptsize Percentages may not total 100\% due to rounding.} \\

    \bottomrule
\end{tabular}
\end{table*}

\subsection{How has team communication and collaboration changed during WFH?~(RQ4)}\label{sec:RQ4Results}
Issues with communication and collaboration among teams were commonly cited by developers in the WFH-Survey.  These are important challenges to recognize because previous research has highlighted the importance of efficient communication flow for developer productivity~\cite{sadowski19}. 
The findings from the WFH-Survey show that communication is (1) often affected by changes in team culture, (2) used by teams to support their members, and (3) a common challenge teams face in reaching milestones. Teams are often being conscious of communication and working hard to make improvements, but it is still one of the biggest roadblocks to reaching milestones. 
This suggests that while communication is a cornerstone challenge for WFH teams, this is not a lack of communication issue, but rather a \textit{quality of communication} issue.

Numerous WFH-Survey respondents mentioned issues communicating in ways that used to be trivial, like brainstorming with team members or the ability to collaborate with team members one-on-one.  One survey participant described the \myquotex{explosion of meeting requests and heightened expectations on availability and response time from emails and online messages. It is much harder to interject in online meetings due to lack of visual cues, so meetings and decisions end up monopolized by a small set of people.}{R_1mDHqlhVOHEYxgf}  Even though most teams report they are still meeting regularly, there are questions regarding the quality and ease of communication.  

\smallskip
In the Team-Survey, we therefore sought to further explore the prevalence of these challenges. When asked about WFH issues faced, 58\% of the Team-Survey respondents reported being less aware of colleagues' work, and 47\% said they had difficulty communicating with colleagues.  

Table~\ref{Q12LikertTable} shows a visualization of the responses to the question \inlinequestion{How has communication with your team members changed since working from home?}{\Qcommunication}.
The communication aspects that have decreased the most are \textit{feeling socially connected to team}~(65\%, CM20),\footnote{The labels CM\# and SC\# refer to the rows in Tables \ref{Q12LikertTable} and \ref{Q13LikertTable} respectively. CM is short for Communication Method; SC is short for Social Connection.}    \textit{ability to brainstorm with colleagues} (57\%, CM17), \textit{communication ease with colleagues} (51\%, CM13), and \textit{knowledge flow within my team} (42\%, CM18). The communication aspects that have increased the most are \textit{frequency of scheduled meetings}~(51\%, CM1), \textit{frequency of team member notifications} (46\%, CM2), and \textit{frequency of ad-hoc meetings}.  

The fact that 65\% of respondents reported that they had experienced a decrease in \textit{feeling socially connected to their team}~(CM20) is alarming. Previous research has shown that team camaraderie is an important factor in developer productivity~\cite{wagner18}. %
The majority of respondents reported a decrease in \textit{communication ease with colleagues} (51\%, CM13) and the same or a higher \textit{frequency of scheduled meetings} (88\%, CM1). This supports the hypothesis that although communication is occurring on teams, there are still issues regarding the quality of the communication.    
 
\begin{table*}[ht]
\setlength\tabcolsep{0.2cm}
\renewcommand{\arraystretch}{1.2}
\caption{How has social interaction with your team members changed since working from home?~(\Qsocialinteraction, 608 responses)}
\vspace{-\baselineskip}
\label{Q13LikertTable}
\begin{center}
    \begin{tabular}{r l c}
    \toprule
    & Social Connection \\ %
    \midrule
    SC1 & Frequency of scheduled social activities & \includegraphics[width=6.5cm]{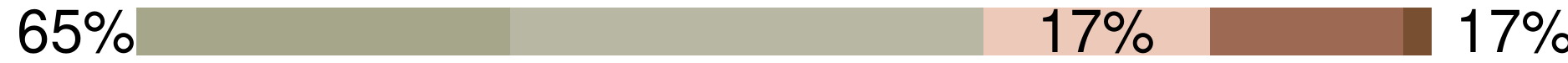}  \\
    SC2 & \begin{tabular}{@{}c@{}}Satisfaction with the social interaction from social activities\end{tabular} & \includegraphics[width=6.5cm]{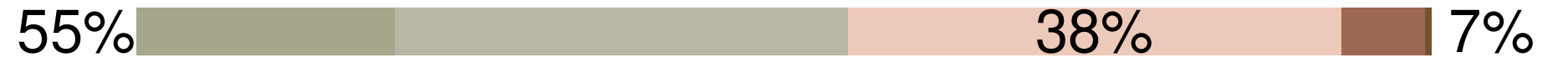}\\
    SC3 & Connection with team members & \includegraphics[width=6.5cm]{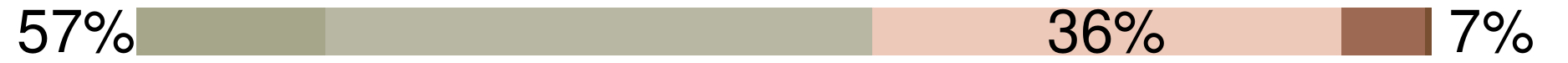}\\
    SC4 & Team's overall sense of connection &  \includegraphics[width=6.5cm]{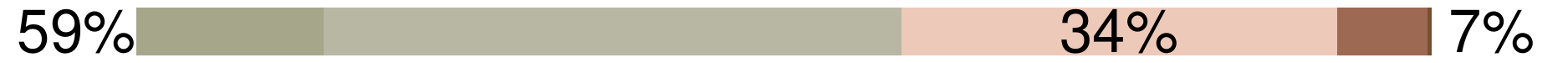}\\
    SC5 & Social connection with team members & \includegraphics[width=6.5cm]{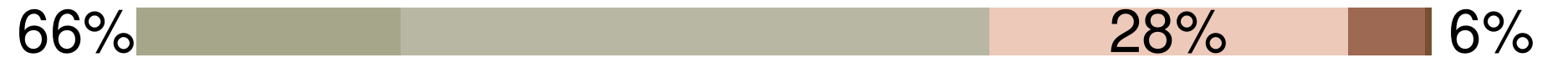}\\
    SC6 & Frequency of impromptu social activities & \includegraphics[width=6.5cm]{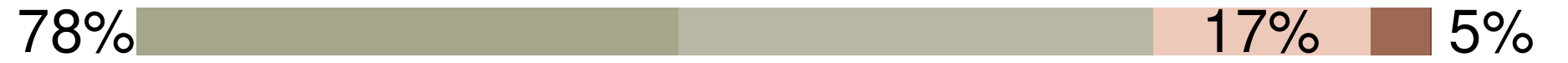}\\ 
    SC7 & Enjoyment of social activities & \includegraphics[width=6.5cm]{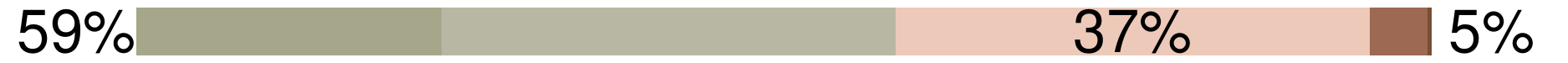}\\
    \midrule
    \multicolumn{2}{l}{\scriptsize \crule[darkGreenAccentColor]{.2cm}{.2cm} Significant Decrease \crule[lightGreenAccentColor]{.2cm}{.2cm} Decrease \crule[neutralAccentColor]{.2cm}{.2cm} About the Same \crule[lightBrownAccentColor]{.2cm}{.2cm} Increase \crule[mediumBrownAccentColor]{.2cm}{.2cm} Significant Increase} &
    {\scriptsize Percentages may not total 100\% due to rounding} \\
    \bottomrule
    \end{tabular}
\end{center}
\vspace{-\baselineskip}
\end{table*}

\begin{table}[t!]\centering

\newlength{\myboxheight}
\settoheight{\myboxheight}{1234567890\%}

\def\mybarchart#1{
\resizebox {\dimexpr#1/3*2} {\myboxheight} {%
\begin{tikzpicture}[]
\definecolor{clr1}{RGB}{99,99,99}
\definecolor{clr2}{RGB}{240,240,240}
\definecolor{mediumBrownAccentColor}{HTML}{784F31}
\definecolor{lightBrownAccentColor}{HTML}{9E6952}
\begin{axis}[
      axis background/.style={fill=gray!10, draw=gray!50},
      axis line style={draw=none},
      tick style={draw=none},
      ytick=\empty,
      xtick=\empty,
      ymin=0, ymax=1, %
      xmin=0, xmax=1]
\addplot [
      ybar interval=.5,
      fill=black,
      draw=none,
]
	coordinates {(1,1) (1,1)}; 
\addplot [
      ybar interval=.5,
      fill=lightBrownAccentColor,
      draw=none,
]
	coordinates {(1,1) (0,1)}; 
\end{axis}%
\end{tikzpicture}%
}%
}

\caption{Changes in perceived team productivity.}
\label{tab:productivity}

\begin{tabular}{rp{2cm}p{2cm}}
\toprule
& \multicolumn{1}{l}{WFH-Survey} & \multicolumn{1}{l}{Team-Survey}  \\
\cmidrule(r){1-1}\cmidrule(lr){2-2}\cmidrule(l){3-3}
Significantly more productive & \mybarchart{3pt} 3\% & \mybarchart{2pt} 2\% \\ 
More productive & \mybarchart{19pt} 19\% & \mybarchart{18pt} 18\% \\ 
\cmidrule(r){1-1}\cmidrule(lr){2-2}\cmidrule(l){3-3}
About the same & \mybarchart{55pt} 55\% & \mybarchart{56pt} 56\% \\ 
\cmidrule(r){1-1}\cmidrule(lr){2-2}\cmidrule(l){3-3}
Less productive & \mybarchart{21pt} 21\% & \mybarchart{21pt} 21\%  \\ 
Significantly less productive & \mybarchart{2pt} 2\% & \mybarchart{2pt} 2\% \\ 
\bottomrule
\end{tabular}

\end{table}

\subsection{How have social interactions within teams changed during WFH?~(RQ5)}\label{sec:RQ5Results}

Based on the findings from the WFH-Survey, challenges regarding \textbf{social connection} and \textbf{social interaction} appear to be a common thread among developers, motivating our further analysis. 
Many respondents are actively missing interactions with their team members, specifically social interactions, and numerous teams are using social engagements as a way to actively support their members.  Nonetheless, plenty of respondents still reported deficiencies in social connection. 
In the Team-Survey, 74\% of respondents reported missing social interaction as a WFH challenge they experienced.  

Many participants in the WFH-Survey also mentioned feeling disconnected from their team members, and feeling like they are not connecting with their team members remotely in the same way they used to in person.  The loss of team lunches, which were a popular low-effort social event, was commonly cited.  One participant described how \myquotex{we used to have lunches together, and had good [camaraderie]. No more lunches together, less [camaraderie], and less incidental `water cooler' information exchange.}{R_3Rsw5b4gmtj0JD1}

\addtocounter{footnote}{-1}

\smallskip
Table \ref{Q13LikertTable} shows the responses to the question \inlinequestion{How has social interaction with your team members changed since working from home?}{\Qsocialinteraction}.
There is a large decrease in social activities. 
Over half of the participants reported experiencing a decrease for all items in \Qsocialinteraction.
The \emph{frequency of impromptu social activities} (78\%, SC7)\footnotemark{} and \emph{scheduled social activities} (65\%, SC1) decreased dramatically.  Furthermore, 59\% of respondents reported a decrease in \textit{enjoyment of social activities}~(SC7) and 66\% of respondents reported a decrease in \textit{social connections with their team members}~(SC5).

\smallskip
For the responses to the question \inlinequestion{How do you stay socially connected with your colleagues (in place of hallway conversations, coffee breaks, etc.?)}{\Qsocialconnection, not displayed in a table}, we could observe that the most frequently utilized social activities were relatively low-effort. The two most frequently utilized social events were \textit{starting or ending meetings with non-work conversations} and \textit{sharing pictures and memes}, with 40\% and 25\% of teams respectively using them at least 2-4 times weekly. Whereas, activities that require more organization and coordination like \textit{external team bonding activities} and \textit{playing online games together} were used less frequently, with 1\% and 2\% respectively using them at least 2-4 times weekly.  

\begin{table}%
\caption{Factors from model that had statistically significant relationship with change in perceived team productivity}
\label{ModelResultsTable}
\centering
\begin{tabular}{l l} 
\toprule
Factor & Coef. \\
\midrule
(Intercept) & \phantom{-}0.10 \\[1mm]

\textbf{Less Awareness of What Colleagues are Working On}  & \\     
\hspace{2mm}Minor Issue &  -0.23*** \\
\hspace{2mm}Major Issue &  -0.40*** \\[1mm]

\textbf{Ability To Brainstorm With Team Members} & \\  
\hspace{2mm}Significantly Decreased & -0.20* \\
\hspace{2mm}Decreased &  -0.17** \\
\hspace{2mm}Increased &  \phantom{-}0.09 \\
\hspace{2mm}Significantly Increased & \phantom{-}0.11 \\[1mm]

\textbf{Difficult to Communicate with Colleagues}  & \\   
\hspace{2mm}Minor Issue &  -0.16* \\
\hspace{2mm}Major Issue &  -0.35** \\[1mm]

\textbf{Impactful Contributions I Make to Team}   & \\ 
\hspace{2mm}Significantly Decreased & -0.35. \\
\hspace{2mm}Decreased &  -0.12 \\
\hspace{2mm}Increased &  \phantom{-}0.23*** \\
\hspace{2mm}Significantly Increased & \phantom{-}0.79*** \\[1mm]

\textbf{Ability to Make Decisions as Team} & \\   
\hspace{2mm}Significantly Decreased &  -0.48** \\
\hspace{2mm}Decreased &  -0.08 \\
\hspace{2mm}Increased &  \phantom{-}0.31** \\
\hspace{2mm}Significantly Increased & \phantom{-}0.11 \\[1mm]

\textbf{Frequency of Scheduled Social Activities}  & \\
\hspace{2mm}Significantly Decreased & \phantom{-}0.22* \\
\hspace{2mm}Decreased &  \phantom{-}0.18* \\
\hspace{2mm}Increased &  \phantom{-}0.24** \\
\hspace{2mm}Significantly Increased & \phantom{-}0.11 \\[1mm]

\textbf{Start or End Meetings with Non-Work Conversations}   & \\  
\hspace{2mm}Monthly &  \phantom{-}0.06 \\
\hspace{2mm}Biweekly &  -0.01 \\
\hspace{2mm}Weekly &  -0.13. \\
\hspace{2mm}2-4 times weekly & -0.24** \\
\hspace{2mm}Daily &  \phantom{-}0.04 \\
\hspace{2mm}Multiple times daily & -0.15 \\[1mm]

\textbf{External Team Bonding Activities} & \\ 
\hspace{2mm}Monthly & \phantom{-}0.16. \\
\hspace{2mm}Biweekly &  \phantom{-}0.24. \\
\hspace{2mm}Weekly &  \phantom{-}0.07 \\
\hspace{2mm}2-4 times weekly & \phantom{-}0.52 \\
\hspace{2mm}Daily &  \phantom{-}1.47*** \\[1mm]

\textbf{My Team Members are Supportive of New Ideas} & \\ 
\hspace{2mm}Strongly disagree &  -0.87** \\
\hspace{2mm}Disagree & -0.04 \\
\hspace{2mm}Agree &  -0.18* \\
\hspace{2mm}Strongly agree & -0.20* \\

\midrule

Adjusted $R^2$ & \phantom{-}0.45 \\
\midrule
***p$<$0.001, **p$<$0.01, *p$<$0.05,.p$<$0.1 \\
\bottomrule
\end{tabular}
\vspace{-0.5\baselineskip}
\end{table}

\begin{table}[!ht]
\caption{Full list of factors included in team productivity model}
\label{BothModelsANOVAResultsTable}
\centering
\begin{tabular}{l l}
\toprule
Productivity Factors & F-Statistic \\
\midrule
Less awareness of what colleagues are working on & 9.86***  \\ %
Impactful Contributions I Make to Team & 9.67*** \\ %
Ability to Make Decisions as Team & 4.52** \\ %
External Team Bonding Activities & 3.97** \\ %
My Team Members are Supportive of New Ideas & 3.26* \\ %
Communication Breakdowns Within my Team & 1.37 \\ %
Difficult to Communicate with Colleagues & 5.89** \\
Ability to Brainstorm with Team Members & 2.93*  \\
Start/End Meetings with Non-Work Talk & 2.94** \\
Frequency of Scheduled Social Activities & 2.48* \\
My Manage is Highly Capable & 1.86 \\
Fun Informal Chats Non-Teams & 1.73 \\
Ability to Ask Questions in Group Meetings & 1.63 \\
Satisfaction with Social Interaction from Social Activities & 1.04 \\
\midrule
***p$<$0.001, **p$<$0.01, *p$<$0.05,.p$<$0.1 \\
\bottomrule
\end{tabular}
\vspace{-0.5\baselineskip}
\end{table}

\subsection{Which factors are associated with a change in team productivity during WFH? (RQ6)}\label{sec:RQ6Results}

When we asked participants in the Team-Survey how their team productivity has changed compared to working in the office, 20\% reported an increase and 23\% reported a decrease (see Table~\ref{tab:productivity}). The numbers were similar to the WFH-Survey.

To understand which factors are associated with a change in team productivity, we built a regression model as described in Section~\ref{subsec:TeamProductivitySurvey} based on the Team-Survey data.
As a reminder, the response variable, \emph{perceived change in team productivity} (\Qteamproductivity), was a 5-point response from \textit{significantly decreased} to \textit{significantly increased}.  The responses were coded as a numerical variable as follows: $-2$ means \textit{significantly decreased}, $-1$ means \textit{decreased}, $0$ means \textit{about the same}, $+1$ means \textit{increased}, and $+2$ means \textit{significantly increased}.
Table \ref{ModelResultsTable} shows the model coefficients for the factors that had a significant relationship with change in team productivity.  
Table \ref{BothModelsANOVAResultsTable} lists the ANOVA F-statistic and p-value for all factors in the reduced model.  

\emph{Interpreting Model Coefficients}.  Since the response variables are categorical, each has \textit{n-1} dummy variables where \textit{n} is the number of categories for a given variable.  
It is important to note the dummy variable coefficients do not represent absolute change, but rather relative change to the baseline category (i.e., the category that is not coded with a dummy variable).  
The dummy variable coefficients represent the difference between the team productivity of a given category relative to the team productivity of the baseline category for that factor.  

\emph{Baseline Categories}. For questions on a 5-point Decreased-Increased, or Disagree-Agree scale, the baseline categories are \textit{about the same} and \textit{neither agree nor disagree}, respectively.  For questions on a 3-point issue scale, the baseline is \textit{not an issue}.  For \textit{External Team Bonding Activities}, the baseline is \textit{rarely or never}.  

\medskip

Next, we discuss some observations from Table \ref{ModelResultsTable}.

\smallskip
\textbf{Less Awareness of What Colleagues are Working On}.  Experiencing this issue was associated with a decrease in team productivity.  This is a concern because 58\% of survey respondents reported experiencing this issue.  When experienced as a major issue, the drop in team productivity was roughly double the drop when experienced as a minor issue.   

\smallskip
\textbf{Ability to Brainstorm with Team Members}.  A decrease in the ability to brainstorm with team members was associated with a drop in team productivity, and similarly, an increase in ability to brainstorm was associated with an increase in team productivity.  
This is an important finding because 57\% of respondents reported a decrease in the ability to brainstorm with team members (see Table \ref{Q12LikertTable}), and we can now report that this is also associated with a decrease in team productivity.  This is also relevant given several respondents made comments in the WFH-Survey about challenges related to collaboration work with team members, particularly brainstorming.  As one participant put it, \myquotex{brainstorming activities that need whiteboard and involve more passionate conversation progresses slower than before and can be frustrating.}{R_1lbimxXMqOnnKkP}  

\smallskip
\textbf{Difficult to Communicate with Colleagues}.  Experiencing the issue of having difficulty communicating with colleagues was associated with a drop in team productivity.
From the RQ1 findings, we know difficulty with communication is common among those working from home, and from Table \ref{Q12LikertTable} we know that 46\% of respondents reported experiencing this challenge. 

\smallskip
\textbf{Additional Significant Factors.} For the factors \emph{Impactful Contributions I Make to Team} and  \emph{Ability to Make Decisions as a Team}, an increase  was associated with an increase in team productivity.  A decrease in these factors was associated with a drop in team productivity.  Other factors that had a statistically significant relationship with change in team productivity were \textit{frequency of scheduled social activities}, \textit{starting or ending meetings with non-work talk}, \textit{external team bonding activities}, and \textit{my team is supportive of new ideas}.

\section{Threats to Validity}\label{sec:threatsToValidity}
We discuss several of the threats to the validity of our work.  
The first point is not a threat but an important disclaimer, this paper does not suggest causal inference of any kind.  All findings are based on survey data it's statistical analysis.  Our findings show factors that are constructive when modeling perceived change in team productivity and factors that have a statistically significant relationship with it, but we are not implying causality of any manner.    

\smallskip
\textbf{External Validity.}
The participants for the surveys were all from Microsoft, a large software company, and all respondents were based in the United States. This limits the generalizability of our findings, which cannot be assumed to generalize to all software developers, especially those who work at smaller companies, in different regions, or in open source. At Microsoft, we collected a large random sample of developers across multiple divisions (called organizations at Microsoft) who worked on a vast number of different kinds of projects and use a variety of development techniques.  Although our findings cannot be generalized to all software developers, we believe our findings are still of value to the community because it has been shown that historically single-case case studies contribute to scientific discovery~\cite{flyvbjerg06}.  To help increase the generalizability of our work in the future, we have created a comprehensive replication package~\cite{courtney_miller_2021_4456041} that can be used to perform our survey and analysis at other companies. 

Previous research has shown that people who are inclined to participate in surveys may have different demeanor and personality traits than the entire population being sampled from~\cite{rogelberg03, marcus05}.  In an attempt to minimize the risk of non-response and volunteer bias, we aimed to lower the opportunity cost of participating in the Team-Survey by making the expected completion time 10 minutes, incentivizing engagement with a donation to a civil rights organization, and anonymizing survey responses. In the WFH-Survey, we incentivized engagement with entry into a raffle for an Amazon gift card and anonymizing survey responses.   

The time period when the participants took the survey may have also impacted their perspective, which could potentially evolve over time.  
It is also important to note that pandemic WFH is not the same thing as WFH during normal times, so there are limits on the generalizability of our findings.  

\smallskip
\textbf{Construct Validity.}
When measuring the change in productivity, we used a single metric, self-reported perceived change in productivity.  It's important to note that there is no single definitive metric for productivity and that we cannot claim that our findings are true for productivity as a whole, because the factors that impact productivity inherently depend on the measures you are using to define productivity~\cite{jaspan19}. 
The team productivity scores are from the perspective of an individual team member. Since the survey was anonymous and no personal identifiable information was collected (to encourage more candid responses), we were unable to triangulate responses from  members of the same team, which would have further improved the validity of the findings. 
While there are advantages and disadvantages to selecting self-reported measures of productivity versus automated technical ones, our choice aligns similar recent work~\cite{storey19, murphy19, johnson19}.  

While we performed a thorough literature review, there may be additional team culture factors that have an impact on team productivity that we did not include in our survey.  Our findings are not meant to be an exhaustive list but to rather highlight some of the team culture factors that \textit{are} constructive when modeling team productivity.  

The first iteration of qualitative analysis coding was done by a single author, but were later reviewed by the other authors, and any responses the first author was unsure of were decided with discussion with another author.  

The response variable for the model, change in team productivity, was coded in the survey as a 5-point Likert scale.  However, for the sake of the data analysis, it was coded numerical values from -2 to 2. Since Likert scales are ordinal categorical scales, there has been a debate within the statistics community about whether it is appropriate to code them as numerical values~\cite{carifio08}. Norman stated that ``parametric statistics can be used with Likert data''~\cite{norman10}, even when the sample sizes are small and the distributions are non-normal.  It has also been suggested that higher p-values be used with Likert scores \cite{carifio08}, so all p-values have been included in our model so readers can interpret results with both a standard p-value of $\alpha=0.05$ or a smaller one. 

\smallskip
\textbf{Internal Validity.}
Due to the nature of surveys, there is a chance some participants found certain survey questions to be ambiguous, confusing, or unclear.  This means participants could misunderstand what we're asking about and respond in a way that does not align with what we meant to ask.  To help prevent this possibility, we performed a small pilot study with the first 200 participants, which included a question at the end of the survey asking about whether there were any ambiguous or unclear questions or other aspects of the survey.

Since the survey responses were anonymous, there is a possibility that participants took the survey multiple times.

The structure of the surveys may have had an impact on the ways the participants responded to questions and could have potentially primed them.  To prevent this from affecting perceived productivity, which was our response variable, we put the   productivity questions at the beginning of the survey.

\section{Related Work}\label{sec:relatedWork}

%

%

\subsection{Working from Home}
In 1973, Jack Niles, a NASA engineer, was sick of Los Angeles traffic.  It was then that he coined the term ``telecommuting,'' and the concept of remote work was conceived \cite{gan:2015}.  Granted, his vision for telecommuting was different than our reality today since his idea was from over a decade before the public internet~\cite{leiner09}.  Nonetheless, the concept of distributed workplaces was present in his proposal for telecommuting~\cite{niles76}, and that idea has evolved into the remote work we know today.  

Research on WFH has increased in the last few years~\cite{felstead17, bailey99, nilles94, baruch00}. There are many benefits to WFH for both the employee and the employer. For example, employees can reduce or eliminate their commute to work and save money on gasoline.  They can also control the environment they work in (e.g., the air conditioning).  Companies can benefit in many ways, such as saving money from overhead and remote work allows companies to hire the best talent from any location.   %

However, employees sometimes find that they are working increased hours when WFH~\cite{felstead17}. This may mean they work unsustainable schedules that may eventually lead to burnout~\cite{maslach08}.  Previous work has also shown that increased work hours can increase the risk of developer disengagement~\cite{miller19}.  Remote work can also make it more difficult for managers to manage their teams effectively.  It can be challenging for managers to remain in touch with their employees and stay on top of their progress~\cite{bailey99}.

Teamwork can be significantly impacted by remote work~\cite{herbsleb01, herbsleb01Second, bird09, ford19}.    
Wagstrom \textit{et al.} found the temporal distribution of teams had a significant negative impact on communication response time, suggesting that teams bifurcated across multiple time zones communicate significantly slower~\cite{wagstrom14}.  
Also, Butler \textit{et al.} explored the challenges faced by developers working from home during the COVID-19 pandemic and the impact that these challenges have had on job satisfaction~\cite{butler20}. They found that having many meetings and feeling overworked were some of the biggest challenges faced. 

In addition, WFH has been found to have an impact on individual software development productivity.  Kazekami found that when remote work hours are too long, worker productivity is reduced~\cite{kazekami20}.  In the GitHub Octoverse report, they found that developers were working longer hours, in some cases up to two additional 8-hour workdays per week~\cite{forsgren20}. %
It is important to note that Bao \textit{et al.} also performed a case study %
and found that the effect of WFH on productivity varies depending on which productivity metrics are used and the characteristics of a project~\cite{bao20} (e.g., its size). Finally, Ralph \textit{et al.} ran a large-scale international survey of developers and found that ``software professionals who are working from home during the pandemic are experiencing diminished emotional well being and productivity, which are closely related~\cite{ralph20}."  The relationship between self-reported productivity and developer satisfaction was also discussed in previous research~\cite{storey19}.

\subsection{Developer Productivity}

Software development is a complex process, and doing it well, on budget, and on time can be hard~\cite{jorgensen06}.  Teams may find challenges in coordination and communication~\cite{brooks95}. %
Therefore, we must study developer productivity, which has been of interest to researchers for decades~\cite{sackman68,lister87, brooks81, meyer19, meyer17}. 

One fundamental challenge in studying developer productivity is defining and operationalizing factors that encapsulate `productivity'~\cite{wagner19Second}.  Generally speaking, there are two overarching types of productivity factors, technical and soft~\cite{wagner18}.

Some popular technical productivity metrics are the number of lines of code~\cite{devanbu96}, the number of builds~\cite{bao20}, and the number of commits per unit of time~\cite{bao20}.  Soft factors that have been shown to influence developer productivity include corporate culture, workplace environment, and team culture factors~\cite{wagner18}.  However, there has been little work focused on team culture factors, which are often more complex and nuanced.  %

Team culture factors are often directly impacted by the distribution of teams~\cite{carmel02, espinosa03}.  
For example, we know that communication for teams that are separated across multiple time zones is often slower than co-located teams~\cite{wagstrom14}.  And previous research also suggests that coordination and communication are some of the biggest challenges faced by developer teams~\cite{brooks95}.
Tang \textit{et al.} found that remote teams customize how they work together based on their unique situations~\cite{tang11}.

\smallskip

The majority of the related work explores individual developer productivity and individual challenges during the COVID-19 pandemic.  However, in our work, we specifically directed our efforts on studying the impact that WFH has had on \textbf{software development teams} and their productivity.  %

\section{Discussion and Recommendations}
\label{sec:Discussion}

Through our survey analysis, we learned that many developers are experiencing social connection and communication challenges with their teams.  
While many of our respondents reported challenges, some reported that the transition to WFH had been a positive experience.  
Additionally, only 23\% of respondents experienced a decrease in productivity during this time.  
In this section, we discuss some of the positive WFH experiences, implications for teams, and provide recommendations for improving social connection and communication.  

\subsection{Benefits of Working from Home}

There are multiple benefits to WFH. For example, WFH can be more accessible for workers with disabilities~\cite{spark17}. WFH can also help create an environment where traditionally marginalized people (e.g., trans developers) can be empowered and have more autonomy over workplace interactions~\cite{ford19}.

In our survey, we found that one of the benefits of WFH included the increased inclusion of all team members.

\myquote{ [Microsoft] Teams has forced people to be more inclusive. Our in person meetings usually have multiple people speaking over each other. Online meetings over Teams on the other hand have enabled people to stop speaking over each other since they cannot see the other person and people usually wait for the other person to complete.}{R_3npWNj1BAXcbkn7}

Another one of the reported benefits of WFH was increased empathy for fellow team members, including increased empathy for teammates who were remote before COVID-19.    

\myquote{I already had several employees who didn't work in Redmond. This has enabled us to build more empathy with working with distant co-workers. }{R_pa4Azj57DrrARtD}

As many teams were no longer hybrid but fully remote, teams were now collectively experiencing similar remote work and WFH problems. This placed teammates on the same playing field during meetings and increased team bonding.

\myquote{Flexibility, forgiveness of missed meetings and challenges in connectivity. More mutual bonding over shared difficulties. Time taken to try to connect outside of work-related tasks.}{R_2TTACcsbp9OcmQx}

Overall, the benefits of WFH provided an opportunity for teams to respect one another's perspective, build empathy, and understanding of unprecedented home challenges that arise---all characteristics that hopefully persist after the pandemic.

\subsection{Recommendations}
  Based on our results, we provide the following recommendations to improve social connection and communication. %

\smallskip

\begin{tcolorbox}[width=0.48\textwidth]
\underline{\textbf{Recommendations for improving social connection:}}
\renewcommand{\labelitemi}{$\blacksquare$}
 \begin{itemize}
   \item Build and maintain team culture.
    \item Include social activities as part of ``work.''
    \item Be mindful of other people's time.
    \item Actively work to be inclusive.
   \end{itemize} 
\end{tcolorbox} 

\smallskip

\textbf{Build and maintain team culture.}
Managers should strive to create and maintain a team culture where small daily social interactions are promoted.  For example, taking 10 minutes at the start and end of meetings for small talk to maintain social connection, as also recommended in~\cite{ford20, heisman20}.  %

\textbf{Include social activities as part of ``work.''} We recommend adding regular social activities to ``work.'' We suggest being creative with these social events; for example, an alternative to virtual lunch could be playing an online game or holding a remote arts and crafts event.   %

\textbf{Be mindful of other's time.} Developers are working more hours~\cite{forsgren20}, and many respondents reported meeting fatigue.  Therefore, we recommend scheduling social events during work hours and avoiding adding meetings during off-hours.  

\textbf{Be inclusive.} We recommend that teams are mindful of events that include hybrid participants. We suggest that hybrid teams choose remote-centric activities to maintain accessibility for team members who remain remote indefinitely~\cite{rodeghero20}. 

\smallskip

\begin{tcolorbox}[width=0.48\textwidth]
\underline{\textbf{Recommendations for improving communication:}}
\renewcommand{\labelitemi}{$\blacksquare$}
 \begin{itemize}
   \item  Managers should stay aware of contributions.
    \item Managers should hold weekly 1:1 meetings.    
    \item Hold regular team meetings.
    \item Create communication standards.
   \end{itemize} 
\end{tcolorbox} 

\smallskip

\textbf{Managers stay aware of contributions.} One commonly cited WFH challenge was effectively communicating in group meetings. 
During meetings, we recommend managers remain aware of each participant's contributions and ensure all voices are given an opportunity to speak.

\textbf{Managers hold weekly 1:1 meetings.} We found that many respondents also reported appreciating check-in 1:1s with managers and team members, which could be a productive way to increase direct communication. We recommend having weekly 1:1s regularly scheduled with managers and their team. %

\textbf{Hold regular team meetings.} We recommend that teams meet regularly.  These meetings should be at least once a week and can even be held daily.  We recommend that these are scheduled meetings that are reoccurring and put on the calendar in advance.  We also recommend that during the meeting, each team member provides status updates and discusses any problems. These meetings allow teams and managers to be aware of the progress and potentially address problems faster.  

\textbf{Create communication standards.} We recommend that teams collectively create standards around communication. These standards include which communication tools to use for specific types of communication and how long a teammate should expect to wait for a response.  This will also help new hires onboarding remotely as they can be given clear guidance on how to communicate with their new team~\cite{rodeghero21}.  %

\section{Conclusion and Future Work}
\label{sec:Conclusion}

In conclusion, through the two surveys of developers (2,265 + 608 responses), we found that software development teams working from home during the COVID-19 pandemic have experienced a radical shift in how they work together.  

Our first survey, WFH-Survey, probed about a wide range of topics related to WFH and provided rich insights on team culture and collaboration factors. The second survey, Team-Survey, investigated how team culture factors and team productivity have been impacted by WFH.  Using the team factors that emerged from our surveys and from related work, we built a model to determine which team factors were most useful when modeling changes in team productivity.  

Just as previous research has associated understanding and predicting individual developer productivity with social interactions, we found that many of the factors associated with team productivity were also of a social nature. 
Social connection and communication were the most commonly cited team culture factors. We found that 65\% of respondents reported a decrease in feelings of social connectedness with their team, and 74\% of respondents citing missing social interaction as a WFH challenge.  In terms of communication, 51\% cited a decrease in communication ease with colleagues, and 57\% said the ability to brainstorm with colleagues has decreased.   

In understanding team productivity, we note most respondents from the Team-Survey reported that they perceived little to no change in team productivity (56\%), which is consistent with the prior WFH-Survey. However, from those affected, we were able to build a model of team productivity factors.

Our model of team productivity factors included \emph{the ability to brainstorm with colleagues}, \emph{having less awareness of what colleagues are working on}, and \emph{having difficulty communicating with colleagues}, and these factors all have a significant relationship with changes in team productivity.  Other factors include \emph{satisfaction with social interaction from social activities}, while \emph{communication breakdowns on teams} are also important factors when modeling change in team productivity.  
Understanding team productivity is a new avenue for research and an important topic as developers are now more distributed during the pandemic and likely will continue to be so after it.  Improving productivity is not just a concern for companies. It is also important from the individual developer perspective as previous research has shown a relationship between productivity and work satisfaction, and between productivity and well-being. 
Although effective engineering processes, collaboration and communication tools and work environments are important for  productivity, understanding the nuances of how developers socialize, communicate, and support each other~\cite{miller19} is just as critical.

\section*{Supplemental Material}
A replication package with both surveys and the analysis codebook is available on Zenodo~\cite{courtney_miller_2021_4456041}. \href{https://doi.org/10.5281/zenodo.4456041}{\includegraphics[width=3cm,trim=0 1mm 0 0]{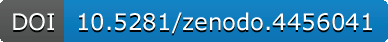}}

\section*{Acknowledgments}

The authors would like to thank the survey participants. \ifIncludeAuthorInfo
Courtney Miller performed this work during a summer internship at Microsoft Research in the Software Analysis and Intelligence Team (http://aka.ms/saintes). Paige Rodeghero and  Margaret-Anne Storey contributed to this work while being visiting researchers at Microsoft.
\else
Additional acknowledgments have been redacted for submission.
\fi

\balance

\bibliographystyle{IEEEtran}
\bibliography{refrences}

\begin{thebibliography}{10}
\providecommand{\url}[1]{#1}
\csname url@samestyle\endcsname
\providecommand{\newblock}{\relax}
\providecommand{\bibinfo}[2]{#2}
\providecommand{\BIBentrySTDinterwordspacing}{\spaceskip=0pt\relax}
\providecommand{\BIBentryALTinterwordstretchfactor}{4}
\providecommand{\BIBentryALTinterwordspacing}{\spaceskip=\fontdimen2\font plus
\BIBentryALTinterwordstretchfactor\fontdimen3\font minus
  \fontdimen4\font\relax}
\providecommand{\BIBforeignlanguage}[2]{{%
\expandafter\ifx\csname l@#1\endcsname\relax
\typeout{** WARNING: IEEEtran.bst: No hyphenation pattern has been}%
\typeout{** loaded for the language `#1'. Using the pattern for}%
\typeout{** the default language instead.}%
\else
\language=\csname l@#1\endcsname
\fi
#2}}
\providecommand{\BIBdecl}{\relax}
\BIBdecl

\bibitem{duffy20}
C.~Duffy, ``Big tech firms ramp up remote working orders to prevent coronavirus
  spread,'' March 2020, retrieved August 17, 2020 from
  \url{https://www.cnn.com/2020/03/10/tech/google-work-from-home-coronavirus/index.html}.

\bibitem{baker07}
E.~Baker, G.~C. Avery, and J.~Crawford, ``Satisfaction and perceived
  productivity when professionals work from home,'' \emph{Research \& Practice
  in Human Resource Management}, 2007.

\bibitem{neufeld05}
D.~J. Neufeld and Y.~Fang, ``Individual, social and situational determinants of
  telecommuter productivity,'' \emph{Information \& Management}, vol.~42,
  no.~7, pp. 1037--1049, 2005.

\bibitem{bao20}
L.~Bao, T.~Li, X.~Xia, K.~Zhu, H.~Li, and X.~Yang, ``How does working from home
  affect developer productivity?--a case study of baidu during covid-19
  pandemic,'' \emph{arXiv preprint arXiv:2005.13167}, 2020.

\bibitem{ralph20}
P.~Ralph, S.~Baltes, G.~Adisaputri, R.~Torkar, V.~Kovalenko, M.~Kalinowski,
  N.~Novielli, S.~Yoo, X.~Devroey, X.~Tan \emph{et~al.}, ``Pandemic
  programming: How covid-19 affects software developers and how their
  organizations can help,'' \emph{arXiv preprint arXiv:2005.01127}, 2020.

\bibitem{wagner19}
S.~Wagner and E.~Murphy-Hill, ``Factors that influence productivity: A
  checklist,'' in \emph{Rethinking Productivity in Software Engineering}.\hskip
  1em plus 0.5em minus 0.4em\relax Springer, 2019, pp. 69--84.

\bibitem{storey19}
M.-A. Storey, T.~Zimmermann, C.~Bird, J.~Czerwonka, B.~Murphy, and
  E.~Kalliamvakou, ``Towards a theory of software developer job satisfaction
  and perceived productivity,'' \emph{IEEE Transactions on Software
  Engineering}, 2019.

\bibitem{hewson96}
C.~M. Hewson, D.~Laurent, and C.~M. Vogel, ``Proper methodologies for
  psychological and sociological studies conducted via the internet,''
  \emph{Behavior Research Methods, Instruments, \& Computers}, vol.~28, no.~2,
  pp. 186--191, 1996.

\bibitem{courtney_miller_2021_4456041}
\BIBentryALTinterwordspacing
\emph{{Supplementary Material for "How Was Your Weekend?" Software Development
  Teams Working From Home During COVID-19}}.\hskip 1em plus 0.5em minus
  0.4em\relax Zenodo, Jan. 2021. [Online]. Available:
  \url{https://doi.org/10.5281/zenodo.4456041}
\BIBentrySTDinterwordspacing

\bibitem{smith13}
E.~Smith, R.~Loftin, E.~Murphy-Hill, C.~Bird, and T.~Zimmermann, ``Improving
  developer participation rates in surveys,'' in \emph{2013 6th International
  Workshop on Cooperative and Human Aspects of Software Engineering
  (CHASE)}.\hskip 1em plus 0.5em minus 0.4em\relax IEEE, 2013, pp. 89--92.

\bibitem{schreier12}
M.~Schreier, \emph{Qualitative content analysis in practice}.\hskip 1em plus
  0.5em minus 0.4em\relax Sage publications, 2012.

\bibitem{murphy19}
E.~Murphy-Hill, C.~Jaspan, C.~Sadowski, D.~Shepherd, M.~Phillips, C.~Winter,
  A.~Knight, E.~Smith, and M.~Jorde, ``What predicts software developers'
  productivity?'' \emph{IEEE Transactions on Software Engineering}, 2019.

\bibitem{wagner18}
S.~Wagner and M.~Ruhe, ``A systematic review of productivity factors in
  software development,'' \emph{arXiv preprint arXiv:1801.06475}, 2018.

\bibitem{blm}
``Black lives matter global network,'' \url{https://blacklivesmatter.com},
  accessed: 2021-01-21.

\bibitem{fox15}
J.~Fox, \emph{Applied regression analysis and generalized linear models}.\hskip
  1em plus 0.5em minus 0.4em\relax Sage Publications, 2015.

\bibitem{montgomery12}
D.~C. Montgomery, E.~A. Peck, and G.~G. Vining, \emph{Introduction to linear
  regression analysis}.\hskip 1em plus 0.5em minus 0.4em\relax John Wiley \&
  Sons, 2012, vol. 821.

\bibitem{norman10}
G.~Norman, ``Likert scales, levels of measurement and the “laws” of
  statistics,'' \emph{Advances in health sciences education}, vol.~15, no.~5,
  pp. 625--632, 2010.

\bibitem{sullivan13}
G.~M. Sullivan and A.~R. Artino~Jr, ``Analyzing and interpreting data from
  likert-type scales,'' \emph{Journal of graduate medical education}, vol.~5,
  no.~4, pp. 541--542, 2013.

\bibitem{akaike74}
H.~Akaike, ``A new look at the statistical model identification,'' \emph{IEEE
  transactions on automatic control}, vol.~19, no.~6, pp. 716--723, 1974.

\bibitem{yamashita07}
T.~Yamashita, K.~Yamashita, and R.~Kamimura, ``A stepwise aic method for
  variable selection in linear regression,'' \emph{Communications in
  Statistics—Theory and Methods}, vol.~36, no.~13, pp. 2395--2403, 2007.

\bibitem{dodge08}
Y.~Dodge, \emph{The concise encyclopedia of statistics}.\hskip 1em plus 0.5em
  minus 0.4em\relax Springer Science \& Business Media, 2008.

\bibitem{scheffe99}
H.~Scheffe, \emph{The analysis of variance}.\hskip 1em plus 0.5em minus
  0.4em\relax John Wiley \& Sons, 1999, vol.~72.

\bibitem{sadowski19}
C.~Sadowski and T.~Zimmermann, \emph{Rethinking productivity in software
  engineering}.\hskip 1em plus 0.5em minus 0.4em\relax Springer Nature, 2019.

\bibitem{flyvbjerg06}
B.~Flyvbjerg, ``Five misunderstandings about case-study research,''
  \emph{Qualitative inquiry}, vol.~12, no.~2, pp. 219--245, 2006.

\bibitem{rogelberg03}
S.~G. Rogelberg, J.~M. Conway, M.~E. Sederburg, C.~Spitzm{\"u}ller, S.~Aziz,
  and W.~E. Knight, ``Profiling active and passive nonrespondents to an
  organizational survey.'' \emph{Journal of Applied Psychology}, vol.~88,
  no.~6, p. 1104, 2003.

\bibitem{marcus05}
B.~Marcus and A.~Sch{\"u}tz, ``Who are the people reluctant to participate in
  research? personality correlates of four different types of nonresponse as
  inferred from self-and observer ratings,'' \emph{Journal of personality},
  vol.~73, no.~4, pp. 959--984, 2005.

\bibitem{jaspan19}
C.~Jaspan and C.~Sadowski, ``No single metric captures productivity,'' in
  \emph{Rethinking Productivity in Software Engineering}.\hskip 1em plus 0.5em
  minus 0.4em\relax Springer, 2019, pp. 13--20.

\bibitem{johnson19}
B.~Johnson, T.~Zimmermann, and C.~Bird, ``The effect of work environments on
  productivity and satisfaction of software engineers,'' \emph{IEEE
  Transactions on Software Engineering}, 2019.

\bibitem{carifio08}
J.~Carifio and R.~Perla, ``Resolving the 50-year debate around using and
  misusing likert scales,'' \emph{Medical education}, vol.~42, no.~12, pp.
  1150--1152, 2008.

\bibitem{gan:2015}
V.~Gan, ``The invention of telecommuting,''
  \url{https://www.bloomberg.com/news/articles/2015-12-01/what-telecommuting-looked-like-in-1973},
  accessed: 2021-01-21.

\bibitem{leiner09}
B.~M. Leiner, V.~G. Cerf, D.~D. Clark, R.~E. Kahn, L.~Kleinrock, D.~C. Lynch,
  J.~Postel, L.~G. Roberts, and S.~Wolff, ``A brief history of the internet,''
  \emph{ACM SIGCOMM Computer Communication Review}, vol.~39, no.~5, pp. 22--31,
  2009.

\bibitem{niles76}
J.~M. Niles, F.~Carlson, P.~Gray, and G.~Hanneman, ``The
  telecommunications-transportation tradeoff,'' \emph{John Willey}, vol.~88,
  1976.

\bibitem{felstead17}
A.~Felstead and G.~Henseke, ``Assessing the growth of remote working and its
  consequences for effort, well-being and work-life balance,'' \emph{New
  Technology, Work and Employment}, vol.~32, no.~3, pp. 195--212, 2017.

\bibitem{bailey99}
N.~Bailey and N.~B. Kurland, ``The advantages and challenges of working here,
  there, anywhere, and anytime,'' \emph{Organizational dynamics}, vol.~28,
  no.~2, pp. 53--68, 1999.

\bibitem{nilles94}
J.~M. Nilles, \emph{Making telecommuting happen: A guide for telemanagers and
  telecommuters}.\hskip 1em plus 0.5em minus 0.4em\relax Van Nostrand Reinhold,
  1994.

\bibitem{baruch00}
Y.~Baruch, ``Teleworking: benefits and pitfalls as perceived by professionals
  and managers,'' \emph{New technology, work and employment}, vol.~15, no.~1,
  pp. 34--49, 2000.

\bibitem{maslach08}
C.~Maslach and M.~P. Leiter, ``Early predictors of job burnout and
  engagement.'' \emph{Journal of applied psychology}, vol.~93, no.~3, p. 498,
  2008.

\bibitem{miller19}
C.~Miller, D.~G. Widder, C.~K{\"a}stner, and B.~Vasilescu, ``Why do people give
  up flossing? a study of contributor disengagement in open source,'' in
  \emph{IFIP International Conference on Open Source Systems}.\hskip 1em plus
  0.5em minus 0.4em\relax Springer, 2019, pp. 116--129.

\bibitem{herbsleb01}
J.~D. Herbsleb, A.~Mockus, T.~A. Finholt, and R.~E. Grinter, ``An empirical
  study of global software development: distance and speed,'' in
  \emph{Proceedings of the 23rd International Conference on Software
  Engineering. ICSE 2001}.\hskip 1em plus 0.5em minus 0.4em\relax IEEE, 2001,
  pp. 81--90.

\bibitem{herbsleb01Second}
J.~D. Herbsleb and D.~Moitra, ``Global software development,'' \emph{IEEE
  software}, vol.~18, no.~2, pp. 16--20, 2001.

\bibitem{bird09}
C.~Bird, N.~Nagappan, P.~Devanbu, H.~Gall, and B.~Murphy, ``Does distributed
  development affect software quality? an empirical case study of windows
  vista,'' in \emph{2009 IEEE 31st International Conference on Software
  Engineering}.\hskip 1em plus 0.5em minus 0.4em\relax IEEE, 2009, pp.
  518--528.

\bibitem{ford19}
D.~Ford, R.~Milewicz, and A.~Serebrenik, ``How remote work can foster a more
  inclusive environment for transgender developers,'' in \emph{2019 IEEE/ACM
  2nd International Workshop on Gender Equality in Software Engineering
  (GE)}.\hskip 1em plus 0.5em minus 0.4em\relax IEEE, 2019, pp. 9--12.

\bibitem{wagstrom14}
P.~Wagstrom and S.~Datta, ``Does latitude hurt while longitude kills?
  geographical and temporal separation in a large scale software development
  project,'' in \emph{Proceedings of the 36th International Conference on
  Software Engineering}, 2014, pp. 199--210.

\bibitem{butler20}
\BIBentryALTinterwordspacing
J.~L. Butler and S.~Jaffe, ``Challenges and gratitude: A diary study of
  software engineers working from home during covid-19 pandemic,'' August 2020,
  {M}icrosoft Research Symposium on the New Future of Work. [Online].
  Available:
  https://www.microsoft.com/en-us/research/publication/challenges-and-gratitude-a-diary-study-of-software-engineers-working-from-home-during-covid-19-pandemic/
\BIBentrySTDinterwordspacing

\bibitem{kazekami20}
S.~Kazekami, ``Mechanisms to improve labor productivity by performing
  telework,'' \emph{Telecommunications Policy}, vol.~44, no.~2, p. 101868,
  2020.

\bibitem{forsgren20}
N.~Forsgren, ``Octoverse spotlight: An analysis of developer productivity, work
  cadence, and collaboration in the early days of covid-19,'' May 2020,
  retrieved June 2, 2020 from
  https://github.blog/2020-05-06-octoverse-spotlight-an-analysis-of-developer-productivity-work-cadence-and-collaboration-in-the-early-days-of-covid-19/.

\bibitem{jorgensen06}
M.~Jorgensen and M.~Shepperd, ``A systematic review of software development
  cost estimation studies,'' \emph{IEEE Transactions on software engineering},
  vol.~33, no.~1, pp. 33--53, 2006.

\bibitem{brooks95}
F.~P. Brooks~Jr, \emph{The mythical man-month: essays on software
  engineering}.\hskip 1em plus 0.5em minus 0.4em\relax Pearson Education, 1995.

\bibitem{sackman68}
H.~Sackman, W.~J. Erikson, and E.~E. Grant, ``Exploratory experimental studies
  comparing online and offline programming performance,'' \emph{Communications
  of the ACM}, vol.~11, no.~1, pp. 3--11, 1968.

\bibitem{lister87}
T.~R. Lister and T.~DeMarco, \emph{Peopleware: Productive projects and
  teams}.\hskip 1em plus 0.5em minus 0.4em\relax Dorset House New York, 1987.

\bibitem{brooks81}
W.~D. Brooks, ``Software technology payoff: Some statistical evidence,''
  \emph{Journal of Systems and Software}, vol.~2, no.~1, pp. 3--9, 1981.

\bibitem{meyer19}
A.~Meyer, E.~T. Barr, C.~Bird, and T.~Zimmermann, ``Today was a good day: The
  daily life of software developers,'' \emph{IEEE Transactions on Software
  Engineering}, 2019.

\bibitem{meyer17}
A.~N. Meyer, L.~E. Barton, G.~C. Murphy, T.~Zimmermann, and T.~Fritz, ``The
  work life of developers: Activities, switches and perceived productivity,''
  \emph{IEEE Transactions on Software Engineering}, vol.~43, no.~12, pp.
  1178--1193, 2017.

\bibitem{wagner19Second}
S.~Wagner and F.~Deissenboeck, ``Defining productivity in software
  engineering,'' in \emph{Rethinking Productivity in Software
  Engineering}.\hskip 1em plus 0.5em minus 0.4em\relax Springer, 2019, pp.
  29--38.

\bibitem{devanbu96}
P.~Devanbu, S.~Karstu, W.~Melo, and W.~Thomas, ``Analytical and empirical
  evaluation of software reuse metrics,'' in \emph{Proceedings of IEEE 18th
  International Conference on Software Engineering}.\hskip 1em plus 0.5em minus
  0.4em\relax IEEE, 1996, pp. 189--199.

\bibitem{carmel02}
E.~Carmel, ``Global software teams: opportunities and challenges of
  technology-enabled work,'' \emph{LERA For Libraries}, vol.~6, no.~2, 2002.

\bibitem{espinosa03}
J.~A. Espinosa and E.~Carmel, ``The impact of time separation on coordination
  in global software teams: a conceptual foundation,'' \emph{Software Process:
  Improvement and Practice}, vol.~8, no.~4, pp. 249--266, 2003.

\bibitem{tang11}
J.~C. Tang, C.~Zhao, X.~Cao, and K.~Inkpen, ``Your time zone or mine? a study
  of globally time zone-shifted collaboration,'' in \emph{Proceedings of the
  ACM 2011 conference on Computer supported cooperative work}, 2011, pp.
  235--244.

\bibitem{spark17}
R.~Spark, ``Accessibility to work from home for the disabled: The need for a
  shift in management style,'' in \emph{Proceedings of the 14th Web for All
  Conference on The Future of Accessible Work}, 2017, pp. 1--4.

\bibitem{ford20}
D.~Ford, M.-A. Storey, T.~Zimmermann, C.~Bird, S.~Jaffe, C.~Maddila, J.~L.
  Butler, B.~Houck, and N.~Nagappan, ``A tale of two cities: Software
  developers working from home during the covid-19 pandemic,'' \emph{arXiv
  preprint arXiv:2008.11147}, 2020.

\bibitem{heisman20}
L.~Heisman, ``Remote work: Reshaping the workplace experience,'' June 2020,
  retrived January 14, 2021 from
  \url{https://github.blog/2020-06-26-remote-work-reshaping-the-workplace-experience/}.

\bibitem{rodeghero20}
\BIBentryALTinterwordspacing
P.~Rodeghero and T.~Hernandez, ``Empowering and supporting remote software
  development team members through a culture of allyship,'' August 2020,
  {M}icrosoft Research Symposium on the New Future of Work. [Online].
  Available:
  https://www.microsoft.com/en-us/research/publication/empowering-and-supporting-remote-software-development-team-members-through-a-culture-of-allyship-2/
\BIBentrySTDinterwordspacing

\bibitem{rodeghero21}
P.~Rodeghero, T.~Zimmermann, B.~Houck, and D.~Ford, ``Please turn your cameras
  on: Remote onboarding of software developers during a pandemic,'' in
  \emph{2021 IEEE/ACM 43th International Conference on Software Engineering:
  Software Engineering in Practice Track (ICSE-SEIP)}.\hskip 1em plus 0.5em
  minus 0.4em\relax IEEE, 2021.

\end{thebibliography}

\end{document}